\documentclass{article}
\usepackage{arxiv}

\usepackage[utf8]{inputenc} % allow utf-8 input
\usepackage[T1]{fontenc}    % use 8-bit T1 fonts
\usepackage{hyperref}       % hyperlinks
\usepackage{url}            % simple URL typesetting
\usepackage{booktabs}       % professional-quality tables
\usepackage{amsfonts}       % blackboard math symbols
\usepackage{nicefrac}       % compact symbols for 1/2, etc.
\usepackage{microtype}      % microtypography
\usepackage{lipsum}		% Can be removed after putting your text content
\usepackage{graphicx}
\usepackage[numbers]{natbib}
\usepackage{doi}
\usepackage{subfigure}
\usepackage{array}
\newcolumntype{K}[1]{>{\centering\arraybackslash}p{#1}}
\usepackage{color}
\usepackage{wrapfig}
\usepackage{blindtext}
\usepackage{titlesec}

\usepackage{relsize}
\usepackage[printonlyused,withpage,smaller]{acronym}

\titleformat{\paragraph}
{\normalfont\normalsize\bfseries}{\theparagraph}{1em}{}
\titlespacing*{\paragraph}
{0pt}{3.25ex plus 1ex minus .2ex}{1.5ex plus .2ex}

\title{Everything You wanted to Know about Smart Agriculture}

\author{ 		\href{https://orcid.org/0000-0002-8796-4819}{\includegraphics[scale=0.06]{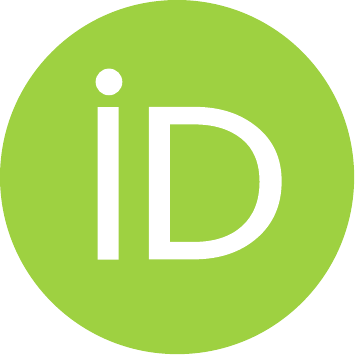}\hspace{1mm}Alakananda Mitra} \\ %\thanks{Use footnote for providing further information about author (webpage, alternative address)---\emph{not} for acknowledging funding agencies.} 
	Dept. of Computer Science and Engineering \\
	University of North Texas, USA \\
	\texttt{alakanandamitra@my.unt.edu} \\
	\And
	Sukrutha L. T. Vangipuram\\
	Dept. of Computer Science and Engineering \\
	University of North Texas, USA \\
	\texttt{lakshmisukruthatirumalavangipuram@my.unt.edu} \\
	\And
	\href{https://orcid.org/0000-0003-4567-7827}{\includegraphics[scale=0.06]{orcid.pdf}\hspace{1mm}Anand K. Bapatla} \\
	Dept. of Computer Science and Engineering \\
	University of North Texas, USA \\
	\texttt{anandkumarbapatla@my.unt.edu} \\
	\And
	Venkata K. V. V. Bathalapalli \\
	Dept. of Computer Science and Engineering \\
	University of North Texas, USA \\
	\texttt{ venkatakarthikvishnuvardbathalapalli@my.unt.edu} \\
	\And
	\href{https://orcid.org/0000-0003-2959-6541}{\includegraphics[scale=0.06]{orcid.pdf}\hspace{1mm}Saraju P. Mohanty} \\
	Dept. of Computer Science and Engineering \\
	University of North Texas, USA \\
	\texttt{ saraju.mohanty@unt.edu} \\
	\And
	\href{https://orcid.org/0000-0002-1616-7628}{\includegraphics[scale=0.06]{orcid.pdf}\hspace{1mm}Elias Kougianos} \\
	Dept. of Electrical Engineering\\
	University of North Texas, USA \\
	\texttt{elias.kougianos@unt.edu} \\
	\And
	Chittaranjan Ray \\
	Dept. of Civil and Environmental Engineering \\
	University of Nebraska-Lincoln, USA \\
	\texttt{cray@nebraska.edu}
}

\begin{document}
	
\maketitle

\begin{abstract}
The world population is anticipated to increase by close to $2$ billion by $2050$ causing a rapid  escalation of food demand. A recent projection shows that the world is lagging behind accomplishing the ``Zero Hunger'' goal, in spite of some advancements.  Socio-economic and well being fallout will affect the food security. Vulnerable groups of people will suffer malnutrition. To cater to the needs of the increasing population, the agricultural industry needs to be modernized, become smart, and automated. Traditional agriculture can be remade to efficient, sustainable, eco-friendly smart agriculture by adopting existing technologies. In this survey paper the authors present the applications, technological trends, available datasets, networking options, and challenges in smart agriculture. How Agro Cyber Physical Systems are built upon the Internet-of-Agro-Things is discussed through various application fields. Agriculture $4.0$ is also discussed as a whole. We focus on the technologies, such as \ac{AI} and \ac{ML} which support the automation, along with the \ac{DLT}  which provides data integrity and security. After an in-depth study of different architectures, we also present a smart agriculture framework which relies on the location of data processing. We have divided open research problems of smart agriculture as future research work in two groups - from a technological perspective and from a networking perspective. \ac{AI}, \ac{ML},  the blockchain as a \ac{DLT}, and \ac{PUF} based hardware security fall under the technology group, whereas any network related attacks, fake data injection and similar threats fall under the network research problem group.  
\end{abstract}

%\begin{IEEEkeywords}
%\end{IEEEkeywords}

%%%%%%%%%%%%%%%%%%%%%%%%%%%%%%%%%%%%%%%%%%%%%%%%%%%%%%%%%%%%%%%%%%%%%%%%%%%
\section{Introduction}
\label{sec:intro}

%https://www.worldgovernmentsummit.org/api/publications/document?id=95df8ac4-e97c-6578-b2f8-ff0000a7ddb6
%https://www.gaia-growth.com/post/why-do-we-need-smart-agriculture

The world population is anticipated to reach $9.7$ billion by $2050$ and could reach $11$ billion by the end of this century \cite{un}. Given these projections, it is anticipated that worldwide food consumption will increase at a rapid pace. The needed increase in food production to serve the future population is a tremendous task. Escalating food supply production is only possible through sustainable and smart agriculture. A goal has been set to end hunger all over the world by 2030. But currently, we are not in a trajectory to reach that goal \cite{food, malnutrition}. Today $800$ million people are undernourished worldwide. Increased population plays a significant role in this issue. More people means more food. By $2050$, $70\%$ more food production is needed to adequately feed the world's population \cite{wgs}.  A number of other factors are aggravating this situation:

\begin{itemize}
	\item Urbanization is changing food habits. People are consuming more animal protein. In $1997$-$1999$ annual  animal protein per capita consumption was $36.4$Kg which will increase to $45.3$Kg by $2030$. 
	\item Natural resources are being depleted. Farming lands are turning into lands unsuitable for agriculture. $25\%$ of the current farming land is highly unsuitable and $44\%$ is moderately unsuitable. Water scarcity has turned $40\%$ of the farming land into barren land.
	\item Deforestation for urban expansion and new farmland is rapidly depleting groundwater. 
	\item Over farming is leading to short fallow periods, lack of crop rotation, and livestock overgrazing causing soil erosion.
	\item Climate change is happening rapidly. It is affecting every aspect of food cultivation. Over the past $50$ years,  greenhouse gas emissions have doubled  which results in unpredictable precipitation and increased occurrence of droughts or floods. 
	\item Food wastage is another contributing factor. $33\%$ to $50\%$ of the food produced is wasted across the globe.  
\end{itemize} 

\begin{figure}[htbp]
	\centering
	\includegraphics[width=0.55\textwidth]{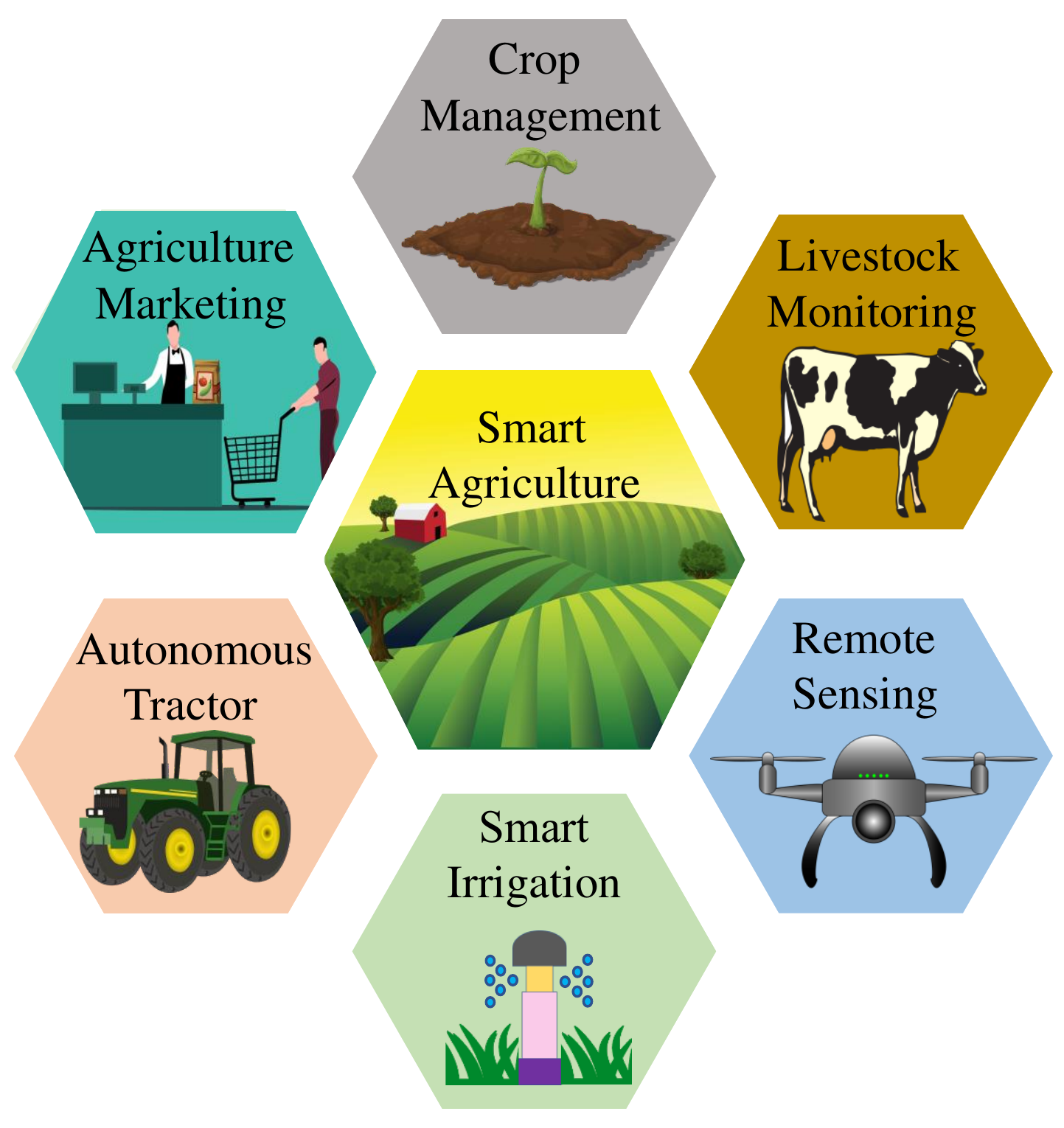}
	\caption{Smart Agriculture Overview.}
	\label{fig:Agri}
\end{figure}

To alleviate these issues, the food and agricultural industries welcome ``Agriculture $4.0$'', a green, smart revolution with science and technology at its core. Fig.\ref{fig:Agri} shows an overview of smart agriculture.

%Among the $17$ goals of the agenda for Sustainable Development by UN when  The growth of population will vary vastly across the globe. Some countries will face a dip in the population but $9$ countries will comprise more than half of the expected growth.  

\begin{figure}[htbp]
	\centering
	\includegraphics[width=0.85\linewidth]{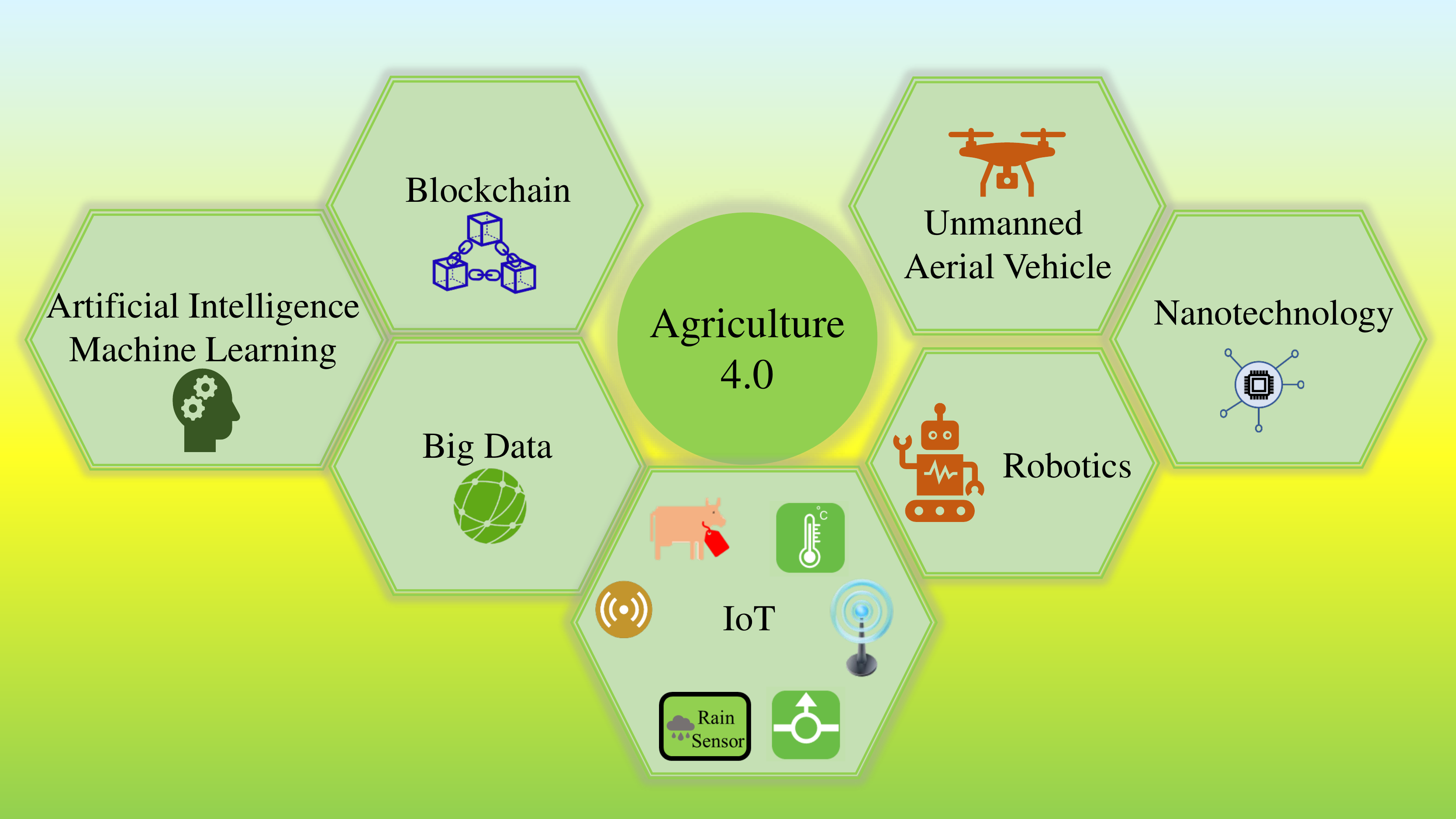}
	\caption{Elements of Agriculture 4.0.}
	\label{fig:agri_4}
\end{figure}

If we travel back through the industrial revolution, we see that it actually started in the Neolithic and Copper Ages when ancient people used wood and rock as instruments and later adopted metals  for farming. But \textit{Industry 1.0} started with the use of the steam engine. Mass production and use of electrical energy initiated \textit{Industry 2.0}. \textit{Industry 3.0} comes with automation and use of information technology whereas \textit{Industry 4.0} connecting the machines and nodes in cyber physical systems through \ac{AI}, \ac{BD}, the \ac{IoT}, robotics, etc. \cite{Liu_2021}. A parallel agricultural revolution also happened - starting with indigenous tools in \textit{Agriculture $1.0$}, use of tractors and fertilizers in \textit{Agriculture $2.0$}, decision and monitoring systems in \textit{Agriculture $3.0$} and smart farming or smart agriculture in \textit{Agriculture $4.0$} \cite{Liu_2021}. 
Agriculture $4.0$ is defined by the amalgamation of various technologies, e.g. the \ac{IoT}, \ac{AI}, the blockchain, the use of \ac{UAV}, nanotechnology, and robotics, as shown in Fig. \ref{fig:agri_4}.

The rest of this survey is organized into eight sections. Section \ref{Sec:Need_SA} presents the importance of smart agriculture. Section \ref{Sec:Architecture} presents the smart agriculture architecture. Internet-of-Agro-Things (IoAT) based Agriculture Cyber Physical Systems (A-CPS) is discussed in Section \ref{Sec:IoT_Smart_Agriculture}. Various applications of smart agriculture are described in  Section \ref{Sec:Applications}. Challenges the industry faced are depicted in Section \ref{Sec:Challenges_Opportunities}. Section \ref{Sec:Technologies} describes different technologies adapted in smart agriculture, whereas Section \ref{Sec:Agriculture_Datasets} discusses available datasets in the agricultural industry. Section \ref{Sec:Research_Problems} talks about the open research problems for the future and finally Section \ref{Sec:Conclusions} concludes the paper. A list of acronyms used in the paper is appended at the end of the paper.%, in Section \ref{sec:acronyms}.

%%%%%%%%%%%%%%%%%%%%%%%%%%%%%%%%%%%%%%%%%%%%%%%%%%%%%%%
\section{Smart Agriculture and Why Do We Need It?}
\label{Sec:Need_SA}

Traditional agriculture with manual labor and low productivity is being transformed into sustainable, intelligent, efficient, and eco-friendly agriculture with the use of technologies such as those depicted in Fig. \ref{fig:agri_4}. Long established, old-world agriculture is transmuting to ``smart'' agriculture. New terminologies are emerging - ``smart farming,'' ``digital farming,'' ``precision farming.'' ``Smart Farming'' is  another name for ``Smart Agriculture.'' In ``Smart Farming'' the focus is accessing data and applying those data to optimize a complex system towards increasing the quality standards and yield of the produce along with reducing human labor. 

``Precision Farming or Agriculture'' and ``Digital Farming'' are mostly the predecessors of ``smart agriculture'' \cite{mohanty2021internet}. When the goal of farming is optimization, accuracy, and customized solutions for a particular field or crop with the help of different technologies, it falls under the label  ``Precision Farming or Agriculture.'' ``Digital Farming'' is the combination of these two. In this paper, we will discuss ``smart agriculture" which addresses ``Agriculture 4.0'' and its future.

\begin{figure}[htbp]
	\centering
	\includegraphics[width=0.90\linewidth]{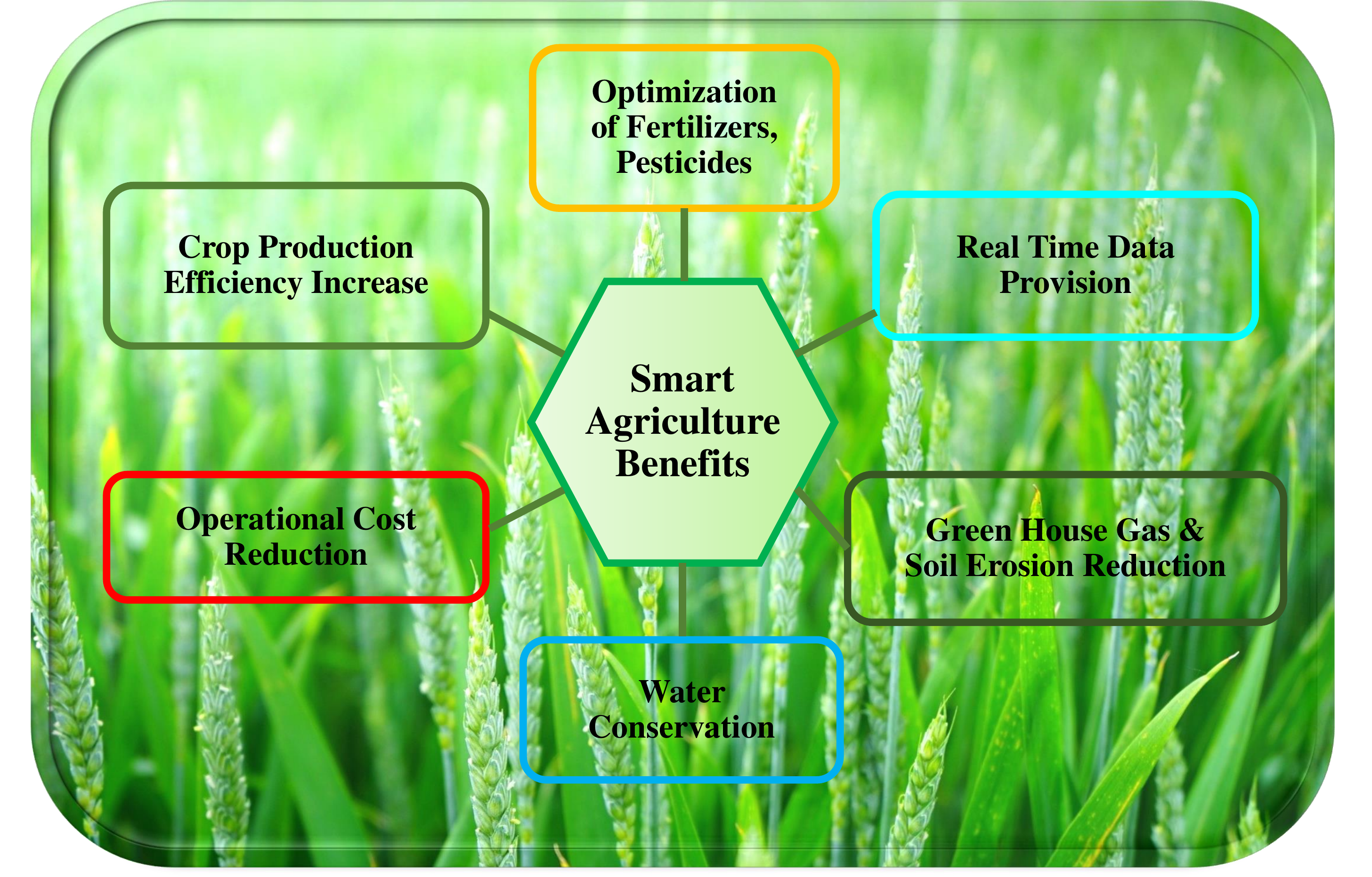}
	\caption{Smart Agriculture Benefits Over Traditional Agriculture.}
	\label{fig:benefits}
\end{figure}

Fig. \ref{fig:benefits} shows the tremendous benefits of smart agriculture compared to traditional agriculture. They are: 
\begin{itemize}
	\item Water conservation. 
	\item Optimization of the use of fertilizers and pesticides. As a result, produce are more toxin free and nutrient rich.
	\item Increased crop production efficiency. 
	\item Reduction of operational costs. 
	\item Opening up of unconventional farming area in cities, deserts.
	\item Lower greenhouse gas emissions.
	\item Reduced soil erosion.
	\item Real time data availability to farmers.	
\end{itemize}

%%%%%%%%%%%%%%%%%%%%%%%%%%%%%%%%%%%%%%%%%%%%%%%%%%%%%%%
\section{Smart Agriculture Architecture}
\label{Sec:Architecture}

\begin{figure}[htbp]
	\centering
	\includegraphics[width=0.90\textwidth]{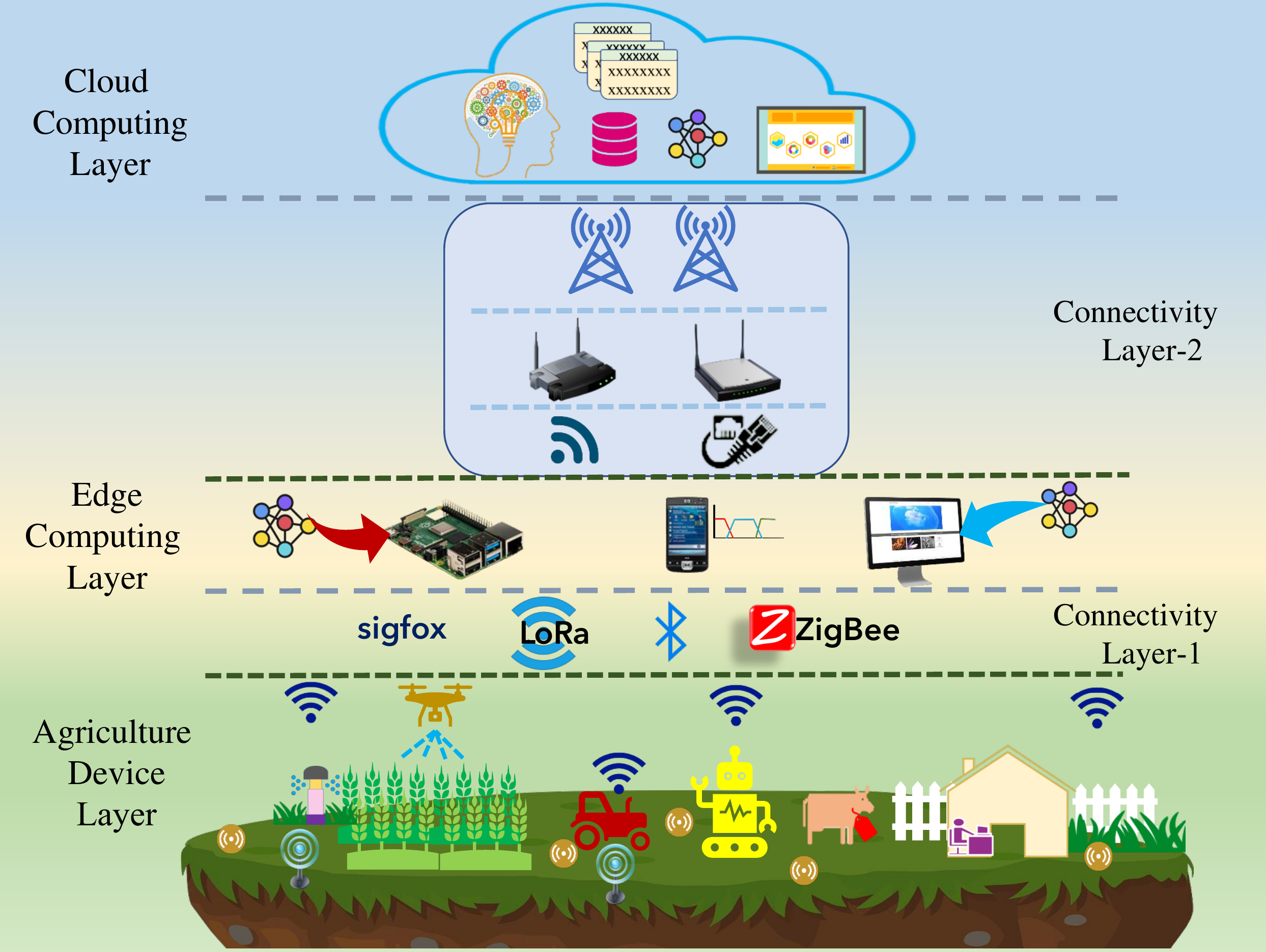}
	\caption{Architecture of Smart Agriculture.}
	\label{fig:smart_agri}
\end{figure}

The \ac{IoT} has  been recently boosting the agricultural industry. Different technologies, protocols and standards are being employed. Depending on the application, the number of associated layers varies in the implementation architecture. Smart agriculture architectures with three layers \cite{Khattab2016ICM,Na2016IOTA,Villa2020Internet}, four layers  \cite{Ferrandez2016Developing,Gupta2020Access}, five layers \cite{Ferrandez2018Precision}, six layers \cite{Ray2017Internet}, and seven layers \cite{Koksal2019Architecture} have been presented in the literature.  Different names and perspectives have been used in those architectures. We adapt a generic architecture, %similar approach like \cite{Ferrandez2016Developing}, 
where layers are defined depending on the location (proximity to the occurrence) of their processing and how are they connected. This smart agriculture architecture is shown in Fig. \ref{fig:smart_agri}. %But we tailor the architecture \cite{Ferrandez2016Developing} with two \textit{connectivity layers} as the network layer in \cite{Gupta2020Access} instead of a spreading communication layer. 

We depict the architecture with three main layers. These layers are connected through two \textit{connectivity layers}. We divide them in two sub layers as both connectivity layers connect different layers with different technologies. As the connectivity layer establishes a bridge among all the layers, %it spreads along the whole structure or in other words 
it is the core layer of smart agriculture architecture through which all the layers work in sync with each other.  

\begin{itemize}
	\item \textbf{Layer-1:} \textit{Agriculture Device Layer} is the base layer of the smart agriculture architecture. It comprises of various things like sensors, laid out through the agricultural land, animal paddocks, green houses, hydroponic systems, tagged livestock, unmanned aerial vehicles, agricultural robots, automated fencing and tractors \cite{Livestock_Sensors, Cropfield_Sensors}. 
	
	These devices or distributed source nodes sense the physical parameters, collect data round the clock in real-time and send them to the gateway node at next layer through the connectivity layer, which is basically a \textit{\ac{WSN}}. Fig. \ref{fig:sensor_parameters} shows the data sensed by different sensors/cameras in various fields of smart agriculture. For example, in a rice crop field, underground soil sensors and on-\ac{UAV} sensors and cameras collect the data  and send them to the edge for further processing. %Once processed at the edge, the prognostics is performed and solution is suggested. 

	\begin{figure}[thbp]
		\centering
		\includegraphics[width=0.98\textwidth]{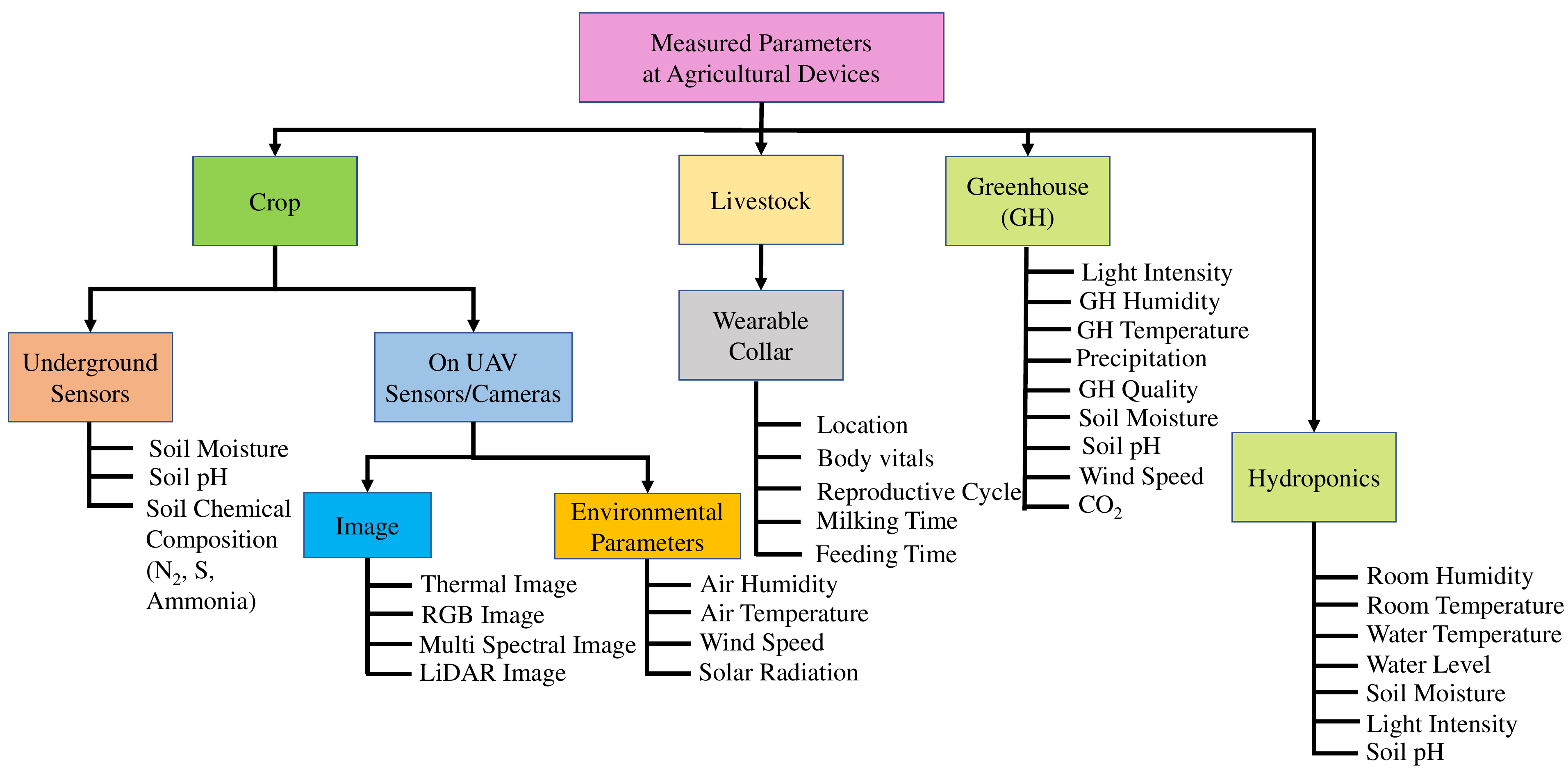}
		\caption{Sensor Parameters in Various Sectors of Smart Agriculture.}
		\label{fig:sensor_parameters}
	\end{figure}

	\item \textbf{Layer-2:} This is the \textit{Edge Computing Layer}. This comprises of a number of edge nodes. The number of nodes depends on the specific smart agricultural system. Data collected at layer-1 are processed, filtered and encrypted here. Previously, the prognostic and solution parts were done in the next layer because of the resource limitations at the edge layer. But with recent advancement of hardware and AI at the edge initiatives, trained machine learning models can perform the prognostics and suggest solutions at this layer. However, if the job is resource expensive or not time sensitive, prognostic and inference can both be done in the next layer. For example, if a cow is outside its supposed territory in a livestock farm or needs milking, the necessary measures are performed at the edge computing layer and the farmer is notified. 
	
	Hardware boards are being used as the edge devices \cite{Friha2021CAA}. To mention a few common boards and applications, the \textit{Arduino UNO}, has been used in \cite{Shirsath2017IoT} for a greenhouse monitoring and controlling system, the \textit{Raspberry Pi} for a hydroponic system \cite{Mehra2018IoT}, the \textit{ESP8266} for connecting smart agricultural components for managing ambient factors \cite{Khoa2019Smart}, the \textit{ESP32} for smart irrigation \cite{Biswas2018Solar}, the \textit{Intel Edison} for vertical agricultural warehouses \cite{Bhowmick2019Application}, and the \textit{BeagleBone} for monitoring of agrochemical processes \cite{Ali2018Precision}.  
	
	\item \textbf{Layer-3:} The Cloud Computing Layer is the third or topmost layer of the bottom up architecture of the smart agriculture system. This virtual layer usually resides in data centers and can be accessed from anywhere in the world through the Internet \cite{Gupta2020Access}. Massive data, collected by the sensors or cameras in agricultural farms need to be processed, analyzed and saved. Until recently, most of the analysis and decision taking were done at the cloud along with storing the huge data sets \cite{Khattab2016ICM,Na2016IOTA,Lopez2017Software}. %The processing was done at the edge level only. 
	The high computing power of the cloud allows it to perform various complex tasks in reasonable time. But there are certain limitations of cloud computing which demand new computing paradigms to emerge. Latency, high band width Internet requirements, security and privacy of data are some of the limiting factors which restrain the time sensitive monitoring and managing of smart agriculture. 
	
	\ac{AI}, recent developments in hardware boards and 5G network have orchestrated a new paradigm, the \textit{Edge AI}. It increases the security and privacy of data as it processes data near its point of origin. So data does not travel to the cloud or is shared at the centralized cloud. Edge AI has reduced the latency and dependence on the Internet.  
	
	\item \textbf{Connectivity Layers:} They bridge the various layers. \textit{Connectivity layer-1} gets the physical parameter data from layer-1 and passes them to layer-2. Processed data from layer-2 are passed to layer-3 by \textit{Connectivity layer -2}. Various transmission range communication networks are used in this layer depending on the area to be connected as in Fig. \ref{fig:connectivity_layer}. When data is transferred from the agriculture device layer to the edge computing layer, near range ZigBee, Wi-Fi, Z-Wave, Bluetooth, \ac{RFID}, and \ac{NFC} are commonly used, whereas for longer ranges SigFox, LoRaWan, and \ac{NB-IoT} are used \cite{Friha2021CAA}. For example, for a smaller farm in a remote village where the network bandwidth is limited low battery consuming Z-Wave is a good choice. But for larger farms, LoRaWan is suitable for its low energy usage and long distance transmission capability. Bluetooth low power has been used for monitoring soil and air along with water management systems in \cite{Hernandez2018Design}, and ZigBee for managing an irrigation system in \cite{Mafuta2013Successful}. \ac{RFID} is extensively used in the smart agricultural industry \cite{Kodali2016Iot, Hamrita2005Development, Peets2009Rfid,Ruiz2011Role, Sjolander2011Wireless, Vellidis2008Real}. LoRa has been used for water management in \cite{Zhao2017Design}. 
	
	When the processed data are sent from the edge computing layer 	to the cloud layer, cellular technologies like \ac{GPRS}, \ac{LTE}, 3G/4G, and 5G are used. The recent 5G technology has low latency, high reliability, large coverage areas, high data rate and new frequency bands \cite{Villa2020Internet}. This can greatly assist smart agriculture to advance.  \ac{GPRS} has been used for irrigation in \cite{Lopez2017Software}. New initiatives have started using 5G \cite{Faraci20185g,Alahmadi2018Wireless}. The successor of the 5G network is 6G cellular technology which is under development. It will be much faster than the existing mobile networks. Flexible decentralized models will propel various areas like edge computing, \ac{AI}, and blockchain which will advance the growth of smart agriculture. 
	
	\begin{figure}[htbp]
		\centering
		\includegraphics[width=0.8\linewidth]{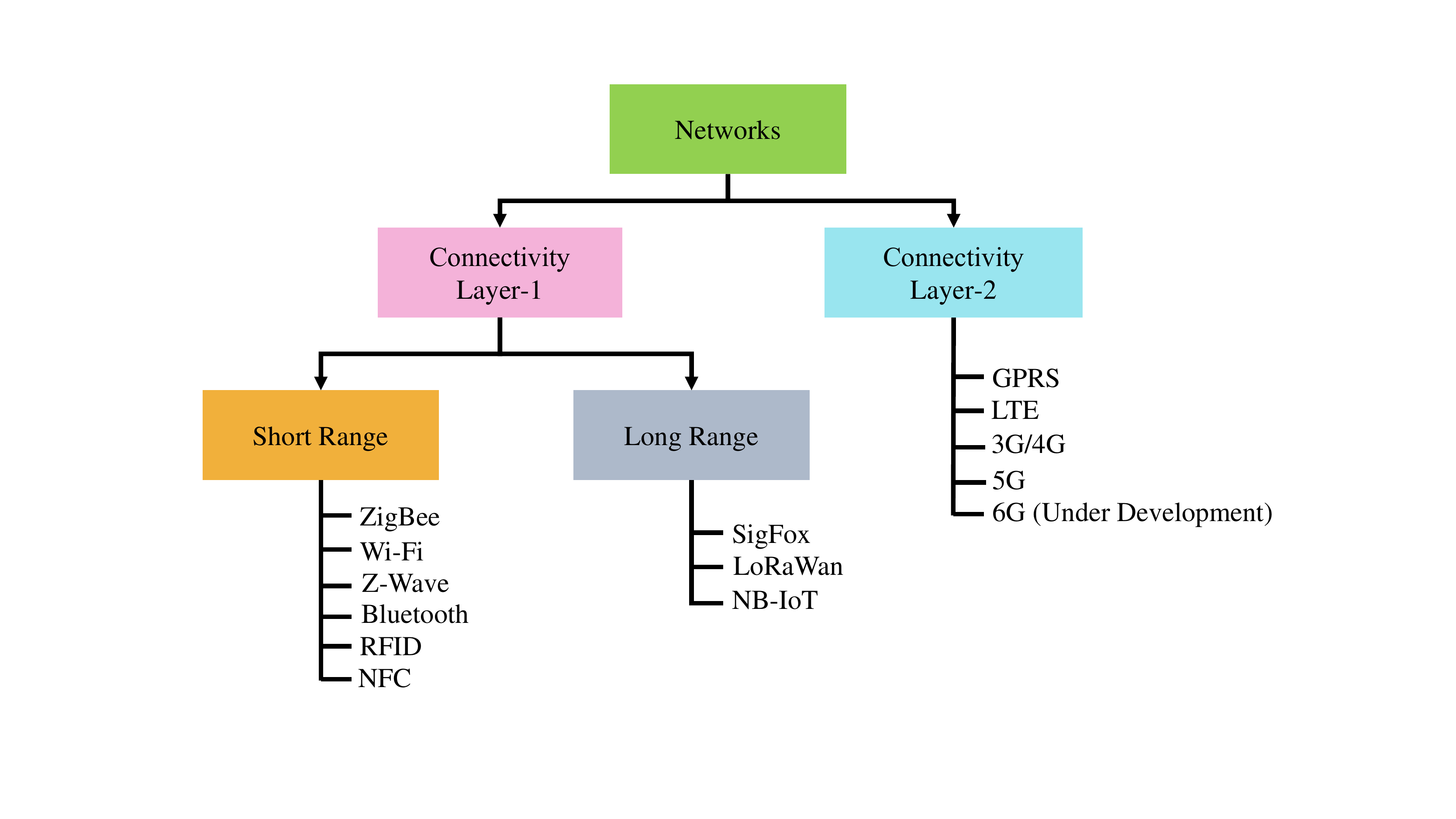}
		\caption{Various Networks for Smart Agriculture.}
		\label{fig:connectivity_layer}
	\end{figure}
	
	%\item Fog Computing Layer (Optional): The term \textit{Fog Computing} is coined by Cisco. The idea is to push the cloud computing part near but not to the edge nodes. This idea is mostly applicable to the enterprise network. 
\end{itemize}

%%%%%%%%%%%%%%%%%%%%%%%%%%%%%%%%%%%%%%%%%%%%%%%%%%%%%%%%
\section{\ac{IoAT} based \ac{A-CPS}}
\label{Sec:IoT_Smart_Agriculture}
The \ac{IoT} is the network of interrelated physical things, devices, objects with unique id for connecting and sharing data with other devices and systems through the Internet. Implementation of the \ac{IoT} in physical systems gives birth to \ac{CPS}. \ac{CPS} are hybrid systems of physical entities and software or computing capabilities. It is a modern way to define an industry. 
Smart cities and smart villages include one or more \ac{CPS} like smart health, smart agriculture, smart energy, smart transportation, smart citizens, renewable energy, etc. \ac{A-CPS} is the core of smart agriculture. It revolutionizes the agricultural industry. As the \ac{IoMT} forms \ac{H-CPS}, the \ac{IoAT} forms \ac{A-CPS} \cite{mohanty2021internet}. 

\begin{figure}[htbp]
	\centering
	\subfigure[t][Before TinyML Era]
	{\includegraphics[width=0.48\linewidth]{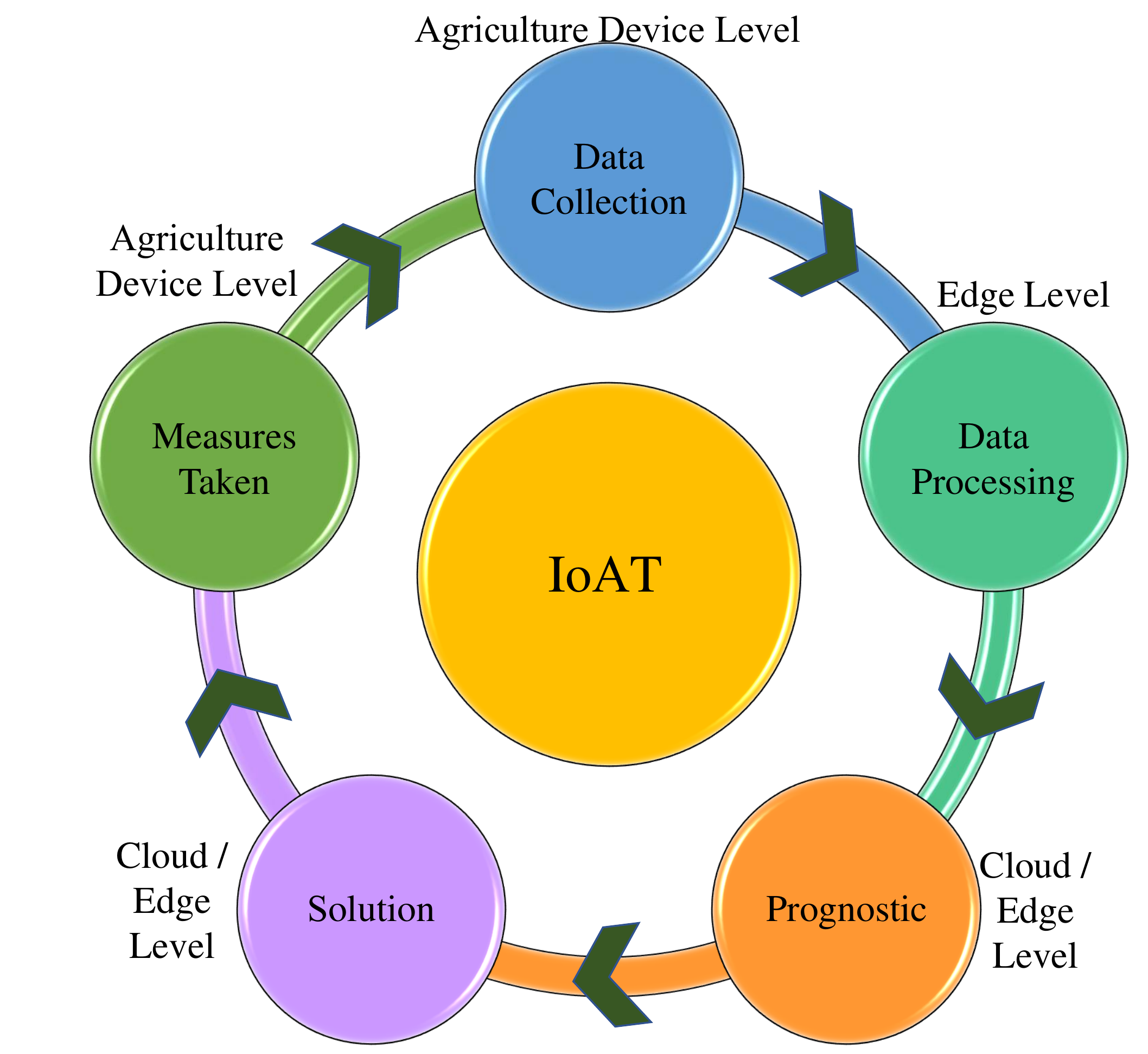}
		\label{fig:cycle}}
	\subfigure[t][After TinyML Era]
	{\includegraphics[width=0.48\linewidth]{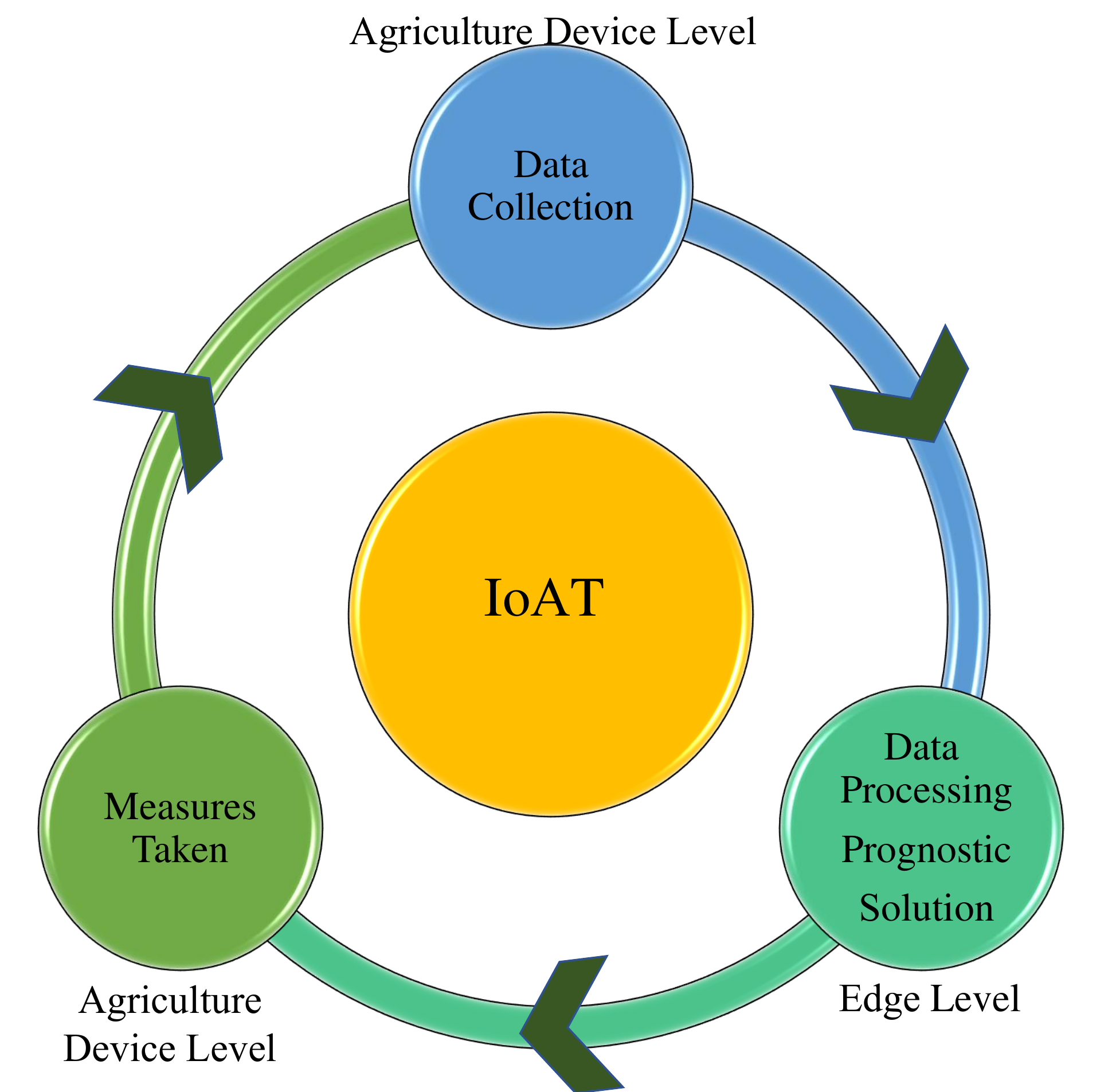}
		\label{fig:tinyml}}
	\caption{IoT Based Smart Agriculture Cycle.}
\end{figure}

The \ac{IoAT} is a data driven system. Constant data collection, processing, and measures are taken to make the workflow highly efficient. Fig. \ref{fig:cycle} shows such an iterative system  workflow which allows  farmers to take actions readily if any issue is observed. The cycle consists of five stages: 
\begin{itemize}
	\item \textbf{Data Collection:} First, various things (``T'') or sensors connected through the Internet (``I'') collect the data at sensor level or end level. 
	
	\item \textbf{Data Processing:} Second, if any data processing is needed to make the data compatible to the model, it is done in this phase at the edge level. E.g. if the sensor data is not in range or the photos taken by an \ac{UAV} are needed to be changed to gray scale or any encryption of data is needed before sending to the cloud, they are performed here. 
	
	\item \textbf{Prognostic:} This is mostly done in the clouds for existing technologies. The data, processed at the edge, are  analyzed here from the predefined rules or models (mostly \ac{ML}, \ac{FL}, and \ac{ANN} based). This is where data is stored for future use. The \textit{edge AI} initiative is transforming the scenario.  
	
	\item \textbf{Solution:} Once the issue is detected in the cloud platform, the solution is suggested. This stage can be done in the cloud or at the edge. E.g. if part of the farm land is dry, this stage suggests which valves of the irrigation system need to release what value and how long to optimally water the dry patch. 
	
	\item \textbf{Measures Taken:} This is the final stage of the cycle where the implementation of the solution is performed. This is performed by the IoT device. In the previously mentioned example, the opening of the valve of the irrigation system is performed here. 	
\end{itemize}

The cycle continues to serve the whole farming process optimally. Fig. \ref{fig:cycle} shows that when \ac{ML} based tasks are performed in the cloud and edge settings, the decision or solution is sent to the \ac{IoT} device for measures taken there. But, TinyML as-a-service is bridging the \ac{ML} and embedded worlds. Instead of ``outsourced'' the decision to the \ac{IoT} device, the \ac{ML} based task is being performed at the limited resource \ac{IoT} device only. In this new era, Fig. \ref{fig:cycle} is changing to Fig. \ref{fig:tinyml}.

%%%%%%%%%%%%%%%%%%%%%%%%%%%%%%%%%%%%%%%%%%%%%%%%%%%%%%%
\section{Smart Agriculture: Applications }
\label{Sec:Applications}

\begin{figure}[htbp]
	\centering
	\includegraphics[width=0.8\linewidth]{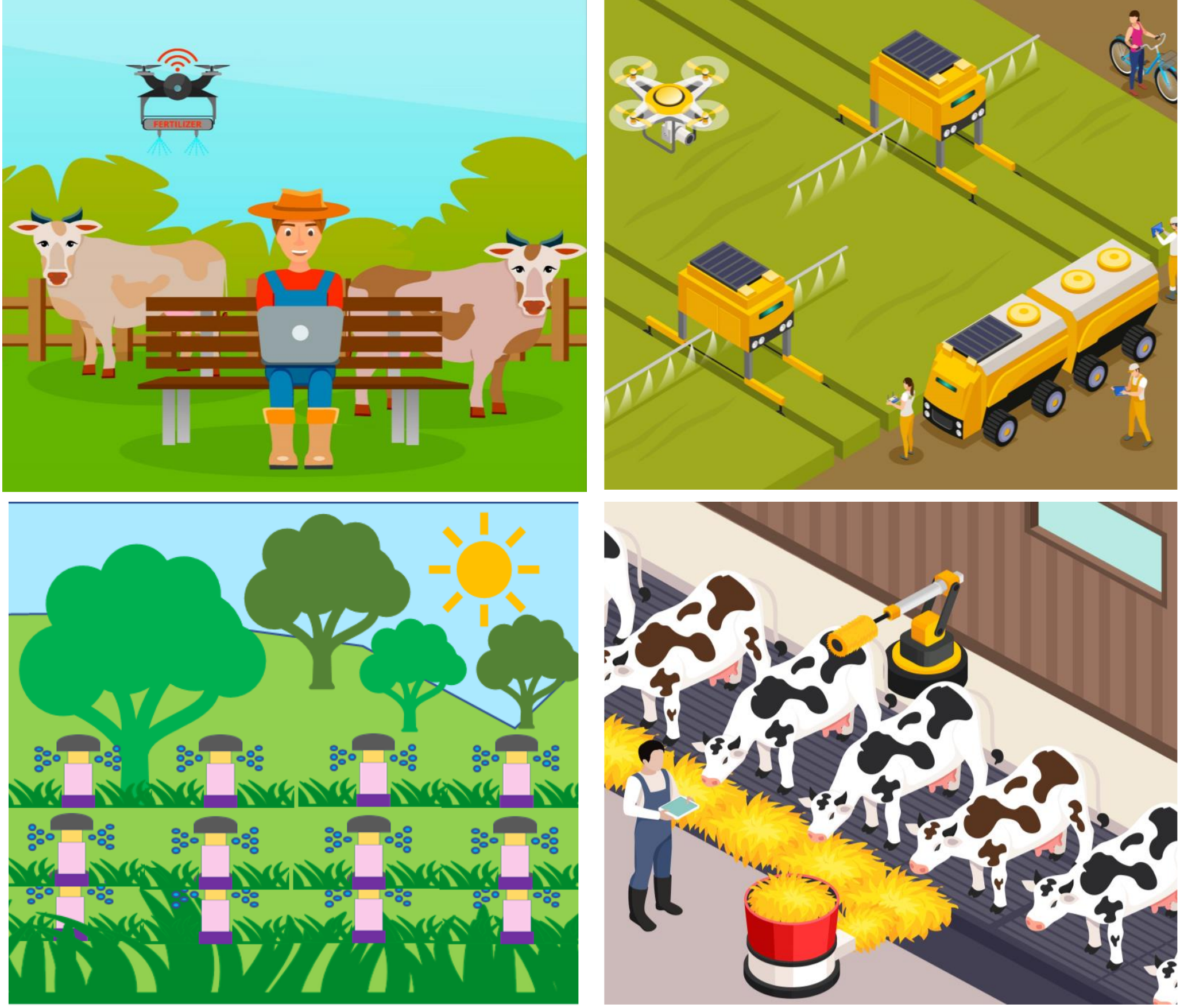}
\caption{Applications of Smart Agriculture - Crop Management, Pest Control,  Smart Irrigation, Livestock Monitoring  \cite{smart_app}.}
	\label{fig:smart_app1}
\end{figure}

\begin{figure}[htbp]
	\centering
	\includegraphics[width=0.8\linewidth]{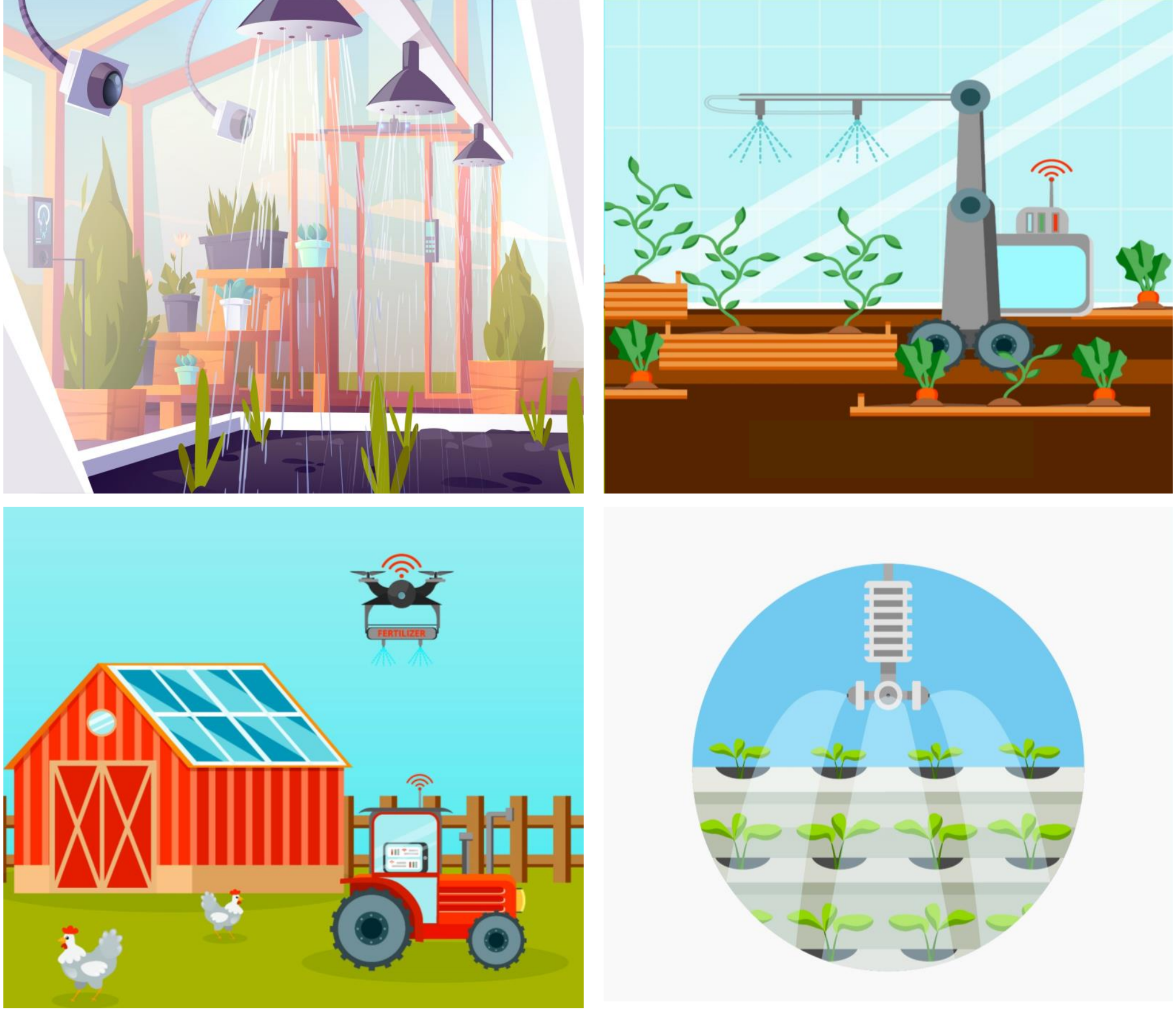}
\caption{Applications of Smart Agriculture - Smart Greenhouse, Agriculture Robot, UAV and Autonomous Tractor, Hydroponic System \cite{smart_app}.}
	\label{fig:smart_app2}
\end{figure}

In this section, application areas of smart agriculture are discussed. Fig. \ref{fig:smart_app1} shows some application areas of smart agriculture, e.g., crop management, smart irrigation, livestock monitoring, and pest control and Fig. \ref{fig:smart_app2} shows some more applications, e.g., smart greenhouse, \ac{UAV} and autonomous tractor, and hydroponic system.

%Smart Farming is an umbrella concept of IoT which uses modern technologies like AI, Machine learning to increase the crop yield with minimal human intervention by using super human accuracy and efficiency of Internet of Things devices. The usage of on field sensor nodes to collect the data and analyze it using super human intelligence and performing the process of actuation by processing the information using satellite imagery, weather data, weed and pest information using the full potential of unmanned Aerial vehicles, autonomous tractors and agricultural robots for accurate decision making and Field monitoring with the objective of achieving hundred percent efficiency in crop yield. Simply, Precision farming is usage of technology with artificial human intelligence for crop cultivation \cite{Bacco2019,Adamides,Kalyani}. An overview of Smart Agriculture is described in the Fig. \ref{fig:Agri}. 

%\begin{figure}[htbp]
%	\centering
%	\subfigure[b][Before TinyML Era]
%	{\includegraphics[width=0.75\linewidth]{Smart_Farming_Applications_1}
%		\label{fig:smart_app1}}
%	\subfigure[b][After TinyML Era]
%	{\includegraphics[width=0.75\linewidth]{Smart_Farming_Applications_2}
%		\label{fig:smart_app2}}
%	\caption{Application Areas of Smart Agriculture \cite{smart_app}}
%\end{figure}

\subsection{Crop Management}
Crop Management is the process of analyzing the economical, ecological and sociological aspects which constitute an important part in the crop selection, cultivation and marketing.

Crop growth, water resources availability, labor, insurance and environmental factors guide cropping patterns. Ecological factors contribute to a change in cropping patterns. For instance, in areas with depleting water resources and groundwater tables, traditional crops like paddy cultivation,  which requires abundant water resources, cannot be sustained. The market for an agricultural product, as well as different countries, export and import policies, also affect crop selection. Once a crop has been selected then crop cultivation is the next important aspect. 

Using the \ac{IoT}, farmers are equipped with the latest technology and sensors placed in-field monitor plant growth. For instance, ultrasonic sensors are placed in the field to monitor the presence of pests and insects affecting plant growth. After identifying the presence of pests, high frequency sound waves are generated to remove the pest and the farmer is also notified of the pests' presence for further help \cite{Vitali2021}.

\begin{figure}[htbp]
	\centering
	\includegraphics[width=0.99\textwidth]{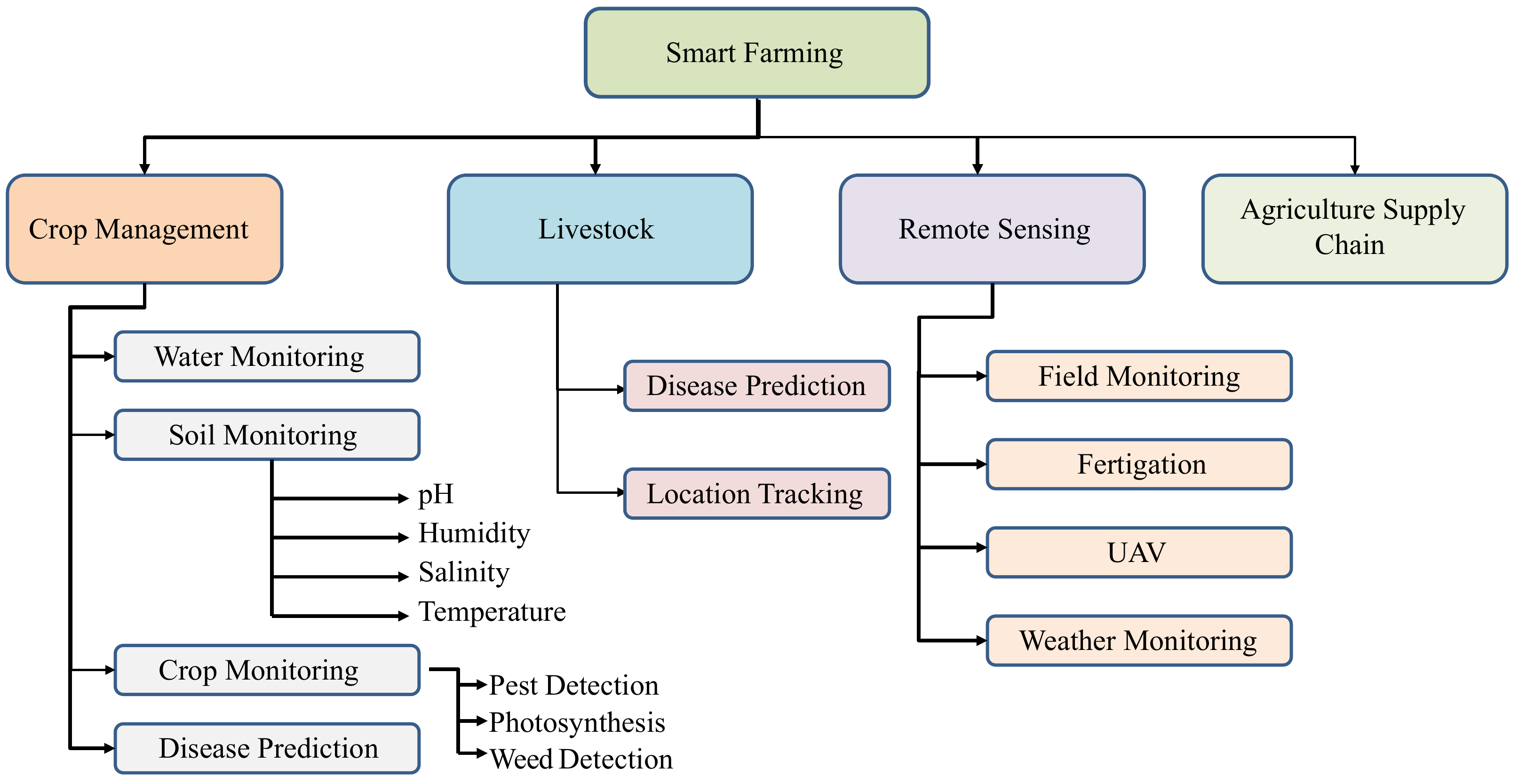}
	\caption{Applications of Smart Agriculture.}
	\label{fig:SF}
\end{figure}

The flowchart illustrating the concept of Smart Farming is detailed in Fig. \ref{fig:SF}.

\subsection{Soil Monitoring}
Soil moisture plays an important role in the overall farming process. It is responsible for photosynthesis, respiration , transpiration and transportation of minerals during the plant growth process \cite{balakrishna2016real}. Soil monitoring constitutes an important role in farm decision making. Cropping patterns depend on various factors like water availability, soil salinity, pests, moisture, pH and humidity. These factors help in assessing soil health. Sensors on the field monitor soil temperature and humidity and the analyzed data are sent to the cloud. Farmers will receive an alert on a range of factors and cropping patterns are analyzed and decided based on salinity content and soil nutrient level, humidity, and temperature. Soil moisture is a vital aspect in plant growth process as water is an important component in photosynthesis, regulating temperature and acting as a carrier of food and essential nutrients for the plant growth. Humidity controls the nutrient supply and regulates the rate of transpiration for optimum plant growth. The ideal humidity for vegetable plants cultivation is 50\% to 60\% \cite{Yaolin2011}. Soil moisture sensors placed inside the root of plants analyze soil moisture level values to facilitate optimum utilization of water resources \cite{Ma2012AgriculturalEI,Radha}.

\subsection{Smart Irrigation}
Smart irrigation is the process of improving the quality and quantity of yield with optimal utilization of water using the latest technologies. It conserves water by optimally watering the plants. There are two types of irrigation systems - weather based and soil moisture sensor based. Weather based irrigation systems receive temperature and rainfall data from a local mini weather station and a controller regulates the irrigation. In soil moisture sensor based irrigation systems, sensors placed inside the turf of trees, accurately determine soil moisture content. In this type of irrigation, accurate values of humidity and air temperature along with weather monitoring and cropping pattern are required to irrigate the field. Data is sent to the cloud and actuators like sprinklers are activated \cite{Kanchana2018}. The soil moisture sensor values guide the irrigation schedule per unit area of farm. The micro level analysis and scheduling of irrigation and efficient actuation ensures optimum crop growth and 100\% efficiency in water utilization \cite{Nagothu}. %and  watering of the field is done based on the commands for either Edge or Cloud.
Farmers can operate the irrigation system from a smartphone based mobile application. This irrigation system is based on data from temperature, humidity, soil moisture and ultrasonic sensors placed in the field \cite{Ogidan2019}. 
The smartphone based mobile application for automatic irrigation  is connected to the cloud for analysis using a user friendly mobile application where farmers can perform actuation by enabling the irrigation pumps to water the farm.

%The important aspects of smart irrigation are weather and soil moisture. Pre programmed irrigation controllers are facilitated with local weather data and depending on local climatic conditions the irrigation schedules are automatically adjusting in the Edge controller. 

\subsection{Livestock Monitoring}
Livestock management constitutes an important part of smart agriculture. %dairy industry and agro-economy. 
An \ac{IoT} enabled livestock health monitoring system enables farmers to monitor the health of cattle herds, track grazing animals, and optimize breeding practices. Cattle health can be monitored automatically by measuring body vitals like heart rate, blood pressure or respiratory rate using a wearable collar or \ac{RFID} tag. This has a two fold advantage -  saving man power and providing time sensitive treatment to the animal which in turn stops spreading of diseases. For this purpose, \ac{GPS} tracking is used \cite{Kanchana2018}. It also can prevent accidents to the animal. \ac{RFID} tags are also used in animal identification and tracking \cite{Wang2018}.  

\subsection{Remote Sensing}
Remote sensing in agriculture can help farmers receive real time data on the crop using drones which record high quality images to map the farm fields. They can also be used to check the crop yield using the information on crop health and the condition of farmland. Remote sensing can be used to map soil conditions and enable farmers to decide which type of soil is better for a particular crop. Weed and pests can be detected and proper pest control mechanisms can be adapted. The most important application of remote sensing is weather forecasting and monitoring. It can be used to track rainfall, drought conditions and in identifying water resources, thereby alerting farmers beforehand on availability of water and on weather so that capital and crop planning can be done in advance \cite{Sishodia2020}. \ac{NVDI} values, which are one of the most important parameters to quantify the crop cultivation process, are used to notify yield prediction and plant growth \cite{Teboh}. Remote sensing instruments on the farm field are used for monitoring abiotic stress agents with the best possible spatial resolution \cite{Jackson1986}. %Timble GreenSeeker is the used to measure NVDI Values using the equation (NIR-R) / (NIR +R) \cite{verhulst2010normalized}.

\subsection{Smart Greenhouse}
In the wake of global climate change, and diminishing natural resources the agricultural industry welcomes technology supported farming techniques. Smart greenhouse is one of them. It is an indoor controlled environment tailored for plants. It is a self-isolated farm monitoring ecosystem integrated with \ac{IoT}, and \ac{AI}/\ac{ML} technologies. It protects the farm from wind, storms, and floods. It increases the efficiency of productivity without manual intervention. 

Solar powered \ac{IoT} sensors are placed inside the greenhouse for monitoring the vitals for vegetables, fruits and other horticultural crops. Automatic drip irrigation can be employed using soil moisture sensors placed inside the root of the tree. If a threshold value is reached, the in-field actuator waters the farm accordingly. Use of LED lighting can better cater to the plants' needs. A controlled illumination with specific wavelength and intensity can revamp the plant growth and all year-round yield. 

Drip fertigation techniques can be used to sprinkle sufficient amounts of minerals like potassium, phosphorus and other minerals required for optimum growth and good health of plants. Smart greenhouse cultivation is increasing as the technologies are at the farmer's disposal and demand is growing for organic fruits and vegetables using smart green techniques \cite{Kodali2016}. %The concept of Smart Greenhouse is illustrated in the Fig. \ref{fig:Greenhouse_Smart}. 
A decision support based IoT friendly smart greenhouse system has been presented in \cite{tripathy2021mygreen} for increasing productivity of rose plants. 

%\iffalse
\begin{figure}[htbp]
	\centering
	\includegraphics[width=0.95\textwidth]{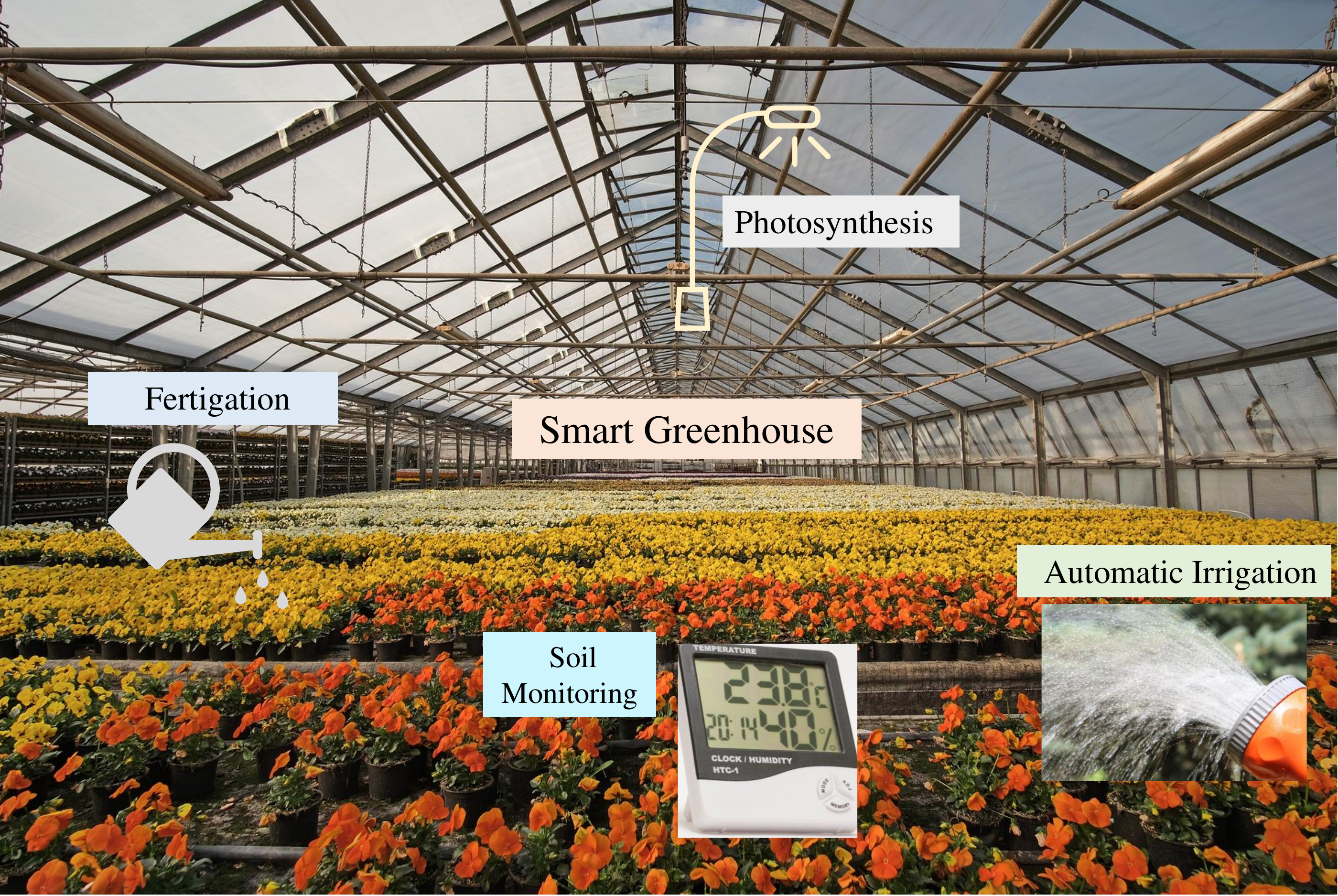}
	\caption{Smart Greenhouse.}
	\label{fig:Greenhouse_Smart}
\end{figure}
%\fi

%\begin{figure}[htbp]
%	\centering
%	\includegraphics[width=0.9\textwidth]{Greenhouse}
%	\caption{Smart Greenhouse Applications}
%	\label{fig:Greenhouse}
%\end{figure}

\subsection{Unmanned Aerial Vehicle}
In the current agricultural industry, the usage of \ac{UAV}, a.k.a drones, has been steadily increasing.  
%Smart Agriculture concept is based on the technological adaption and information interpretation and decision making. Climate change has propelled the advancement and adaption of Information and communication technologies into farming. The revolution of Drones is a perfect example which is a programmed autonomous, uncrewed aircraft used 
They are being used for crop mapping, field monitoring, remote sensing, fertigation, and weed detection. Drones can be a savior for taking photos in large farming areas, mountainous regions or remote areas.  %been widely used to take photos in civil and military applications also due to its ease of use which is the most vital aspect in the success of any technology. 
The \ac{NVDI} is calculated from the drone taken images to asses the crop health. It determines water level, stress condition, plant nutrition, and pest infestation. It can guide the entire crop cultivation process \cite{RadoglouGrammatikis2020, Friha2021, Muniasamy2020}.

\subsection{Autonomous Tractor}
Cutting edge technologies are changing the agriculture industry. The \ac{IIoT} has propelled from crop management, soil monitoring, smart irrigation to pest control, livestock management or agro marketing. We can expect in the near future farming with autonomous, intelligent, and smart instruments. An autonomous tractor is an important part of these instruments. It is a programmable self-driving vehicle. It can perform tillage, and spraying fertilizers. They are equipped with \ac{GPS}, lasers and cameras and can function on their own without requiring farmers to monitor them. Autonomous drones are used along with these smart tractors and are used in weed detection, pesticide spraying, field monitoring and surveillance for sustainable agriculture \cite{Gorli}. The autonomous tractors used for spraying and mowing in orchards have perception systems to detect obstacles and remote aided guide for performing agricultural tasks. The perception system uses cameras for  geometry based obstacle detection and path identification  \cite{moorehead2012automating,Virik}.

\subsection{Urban Farming}
The increased urbanization rate poses an alarming situation in densely populated cities. A new approach to farming has emerged to offer a sustainable farming solution in those areas.  
%As the urbanization is increasing at an unprecedented rate, the cultivable area is also reducing. 
Hence the practice of urban or vertical farming has gained prominence among urban %health conscious   
populations. It takes up 3-D space for farming with controlled water, nutrients, minimal pesticides, and artificial lighting sources. The practical limitation of vertical farming system is generation of artificial light sources for plant growth and the large costs involved \cite{Haris2019}.

\begin{itemize}
	
	\item As the name suggests, hydroponics is a water based system where plants get all the nutrients from the nutrient rich water solution \cite{article}. In hydroponic systems, the nutrient supply needs to be continuous. These systems can be operated through mobile apps. In \cite{Vidhya2018} such a mobile app controls an Arduino controller for watering the plants to manage the hydroponic system. 
	
	\item Aeroponics is a similar system but instead of submerging the roots in the water, the roots are misted. Research shows that aeroponics plants have more nutrients than hydroponics plants \cite{Vertical}. In the Internation Space Station, this technology is used for growing plants.
	
	\item  Another recent farming system is aquaponics which is essentially a hydroponic system but the nutrients (phosphorous, nitrogen) are not mixed from the outside. Fish in the same tank generate those nutrients. 
\end{itemize}

\subsection{Agriculture Marketing}
Proper marketing of the produce is an important aspect of the economic growth of a society. The presence of middle men causes inflation and both consumers and farmers lose.  %are buying at a lower price from the farmers and are selling it to the end consumers at very high price this is causing inflation and ultimately not benefiting the interests of either farmers or end consumers.
Smart agriculture changes this scenario. Farmers can sell the product directly to the consumers using various agro-marketing apps. Ethereum based blockchain  has been used as a platform for trade negotiations between farmers and end consumers \cite{Revathy2020}.
A food supply chain has been implemented with the help of blockchain \cite{Caro2018a} from updating the distributed ledger at the production phase to the final distribution phase.
%Digitization of Food Supply chain has become one of the most important applications of Blockchain in Precision Farming. The product goes from Production phase, where details on crop cultivation process, weather conditions and about fertilizers and pesticides used in the process are updated in the distributed ledger. The second phase is product labeling and packaging process where each crop is assigned a product label and is updated in the blockchain for future reference and the final phase is distribution where the buyer information is updated in the ledger thereby more transparent and efficient agricultural supply chain system is developed using the decentralized and immutable characteristics of Blockchain technology \cite{Caro2018a}.

%%%%%%%%%%%%%%%%%%%%%%%%%%%%%%%%%%%%%%%%%%%%%%%%%%%%%%%
%%%%%%%%%%%%%%%%%%%%%%%%%%%%%%%%%%%%%%%%%%%%%%%%%%%%%%%
\section{Smart Agriculture: Challenges}
\label{Sec:Challenges_Opportunities}

\begin{figure}[htbp]
	\centering
	\includegraphics[width=0.90\textwidth]{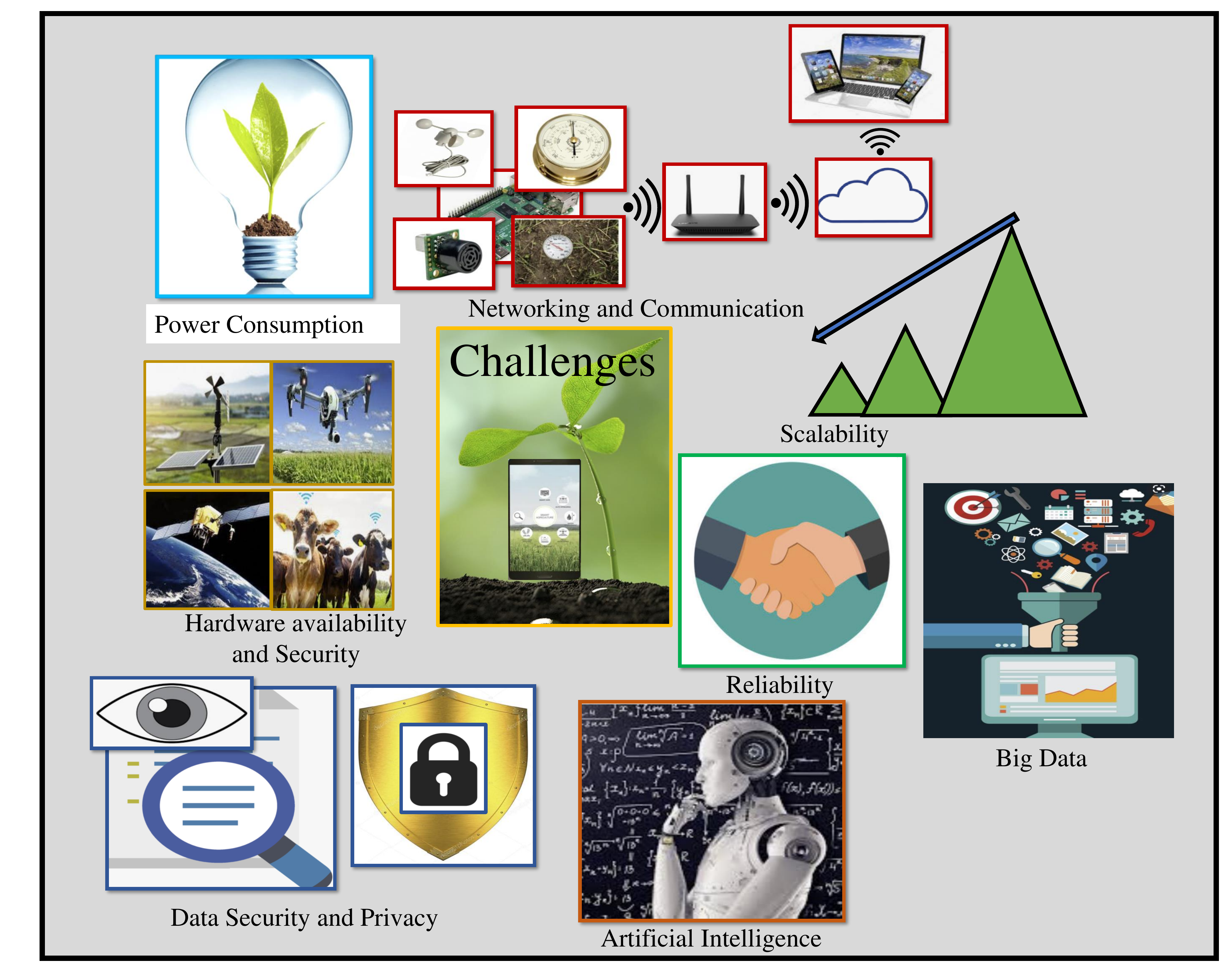}
	\caption{Major Challenges  in Smart Agriculture.}
	\label{fig:S_A_Challenges}
\end{figure}

Traditional Agriculture has been modernized and eased by Smart Agriculture processes. But there are still many challenges needed to be addressed for the adoption of technologies to scale. These issues are associated with a variety of aspects which are discussed in the current Section. Fig. \ref{fig:S_A_Challenges} shows some of  the major challenges of Smart Agriculture.

\subsection{Power Issues }
Most smart agriculture activities utilize large machine automation which requires high amounts of power to operate. As farms are generally vast in area and require many electronic components, it is not unusual to have very high power requirements. This has been a bottleneck for wide adoption of such automation processes in large farms. Some of the solutions propose the use of clean energy from renewable sources like solar, wind, and hydro and provide continuous uninterrupted power to the machinery \cite{Liu_CSEE_2018}. This has been an area of interest for many researchers and research is ongoing to implement and improve such renewable energy sources for smart farming \cite{Ram2021,Huang2020}.  Some of the issues with these alternative power options are the storage and transmission of power, along with uneven energy requirements at different location of the farm. An efficient microgrid architecture is required to overcome such issues and research has been done in this area to propose an efficient smart microgrid working along with renewable power sources in \cite{Desai_PVSC_2021,EBRAHIMI_IST_2019}.  

\subsection{Power Consumption} For a seamless, reliable and sustainable operation of the smart agriculture farm, as \ac{IoAT} devices are needed to be operated by alternative power sources, the deployed models are required to be less power hungry. They should be capable of working in a low resource setting. 

\subsection{Hardware Availability}
Smart Agriculture requires different sensors and devices for sensing different environmental and system parameters. After acquiring the data, the devices act upon those signals to give better predictable yield. Availability of specific hardware is a bottleneck in this scenario. % and designing such high precision hardware equipment is a challenging aspect which needs to be solved for providing better solutions to different processes in agriculture. 

\subsection{Hardware Security}
By 2020, the number of \ac{IoT} connected devices are believed to be $50$ billion \cite{Evans2011HowTN}. These \ac{IoT} devices are needed to be robust and resilient against various attacks. But demand of simple hardware with low price compromises hardware security. % another challenge. IoAT makes traditional agriculture automated and smart.  %With wide adoption of this technology  believed to be 50 billion by 2020 .Due to the price and simplicity constraints on IoT devices has led to some of the hardware security issues which can be used as entry point into the network for adversaries.
Hardware Trojan and Side Channel Attacks are the most common hardware security threats for \ac{IoT} devices, consequently limiting wide adoption of \ac{IoT} network in critical applications.  Hardware Trojans make use of malicious hardware modifications by the adversary which can be used as a backdoor to control the system and to perform attacks. These are very hard to detect and some methods include performing electronic microscope scanning on de-metalized chips \cite{Courbon_DATE_2015} and studying power and delays within the circuit and also inspecting the \ac{PUF} which acts as signature of these electronic devices \cite{Sankaran_ISES_2018}. Side Channel Attacks are another common hardware security threat which make use of side channel signals to retrieve confidential information like cryptography keys. Some of these side channel signals include electromagnetic emanation, power profiling and timing analysis \cite{Tang2018}. As \ac{IoT} networks are more prone to these issues, many solutions have been proposed in \cite{Ju_CSCI_2015,Chakraborty_ICCAD_2009, Bathalapalli2021}.

\subsection{Networking and Communication}
\ac{M2M} interaction is one of the most common aspects in smart agriculture. This makes use of different network and communication protocols to share data and work collaboratively towards a common task. Most of the applications make use of many different communication networks like ZigBee, Wi-Fi, LoRA, SigFox, and \ac{GPRS}. Establishing and maintaining such huge networks is expensive and not a viable option in large, open farms due to physical damages and threats. Research directions have been explored and some solutions for efficient communication networks have been proposed \cite{Concepcion_WiSNet_2014, Zhang_2011,Sahitya_iCATccT_2016}. Additionally, some research has integrated communication equipment with other smart devices making it viable for uninterrupted communication such as \ac{SIL} and \ac{WSN} to create a novel agriculture thing, SIL-IoT \cite{Yang_IoT_2020}. The need for more secure and robust communication is very high in smart agriculture applications and requires further research and new affordable technologies. 

\subsection{Connectivity Issues}
In many rural areas across the globe reliable high bandwidth Internet connection is not available, which stalls the existing cloud based computing and prevents the advancement of smart agriculture. Tall trees or hills can also stop the line-of-sight \ac{GPS} communication \cite{challenges}.

\subsection{Data Security and Privacy}
To maintain data privacy and security during data transmission, robust cryptography techniques and security measures are needed. However, due to the minimalist design of IoAT sensor nodes and underlying protocols they are not resource intensive. Practicing security measures in a resource limited device is difficult in today's existing technologies. Thus data privacy and security has become a serious challenge in smart agriculture. As most of the processes in smart agriculture are automated, an adversary can manipulate these processes to create chaos in the network. This may lead to very serious consequences for yield and overall quality of farm production.

\subsection{Scalability and Reliability}
Agricultural farms vary in their size from smaller individual farms to larger commercial farms. They need different quantities of field sensors. These sensors generate a varied amount of data. Hence, any agriculture technology is needed to be scalable. The devices are required to be reliable, so that the number of redundant devices to accommodate fault tolerance can be lower. It will significantly reduce the cost.   

\subsection{Big Data Challenge} 
\begin{figure}[htbp]
	\centering
	\includegraphics[width=0.99\linewidth]{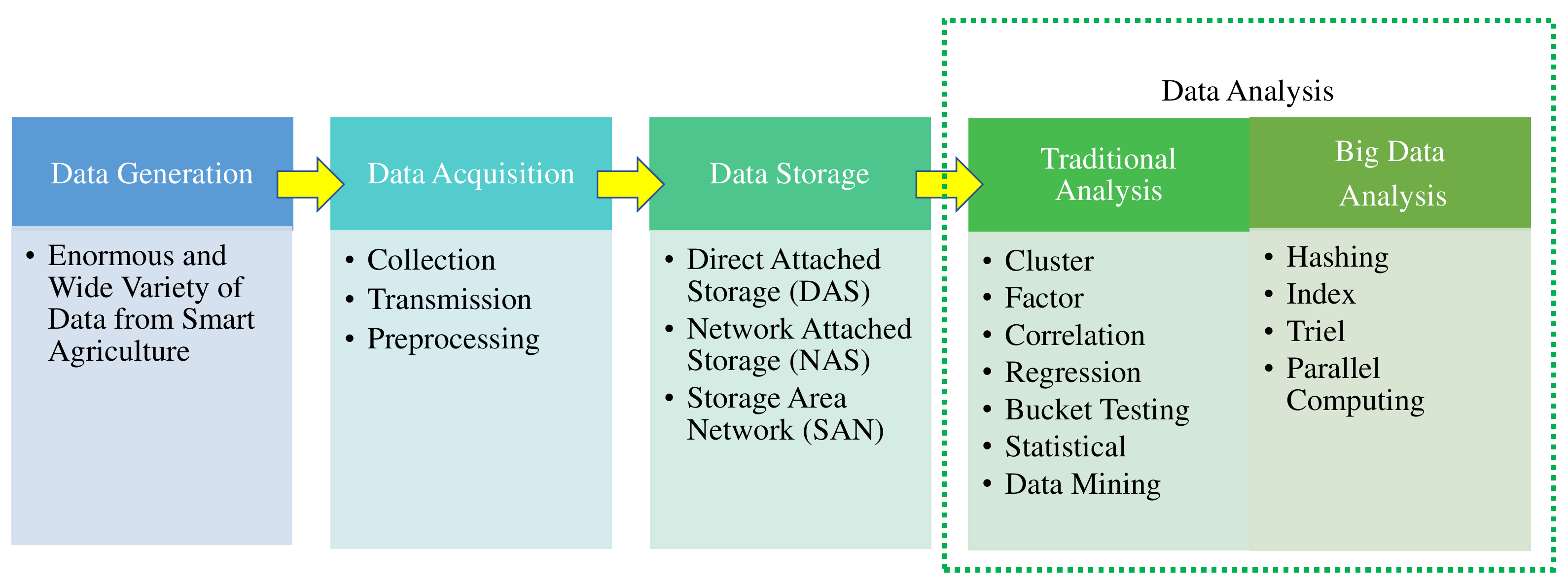}
	\caption{Big Data Work Flow in Context of Smart Agriculture.}
	\label{fig:big_data}
\end{figure}

Massive amounts of heterogeneous data is collected by the sensor nodes or cameras in smart agriculture. Traditional ways of processing this enormous amount of data are insufficient and \ac{BD} analysis comes into play. Big data has the capacity to explore massive datasets. It improves the efficiency of the end-to-end supply chain in smart agricultural systems, mitigates food security issues \cite{Chen2014Big}, provides predictive analysis, real time decision, and introduces new business models \cite{Poppe2015European, Wolfert2017Big}. \ac{SVM} and \ac{ANN} have been utilized to integrate big data platforms 
for milk production chain security \cite{Kempenaar2016Big}. Fig. \ref{fig:big_data} shows the big data workflow in smart agriculture systems based on \cite{Chen2014Big, Wolfert2017Big}. It starts with the data collection at various sensor nodes and ends with the various data analysis methods including both traditional and big data analysis. 

\subsection{Challenges of AI} 
Though \ac{AI} is a logical step toward smart agriculture for sustainable, efficient, and cost effective farming, there are some constraint factors which pose big challenges to applying \ac{AI} in the agricultural industry: 

%{\color{blue}{Add some citations }}
\begin{itemize}
	\item There is a lack of connection between the agricultural industry and AI research field. So, the problems faced by farmers are not well known to \ac{AI} researchers and similarly farmers are not well aware of the existing \ac{AI} technologies. More interdisciplinary collaboration is needed to solve this two-fold problem.
	
	\item As \ac{AI} applications in agriculture are emerging, there are no well established policies and regulations. Thus, many legal aspects of smart farming are unanswered. Until recently, most of the existing \ac{AI}-\ac{IoT}  solutions were cloud based and therefore cyber attacks, data security, and privacy concerns kept farmers away from embracing \ac{AI} techniques. To mitigate this issue, a new \ac{IoT} setting ``Edge AI'' has emerged. Edge AI processes sensor data at the local level, and it provides higher security and privacy in data along with lower latency and cost. 
	
	\item Another challenge for \ac{AI} in agriculture is lack of data. \ac{AI} is a data-driven technology. The unavailability of proper data is a barrier to applying various \ac{AI} techniques. 
	
	\item In remote rural areas where higher bandwidth mobile networks are not available but agriculture is the main industry, Edge AI can be a game changer there. It expands the possibilities of smart agriculture. \ac{CNN} have been used in \cite{Gia2019Edge} at the edge layer to compress the sensor image data and then the compressed data has been sent to the fog layer using \ac{LPWAN} technology.

	%	\item Change of Conditions: Conditions of agricultural field change so the AI model  e.g. the infested pest in a rice field changes from the initial one and the AI model which was working fine earlier might not work for the new pest if it s not trained with that category of pest.   
\end{itemize} 

\subsection{Technical Malfunction} Technical malfunction, e.g., sensor damage can disrupt the technology. A huge amount of loss from wrong decision-making of devices can introduce multi domain damage. For a paddy field, if the sensors are damaged by hail, they will not predict the water content of the soil correctly which in turn can damage crops, impact food supply chain, and cause rice price imbalance. 

\subsection{Lack of Initial Capital Investment} In rural areas of developing countries where farmers work with a very meager profit margin, initial investment for advanced technologies is not always available. It can decelerate the mass scale use of smart technologies. 

\subsection{Unavailability of Uniform Standards} Different countries use different standards of units and technologies which demand customized solution. This increases the price. A uniform standard across the world will solve the problem \cite{challenges}. 

%%%%%%%%%%%%%%%%%%%%%%%%%%%%%%%%%%%%%%%%%%%%%%%%%%%%%%%%
\section{Technologies for Smart Agriculture}
\label{Sec:Technologies}

\begin{figure}[htbp]
	\centering
	\includegraphics[width=0.90\linewidth]{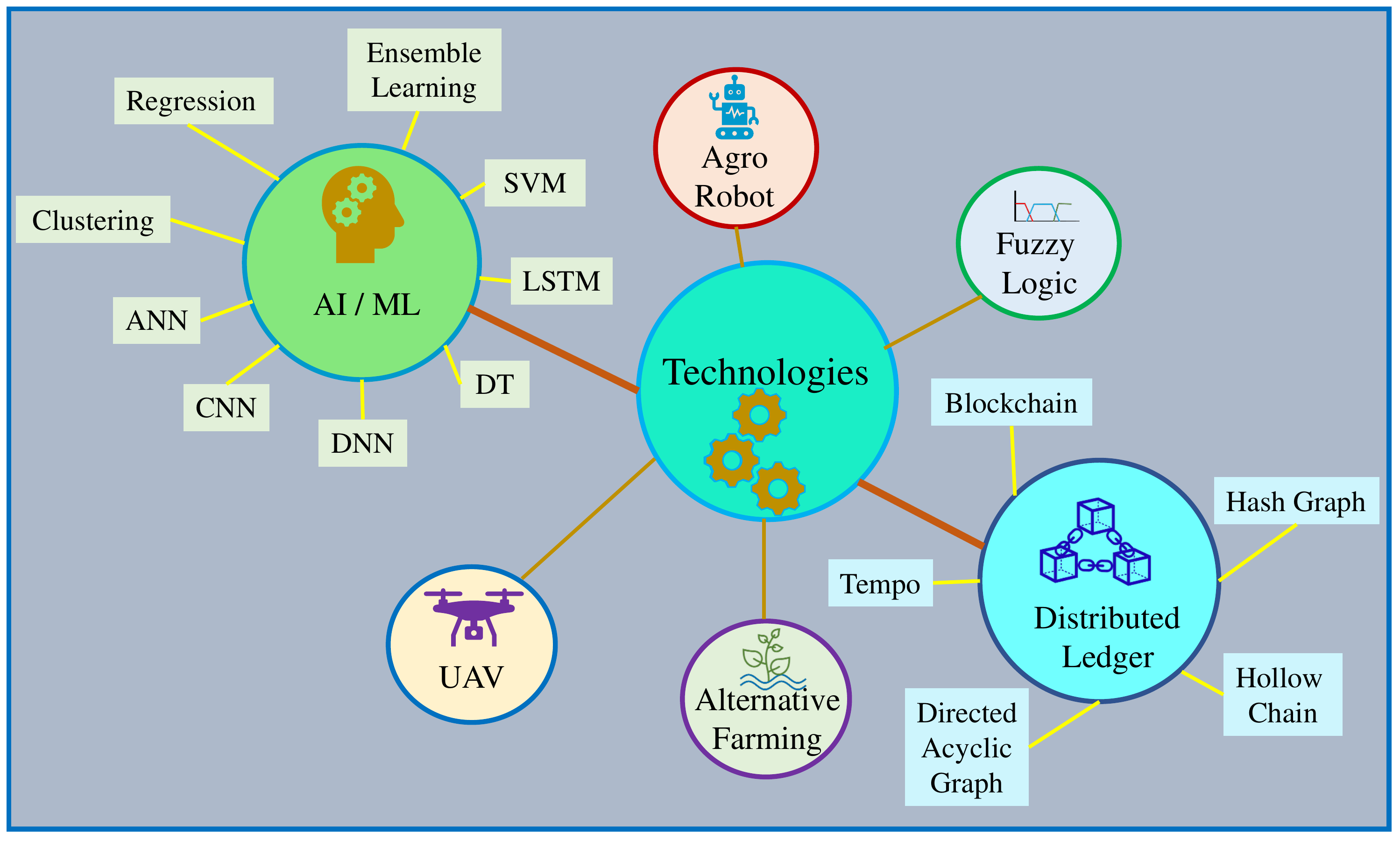}
\caption{Technologies in Smart Agriculture.}
	\label{fig:tech}
\end{figure}

$2021$ has been marked as the start of the \textit{Industry $5.0$} era. It arrived at the right moment when various industry sectors are welcoming digital, smart, green and sustainable ecosystems to cope with COVID-19 challenges. It redefines the relationship between ``man'' and ``machine'' \cite{diff_4_5}. In agriculture, the \textit{Industry $5.0$} era will accelerate the arrival of \textit{Agriculture $5.0$}. Mainly \ac{AI}/\ac{ML} and \ac{DLT} will orchestrate the advancement along with \ac{FL}, \ac{UAV}, agricultural robotics, and alternative farming, as shown in Fig. \ref{fig:tech}. In this section the two main technologies are discussed.   

\subsection{Artificial Intelligence and Machine Learning}
\begin{figure}[htbp]
	\centering
	\includegraphics[width=0.99\linewidth]{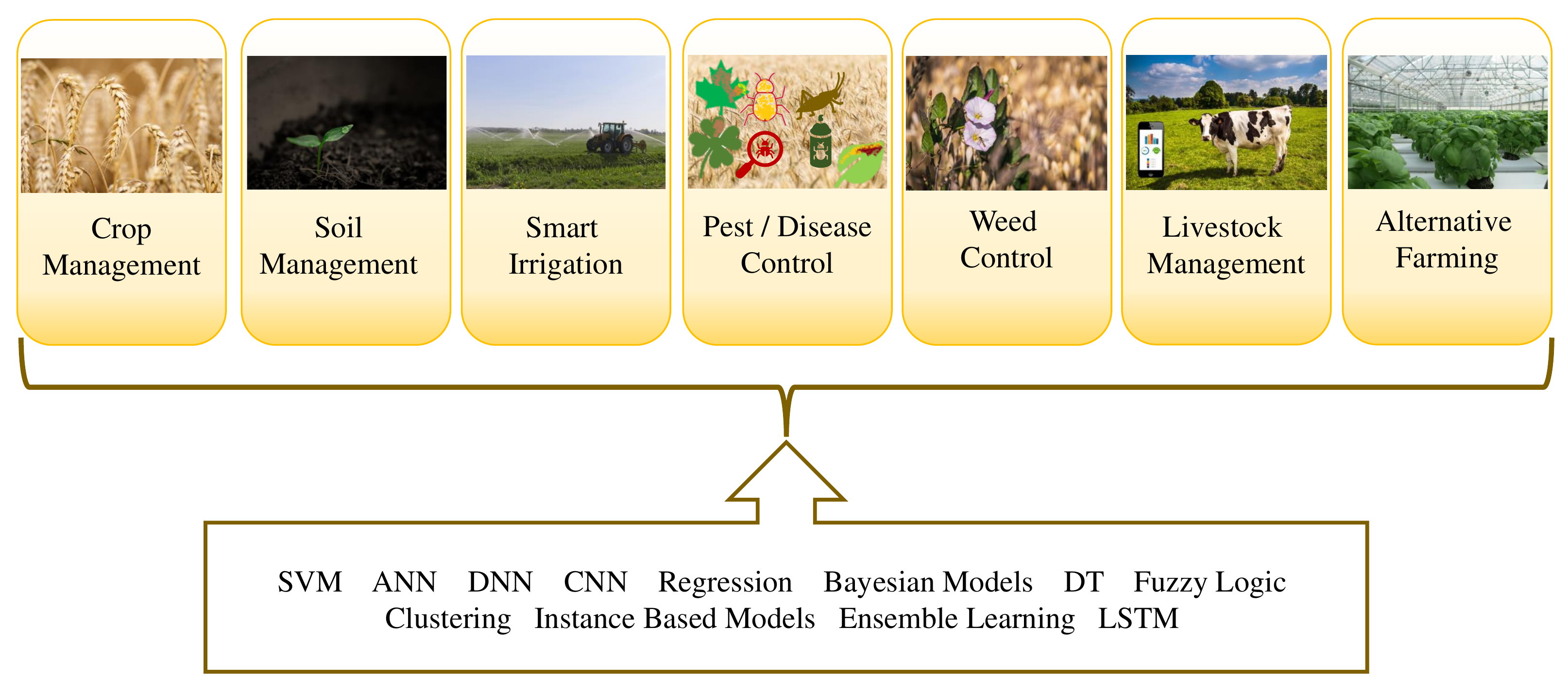}
\caption{AI Tools for Smart Agriculture \cite{Liakos2018Machine}.}
	\label{fig:ai_agri}
\end{figure}

Artificial intelligence is the intelligence displayed by machines that resembles human intelligence. 
% and is used to solve various problems in different domains e-commerce and marketing, human resource, computer vision, multimedia forensics, healthcare, social media, gaming, automobiles, and agriculture.
Advancements in \ac{AI} and \ac{ML} have shown a lot of promise in various domains such as e-commerce and marketing \cite{soni2020emerging}, human resources \cite{strohmeier2015artificial}, computer vision \cite{MitraMCKOCIT21}, multimedia forensics \cite{MitraMCKIFIP21, MitraMCKiSES21}, healthcare \cite{jiang2017artificial}, social media \cite{MitraMCKiSES20, MitraMCKSN21}, gaming \cite{palaus2017neural,Skinner2019}, automobiles, and agriculture. In agriculture, \ac{AI} is used in increasing efficiency, crop yield and profitability, monitoring crop health, monitoring and forecasting climate, optimizing supply chain, managing irrigation systems, pesticide and fertilizer management, weed control, smart sensing and mapping, livestock tracking and geofencing. Researchers are applying Fuzzy logic, various \ac{AI}/\ac{ML} techniques including classification, and logistic regression as well as Neuro-Fuzzy logic to agricultural predictive analytics, decision making systems, agricultural robotics and mobile expert systems \cite{Liu2021Tran}. Fig. \ref{fig:ai_agri} shows the \ac{AI} tools presented in various literature works on smart agriculture.

\subsubsection{Crop Management} 
Crop management consists of crop production or yield prediction, estimation, and crop supply chain management. Various \ac{ML} tools have been used in different sectors of crop management. To count the number of coffee fruits on a coffee plant branch \cite{ramos2017automatic} and identify the green immature citrus fruits \cite{sengupta2014identification}, \ac{SVM} have been used. \ac{SVM} have also been used for rice crop yield prediction \cite{su2017support}. The branches with full of cherries have been estimated with Gaussian Naive Bayes \cite{amatya2016detection}. \ac{ANN} have been used to evaluate grassland biomass \cite{ali2016modeling} and wheat yield prediction \cite{pantazi2016wheat}. Corn and soybean yield prediction has been done in \cite{kaul2005artificial} using \ac{ANN} with better accuracy than regression models. \ac{ANN} with back propagation have also been used to predict yields from soil parameters \cite{liu2005artificial}. \ac{ANN} have also been used to predict corn yield \cite{uno2005artificial}, rice yield in a mountainous region \cite{ji2007artificial}, cotton yield \cite{zhang2008neural}, wheat yield \cite{russ2008data}, maize crop yield \cite{singh2008artificial}, tea yield \cite{soheili2015forecasting}, and general crop yield \cite{dahikar2014agricultural}. \ac{ANN} have also been used in detecting nutrition disorder for crops \cite{song2005crop} and predicting the reaction of crops over soil salinity and water content \cite{dai2011simulation}. From \ac{UAV} imagery, tomatoes have been detected using clustering \cite{senthilnath2016detection}. Crop growth has been monitored in \cite{kumar2019gcrop}.

\subsubsection{Soil Management} 
Soil property management such as soil moisture, temperature, and nutrient content is an important part of smart agricultural systems. Its benefits are two fold - increasing crop yield and preserving soil resources \cite{Eli2019Applications}. But the process is time consuming and costly. So, various inexpensive and autonomous \ac{ML} techniques are being proposed to have a reliable soil management system \cite{Liakos2018Machine}. Mostly, the data from sensors, satellite images or images taken by \ac{UAV} are used as the input of the \ac{ML} models. \ac{ANN}, \ac{SVM}, and autoencoders have been used in predictive analysis. \ac{ANN} and \ac{MLP} have been used for suitability of soil evaluation \cite{Vincent2019MDPI}. Phosphorous in soil has been predicted using various \ac{ML} models \cite{Dong2018Digital}. \ac{DNN} have been utilized to extract geo-parcels from high resolution images and \ac{MLP} have been employed to predict the phosphorous content. Radial basis function neural network have been applied to predict the water retention capacity of soil in Brazilian coastal areas \cite{Carvalho2015RBF}. Soil moisture is also predicted with \ac{BRT} from \ac{UAV}-taken images \cite{Arya2020Retention} and with \ac{ANN} in \cite{Arif2013Estimation}.  Health and condition of soil moisture sensors have been predicted using \ac{SVM} along with the stage of the degradation by using Naive Bayes classification \cite{Jain2020Health}. Autoencoder and \ac{SVM} have been used to predict the soil salinity from satellite images \cite{Klibi2020Salinity}.     

\subsubsection{Smart Irrigation} 
Water management is an integral part of smart agricultural systems. Rainfall patterns are changing worldwide due to climate change. Evapotranspiration plays a vital role to assess water resources. Various \ac{AI} methods have been utilized in smart water management. Deep reinforcement learning has been used for smart water management in a crop field \cite{Bu2019Smart}. A multiple linear regression algorithm has been applied to calculate the water needed for greenhouse organic crops and then water valves have been operated automatically with a LoRa \ac{P2P} network \cite{Chang2019LoRa}. An \ac{ANN} system has been proposed in \cite{Nema2017Application} to predict evapotranspiration by carrying out a study in Dehradun, India. \ac{ANN} and the Penman-Monteith equation have been utilized to predict daily evapotranspiration \cite{Antonopoulos2017Daily}.  A smart irrigation system has been proposed using \ac{LSTM} and \ac{GRU} based models in an Edge-Fog-Cloud setting \cite{Dahane2020Irrigation}. Spatial water distribution has been predicted with \ac{ANN} in \cite{Hinnell2010Neuro} for a neuro-drip irrigation system. 

\subsubsection{Pest/Disease Control} 
To have optimal yield from a crop field, disease, pest, and weed control are necessary. An automated efficient system can save time and cost. From that perspective, \ac{AI} techniques are being proposed in various publications. The advancement started with a rule based system \cite{Sarma2010Expert,Balleda2014Agpest, Pasqual1988Development, Banerjee2017Radial, Mahaman2003Diares} in the last decade and evolved through \ac{FL} systems \cite{Tilva2013Weather, Siraj2006Integrated, Peixoto2015Approach, Van1998Indicator}. Various \ac{ANN} have been used for different diseases in different crops \cite{Francl1997Artificial, Karmokar2015Tea, Sladojevic2016Deep, Hahn2004Spectral} or for pest detection, e.g., a channel–spatial attention module, integrated with a backbone \ac{CNN} and a \ac{RPN} have been used to detect various pests in a crop field \cite{Liu2019Access} and apple leaf disease is detected in \cite{Jiang2019Real} using the GoogleNet Inception network and Rainbow concatenation. An incremental back propagation network  has been used with \ac{CFS} to detect pests in a tea plant. The \ac{CNN} based object detection model YOLOv3 has been utilized to localize the pest Tessaratoma Papillosa and by analyzing the environmental information by \ac{LSTM}, pest occurrence is predicted with $90\%$ accuracy \cite{Chen2020Access}. YOLOv3 and YOLOv3-Dense models have also been employed to detect anthrax on apple surface in an apple orchard \cite{Wang2013Mobile}. \ac{SSD} has been applied with $84\%$ accuracy in detecting pests and with $86\%$ accuracy in classifying pests \cite{Martin2008Early}. Pest detection and recognition have been performed through k-means clustering and correspondence filter \cite{Fina2013Automatic}. \ac{CNN} based models have been used in \cite{udutalapally2020scrop} and in \cite{pallagani2019dcrop} in crop disease detection. 

\subsubsection{Weed Control} Weed affects yield negatively. So, weed control is another important area in smart agriculture. Weeds are sometimes hard to distinguish from crops. The application of \ac{AI} in weed control started in the early 2000s. \ac{ANN} have been employed with Hebbian synaptic modification for distinguishing weeds from crops \cite{aitkenhead2003weed} and the accuracy achieved was reasonable based on the available hardware at that time. YOLOv3 has been used for low cost precision weed management in \cite{partel2019development}. Counter Propagation (CP)-ANN with multi-spectral images \cite{pantazi2017evaluation} and a combination of auto encoder and \ac{SVM} along with hyper spectral images \cite{pantazi2016active} have been utilized to detect weeds. \ac{SVM} has been used in \cite{binch2017controlled} to detect weeds in grassland cropping. 

\subsubsection{Livestock Management} \ac{AI}/\ac{ML} techniques have been used in livestock management in two ways: animal welfare and livestock production \cite{Liakos2018Machine}. Animal welfare, or the well-being of the animals has been addressed in \cite{Dutta2015Dynamic} for cattle using bagging ensemble learning, for calf using decision tree and C4.5 algorithm \cite{Pegorini2015Vivo}, and for pigs using Gaussian Mixture Models \cite{Matthews2017Automated}. 
\ac{AI} helps to optimize the efficiency of livestock production. \ac{ANN} with back propagation has been used in \cite{Craninx2008Artificial} to predict cattle rumen fermentation patterns from milk fatty acids. Pigs' faces have been detected with \ac{CNN} with $97\%$ accuracy in
\cite{Hansen2018Towards}. \ac{SVM} have been used for problem detection and warnings in egg production for commercial hen production \cite{Morales2016Early}, to estimate cattle weight trajectories for evolution \cite{Alonso2015Improved}, and to predict skeleton weight of the beef cattle \cite{Alonso2013Support}. \ac{ANN} with Bayesian Regularization has been used to predict quality milk production and to reduce the heat stress levels of the cows in a robotic cow farm \cite{Fuentes2020Artificial}. A fully connected neural network has been used to predict cow diseases in  \cite{Chatterjee2021CE}.

\subsubsection{Alternative Farming} Alternative farming consists of greenhouse farming, and hydroponics. \ac{ML} and deep learning techniques are used in those systems for better and precise control with less manpower. Greenhouse air temperature is forecast using fully connected \ac{ANN} and \ac{RMSE} \cite{Codeluppi2020Green}. \ac{ANN} have been used for greenhouse tomato yield and growth \cite{ehret2011neural}, greenhouse basil yield \cite{pahlavan2012energy}, greenhouse gas emission and energy consumption of wheat yield \cite{khoshnevisan2013modeling} and that of watermelon \cite{nabavi2016neural}. An \ac{RNN} with back propagation has been used to predict the humidity and the temperature of a greenhouse, powered by solar energy \cite{Hongkang2018Recurrent} and \ac{RNN}-\ac{LSTM} in \cite{Jung2020Time} for climate  (humidity, temperature and CO$_2$) prediction. \ac{ANN} and Bayesian Networks have been used in hydroponic systems to predict the needed action  \cite{Mehra2018IoT}.% \cite{verma2021machine} 

Various \ac{AI} technologies are proposed depending on the location of the computation. For \textit{edge AI} settings, where the \ac{AI} model runs on the limited resource embedded system itself, research is ongoing to design deep neural network models which have higher accuracy but fewer parameters to train \cite{Merenda2020Edge}. MobileNet \cite{mobilenet}, SqueezeNet \cite{iandola2016squeezenet}, EfficientNet \cite{tan2019efficientnet} are such networks where depth wise convolution, down-sampling of data and uniform scaling down of the model are performed respectively. Quantization \cite{zhou2018adaptive, yang2019quantization, choi2016towards, jin2020adabits} and pruning \cite{Yao2017Deepiot, molchanov2016pruning, anwar2016compact, yang2017designing, guo2016dynamic, hinton2015distilling, Han1510Compressing} are used to reduce the \ac{DNN} size. Proper choice of hardware is equally important as the algorithms. 
%Certain edge computing hardware are - single board computers like Raspberry pi,  NVIDIA Jetson Nano, Arduino Mega 2560, Intel Edison, Beagle Bone, micro-controllers like Arduino Nano 33 BLE Sense, STM32 microcontroller, Adafruit EdgeBadge, and system-on-chips like NodeMCU, ESP 8266. 

\subsection{Blockchain and Distributed Ledger Technology}

\subsubsection{Blockchain as a Digital Technology}
\begin{figure}[htbp]
	\centering
	\includegraphics[width=\textwidth]{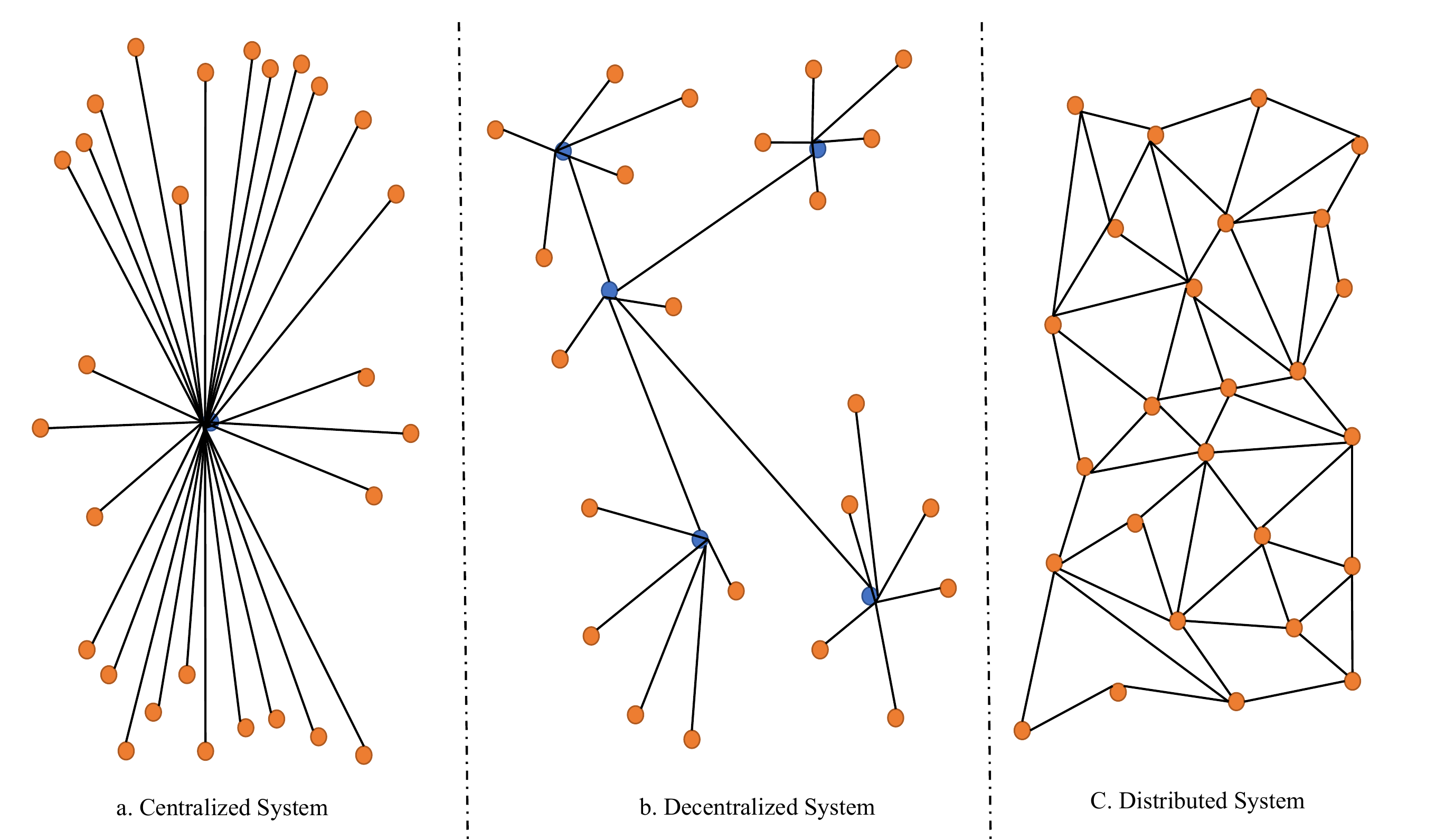}
	\caption{Types Of Networks (a) Centralized network which has single point of information sharing represented by blue sphere and multiple clients represented by orange spheres (b) Decentralized network which has multiple replicated central nodes represented by blue spheres and multiple clients represented by orange spheres (c) Distributed network where there is no central entity}
	\label{FIG:TypesOfNetworks}
\end{figure}

Blockchain is one of the recent technologies with promising applications in different fields which include peer-to-peer financial systems \cite{Nakamoto}\cite{Buterin_2015_ANS}, Real-time Secure \ac{IoT} systems \cite{Singh_WFIOT_2018}, Smart Governance applications \cite{Hjalmarsson_CLOUD_2018, Geng_PMIS_2021}, Digital Asset Copyright technologies \cite{Zeng_SmartBlock_2020, Xiao_Access_2020}, Smart Healthcare \cite{Rachakonda_2021_TCE, Azaria_OBD_2016}, Smart Agriculture and many other industries. The blockchain can be simplistically defined as a peer-to-peer distributed ledger which processes incoming transaction data and updates the shared ledger chronologically based on a set of rules known as Consensus mechanism that are accepted by peers across the network. The main idea behind creating such peer-to-peer networks is to create a reliable and verifiable communication and data storage between un-trusted entities which need to share data and work collectively as a single system. The most commonly used decentralized application structure in the past few decades is the client-server model where instead of housing data on a single central entity it is replicated and partitioned onto multiple servers easily accessed by clients from multiple locations. Even though this model has successfully addressed centralized system problems, it is still prone to security and privacy attacks which can be efficiently addressed using distributed networks. Fig. \ref{FIG:TypesOfNetworks} shows different network configurations.

Centralized systems have all the network data housed at a single location which is controlled and maintained by a network owner. The main drawbacks of this system are \ac{SPoF} and latency in data accessing from long distances. These drawbacks can be avoided by introducing a decentralized system where the data is replicated among multiple central servers which serve different locations effectively even when there is a failure at one of the central nodes. Even though this solved most of the problems, data is still controlled by a third party owner who is responsible for maintaining and storing the client information and interacting with them which may lead to several security and privacy issues. Another disadvantage with such an architecture is lack of data ownership and control on data from clients while interacting with such decentralized systems. Distributed networks can solve these issues by removing the need for central authorities to monitor and verify the network traffic. The \ac{IoT} has sensor and edge devices which form a distributed network. The data sharing and collective working of such devices can be improved by blockchain technology. Main components of the blockchain include Shared Ledger, Node, Transaction and consensus mechanism.

\begin{figure}[htbp]
	\centering
	\includegraphics[width=\textwidth]{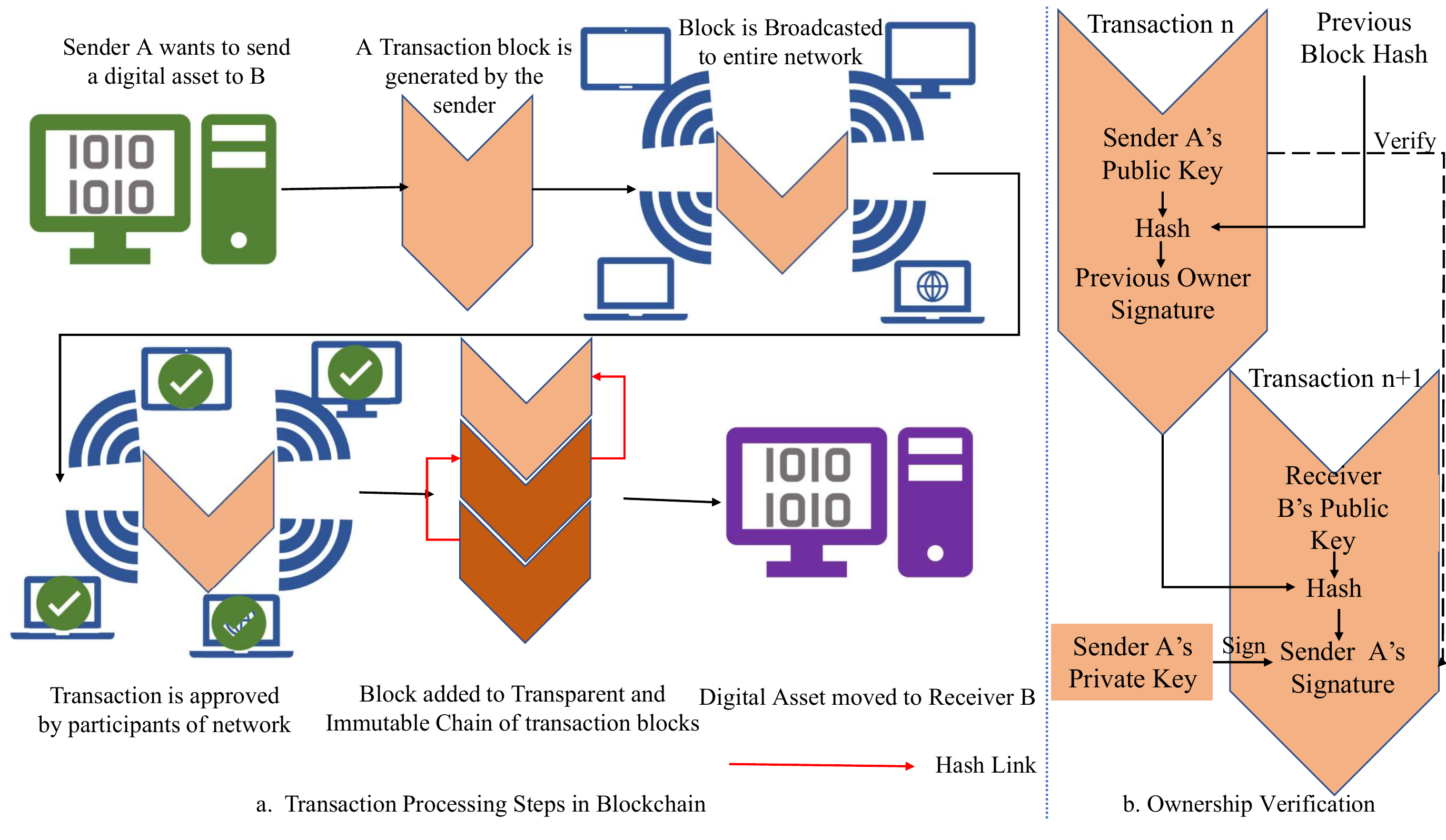}
\caption{Blockchain Transaction Steps and Digital Asset Ownership Verification.}
	\label{fig::Blockchain Transaction Steps and Digital Asset Ownership Verification}
\end{figure}

The blockchain shared ledger is a chronologically connected sequence of blocks of approved transactions. Each block consists of transactions along with the metadata which can be used to verify the integrity and authenticity of the transaction information within. Every node participating in the network will have its own copy of the ledger which will be updated periodically and helps to act as a single point of truth for the network. The ledger is replicated across the nodes in the network to avoid double spending of digital assets. Nodes are the participants of the network which are capable of performing transactions and also participate in network operations. Based on the roles they perform nodes can be peer nodes, full nodes and miners. Peer nodes are less computationally capable and are mainly responsible for generating transactions which utilize the blockchain network to process and handle transactions. Full nodes are nodes with large storage and are responsible for storing the entire trail of transactions. There is no incentive associated with full nodes but these nodes maintain the complete ledger to verify the transactions coming in. Miner nodes are responsible for performing the consensus mechanism where blocks generated are processed based on the pre-defined set of rules called consensus mechanism. These nodes are computationally capable and incentives are given for each block generated by the node. Most popular consensus mechanisms used are \ac{PoW} and \ac{PoS}, among which \ac{PoW} makes use of computationally hard cryptography puzzles to select the miner whereas \ac{PoS} makes use of staking and age of staking into consideration while selecting a miner to generate new blocks. Fig. \ref{fig::Blockchain Transaction Steps and Digital Asset Ownership Verification} shows the transactions steps and digital asset verification process in detail. 

\subsubsection{Relevance of Blockchain Technology in Smart Agriculture}
\begin{figure}[htbp]
	\centering
	\includegraphics[width=\textwidth]{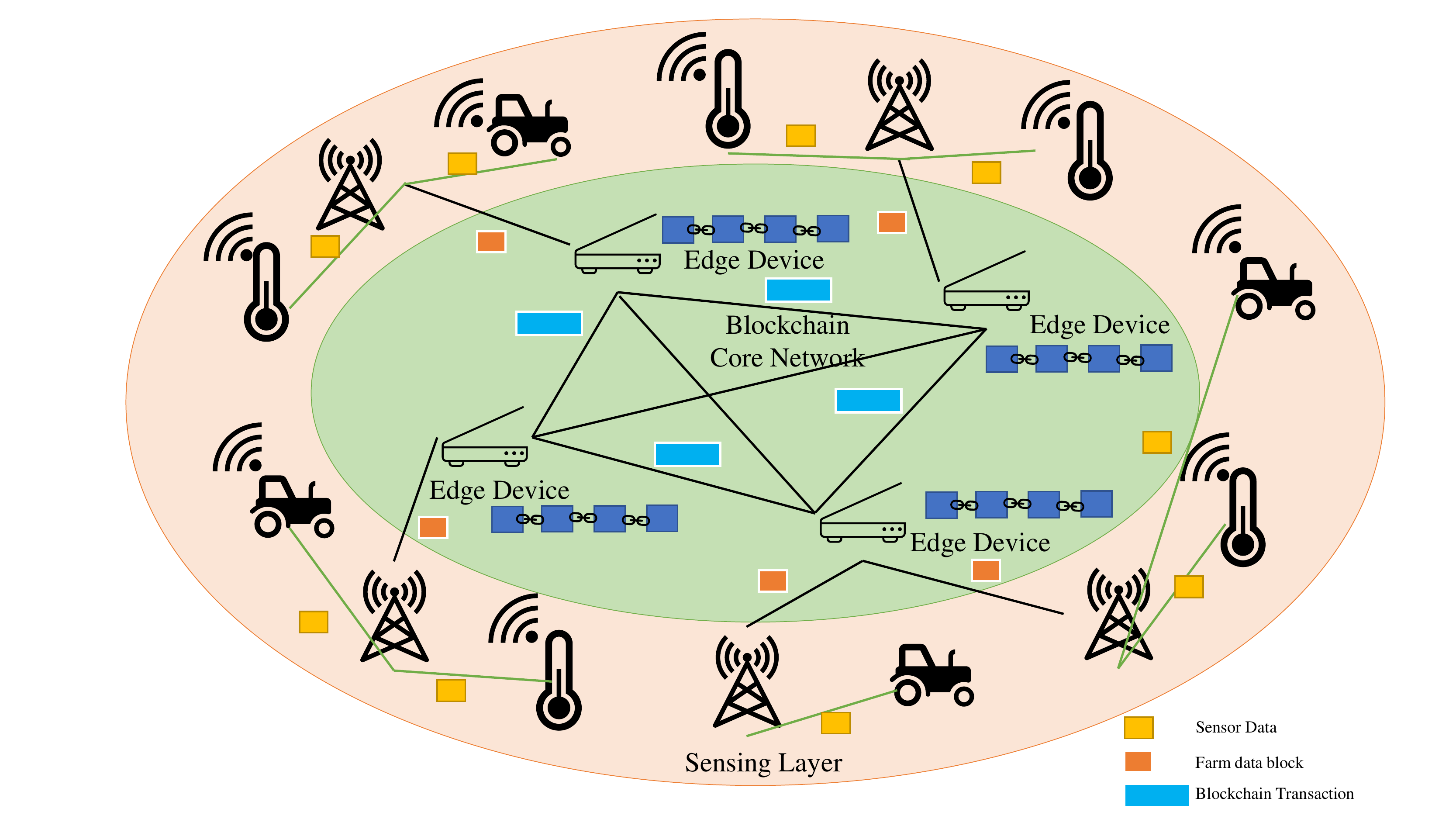}
	\caption{Analogy Between Smart Agriculture IoT Network and Blockchain.}
	\label{fig:BlockchainAnalogy}
\end{figure}

The agriculture sector has evolved by adapting various new technologies to modify farming practices for efficient and better yield of crops \cite{Liu_2021}. One such enabling technology is the \ac{IoT} which provides solutions for automating many anthropocentric tasks in farming. In the layered architecture used in the \ac{IoT} environment in agriculture, \textit{layer-2} or the edge computing layer consists of many \ac{EDC}) which form distributed networks with a critical need to communicate and share data between each other to work collectively \cite{Puthal_ICCE_2019}, as in Fig. \ref{fig:BlockchainAnalogy}. In order to make these Machine-to-Machine communications more secure, there is a need for central authorities to monitor the data and deploy some cryptography techniques to maintain data integrity and privacy. This can be a challenging task with the number of computing edge devices that are needed while monitoring and controlling a large farm. Furthermore, using such central monitoring system can lead to centralization and other problems like single point of failure and latency issues. These issues can adversely affect farms as automated systems will not behave as expected and result in the reduction in yield or quality of crop. 

\subsubsection{Blockchain Applications in Agriculture} 
The blockchain has a large set of applications in smart agriculture and deals with different aspects of farm activities. Some of the major applications and related solutions are discussed below.  

\paragraph{Secure Real-time Data Sharing}
Data security and privacy is one important aspect in smart agriculture which needs to be addressed for efficient functioning of autonomous processes. The blockchain makes use of cryptography techniques and processes transactions in chronological order to maintain integrity of data and avoid adversary attacks such as \ac{DoS} and False Data Injection. Apart from data privacy, data ownership and monetization are also problems. Unlike in centralized applications where the data is monetized by a central authority, blockchain based applications can help farmers control the data access at granular levels and can help in monetizing the data on their own. A typical \ac{IoT} architecture consists of a cloud layer where data from the edge layer is  stored and processed to perform automated tasks.  One of the main drawbacks with such networks is that the latency and access times vary based on network availability and number of access requests going to the server at a given time. As real-time operation is critical in making decisions, the blockchain can help in developing an efficient real-time data sharing model. Some of the secure models using blockchain for secure data sharing are proposed in \cite{Wu2019,Sharma2017,Zhou2018,Ma2019}. \cite{Wu2019} established an identity managed authentication mechanism which makes use of private blockchain and provides a secure information sharing mechanism eliminating \ac{DDoS} attacks. \cite{Sharma2017} has proposed a system which is a combination of \ac{SDN} technology along with blockchain to detect and prevent attacks in the \ac{IoT} environment with only minimal overhead. This can be an optimal solution in resource constrained environments like \ac{IoAT}. \cite{Zhou2018} makes use of homomorphic computation on encrypted data and follows a similar approach to \ac{PBFT} and is based on the threshold number of correct responses from the server relevant smart contracts will be run. For the implementation, the Ethereum blockchain was used and the response times were computed to be 22 sec as the block generation time of Ethereum is fixed at 15 sec and can be further improved by adapting second generation blockchains with shorter block times. \cite{Ma2019} has proposed a novel key management architecture which can address issues of centralized systems using blockchain while increasing scalability and reliability. \cite{Anand_IFIP_2021} made use of distributed ledgers instead of traditional blockchains which are resource intensive to increase the scalability and real-time data availability in Smart Agriculture systems.

\begin{figure}[htbp]
	\centering
	\includegraphics[width=0.90\textwidth]{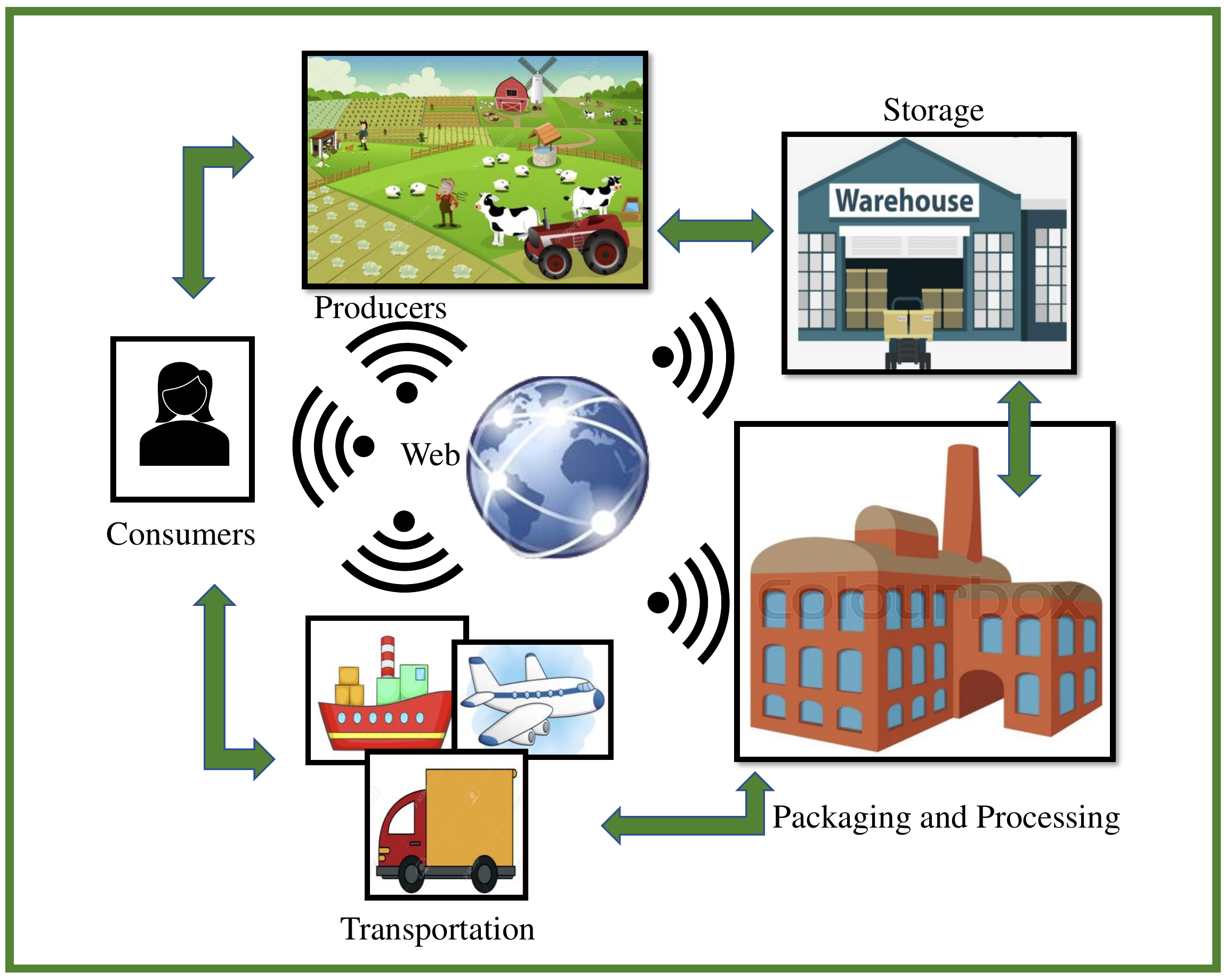}
	\caption{Supply Chain Traceability in Smart Agriculture.}
	\label{fig:supply_chain}
\end{figure}

\paragraph{Community Farming and Local Markets}
Community farming needs collective intelligence and transparent sharing of weather, crop disease or product demand data to help crop choice and achieve better yield. Along with this, local markets will enable the farmers to monetize their product more efficiently with greater profits. The blockchain can help in organizing and managing such applications for farmers to enable them to both produce and realize better profitability. Some of the works towards this area have been presented in \cite{Paul2019} which made use of the Ethereum platform to remove the intermediaries to establish an ethical supply chain and provide deserved profits to farmers. Fig. \ref{fig:supply_chain} shows the logistics present in the Supply Chain Traceability.

\paragraph{Supply-Chain Traceability}
Globalization is the trend which enabled product availability at even remote places. This has made global food supply chains more complex by involving multiple entities working together throughout the process. One of the major problems with such complex food supply chains is traceability and consumer confidence. It is very common to see food-borne disease outbreaks. In such scenarios, the most common approach is to dispose of the entire inventory as testing each product for infestation is not possible. Instead, tracing the product back to the farm from where it was produced can help in determining which products are affected and reduce food wastage. As an end user, a clear and transparent supply chain can help building  customer confidence in the authenticity of food. The blockchain can help in this aspect by building transparent supply chains where traceability and authenticity of the food can be easily verified. Many research works have been proposed. \cite{Wu2019a,Malik2019} makes use of HyperLedger Fabric blockchain to perform a case study on the blockchain based supply chain and discusses limitations. A solution based on \ac{RFID} technology integrated with blockchain is proposed in \cite{Tian2016}. \cite{Kaid2018} proposed a smart contract based financial solution in supply chains to help in solving the issue of \ac{SPoF} in traditional \ac{ERP} systems. \cite{Basnayake2019} has proposed Ethereum based decentralized applications for tracing the supply chain in organic foods to build trust and confidence from the consumer towards suppliers. \cite{Lin2019} proposed an efficient supply chain tracking system integrating blockchain with \ac{EPCIS} and making use of Ethereum smart Contracts. 
\paragraph{Farm Insurance}
Farms are more prone to weather changes and the damages due to weather conditions will lead to financial instability for farmers. Agriculture insurance is based on a farmer paying a fixed amount of premiums before the cropping cycle begins and receiving a payout based on the damage caused by the weather conditions. The problem arises when there is no index available to calculate the amount of damage, hence weather data is used and analyzed by the insurance provider to evaluate an index which forms the baseline for farmers and makes it easy to process these farm insurances. The most common setup is for the insurance provider to make use of weather station data recorded remotely and presented to farmers. Blockchain can help in assessing and accepting premium payments from the farmer using automated smart contracts. Together with that, weather index data can also be made available to farmers with greater reliability. \cite{Nath2016} proposes a blockchain based solution for avoiding fraud in insurance. \cite{Nguyen2019} made use of the NEO platform to build a system for drought based insurance. \cite{Aleksieva2020} proposed an Ethereum blockchain and hyperledger private blockchain based solution for insurance services using smart contracts. 

\subsubsection{Limitations of Blockchain}
Even though the blockchain has many potential applications in smart agriculture to enhance data security and integrity, there are still challenges which need to be addressed before wide adoption of this technology into the agriculture space. \ac{IoT} technology used in smart agriculture is resource constrained both in terms of power and computations whereas the consensus mechanism and cryptography components of the blockchain require large amounts of power and computation. The blockchain as such cannot be an efficient solution, hence research is being done to propose various efficient consensus mechanisms which can be implemented in resource constrained environments as in smart agriculture. \cite{Puthal2019} has proposed a consensus mechanism based on cryptographic authentication and \ac{MAC} address verification which has reduced the computational requirements of the consensus mechanism and has increased the transaction times significantly. Data is another significant problem which needs to be addressed for wide adoption. As the size of each block in the blockchain is predefined and limited, large amounts of data like images are not viable to be stored on-chain. Hence, many researchers are working on making the data stored off-chain while the transaction and access information along with the data are stored on-chain for secure access and integrity. \cite{Vangipuram2021} proposed a system which makes use of the \ac{IPFS} along with Ethereum smart contracts to share COVID-19 related patient data which can help in enforcing social distancing practices. This can be adopted into smart agriculture for storage of large chunks of data. Multi level access management is also a vital aspect which needs to be addressed. \cite{Biswas2021} proposed a blockchain system which can operate at multiple levels with different access policies for an efficient and controlled data management process which can be adopted to smart agriculture environments.

%%%%%%%%%%%%%%%%%%%%%%%%%%%%%%%%%%%%%%%%%%%%%%%%%%%
\section{Datasets for Smart Agriculture Research}
\label{Sec:Agriculture_Datasets}
Smart agriculture makes use of intelligent devices to collect data to analyze crop yields, livestock management, and economics related to supply. The stored data can help further research into the availability of resources in farming for next generations. Table \ref{tab:datasets} shows different datasets of multiple formats that we have studied and collected for the current survey paper.

\begin{table*}[htbp]
	\renewcommand{\arraystretch}{1.7}
	\centering
	\caption{Datasets for Smart Agricuture.}
	\label{tab:datasets}
	\begin{tabular}{|p{4.5cm} p{2.5cm} p{1.5 cm} p{6.5cm}| }
		\hline
		%\hline
		\textbf{Dataset} & \textbf{Source} & \textbf{Dataset Format } & \textbf{Link}
		\\ \hline
		%\hline
		Crop Yield \& Production &  USDA \& NASS & .php &  \url{https://www.nass.usda.gov/Charts_and_Maps/}    \\
		%\hline
		
		Crop Condition \& Soil Moisture & Crop-CASMA & .gis  &  \url{https://nassgeo.csiss.gmu.edu/CropCASMA/} \\
		%\hline
		
		Plant Diseases   & Kaggle & .csv &  \url{https://www.kaggle.com/saroz014/plant-diseases} \\
		%\hline
		
		Soil Health \& Characterization  & NCSS & .mdb &  \url{https://new.cloudvault.usda.gov/index.php/s/7iknp275KdTKwCA} \\
		%\hline
		
		Pesticide use in Agriculture  & USGS & 	.php, .txt &  \url{https://water.usgs.gov/nawqa/pnsp/usage/maps/} \\
		%\hline
		
		Water use in Agriculture  & USGS &  	Tableau & \url{https://labs.waterdata.usgs.gov/visualizations/water-use-15}\\
		
		%\hline
		
		Groundwater Nitrate Contamination  & USGS & .jpeg & \url{https://prd-wret.s3.us-west-2.amazonaws.com/assets/palladium/production/s3fs-public/thumbnails/image/wss-nitrogen-map-us-risk-areas.jpg} \\
		
		%\hline
		
		Disaster Analysis  & USDA \& NASS &  .png, .pdf & \url{https://www.nass.usda.gov/Research_and_Science/Disaster-Analysis/}\\
		\hline
		
	%%	\multicolumn{4}{l}{USDA $^*$ $\rightarrow$ U.S. Department of Agriculture. NASS $^*$ $\rightarrow$ National Agricultural Statistics Service.}\\
		
	%	\multicolumn{4}{l}{NCSS$^*$ $\rightarrow$ National Cooperative soil Survey. USGS $^*$ $\rightarrow$ U.S.Geological Survey.} \\
		%	\hline
	\end{tabular}
\end{table*}%

\subsection{Crop Yield and Production}
\begin{figure*}[htbp]
	\centering
	\subfigure[Small Red Bean Production.] {\includegraphics[width=0.48\linewidth]{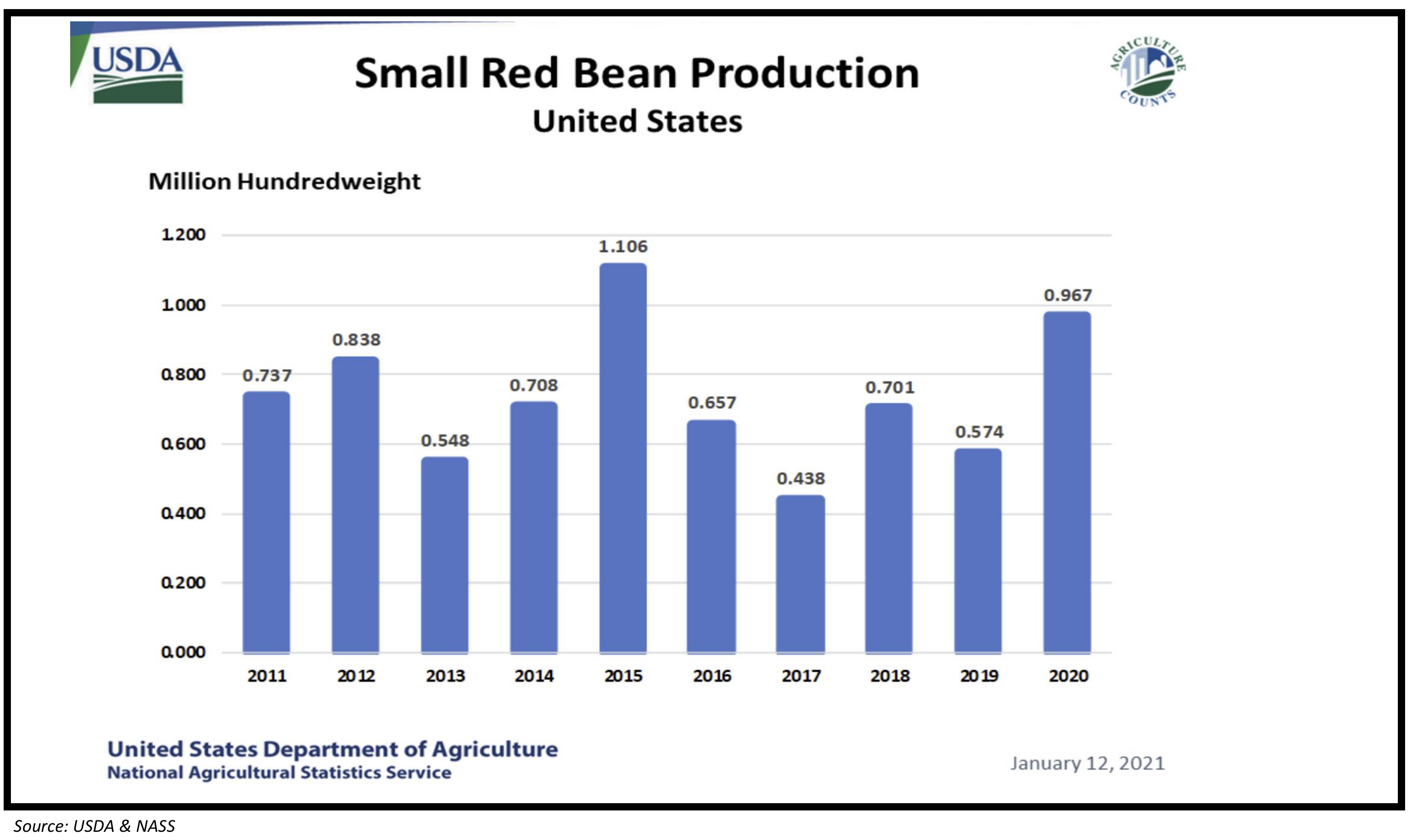}\label{FIG:Production_One}}
	\subfigure[Dry Pea Production.] {\includegraphics[width=0.48\linewidth]{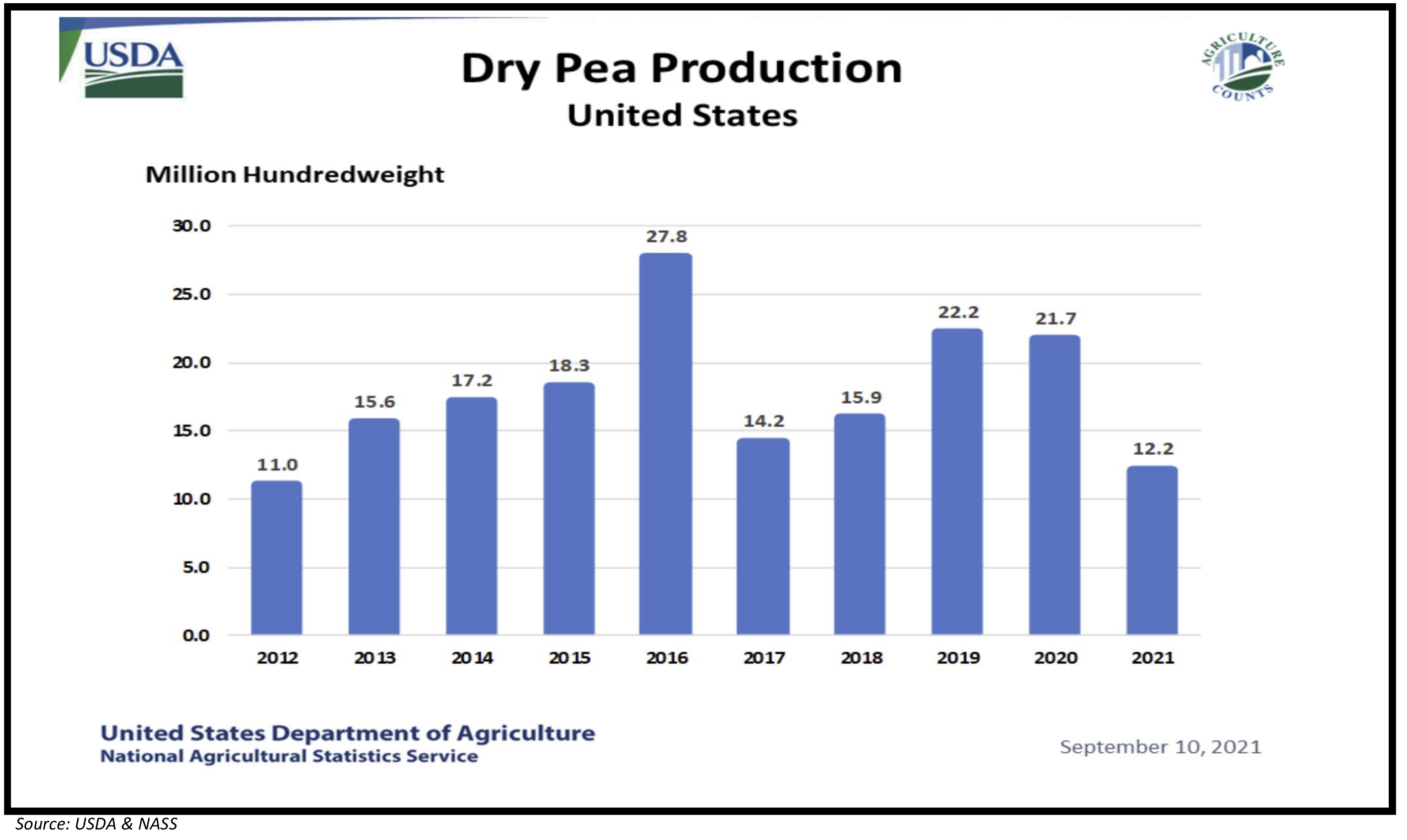}\label{FIG:Production_Two}}
	\subfigure[All Cows and Calves Inventory.]
	{\includegraphics[width=0.48\linewidth]{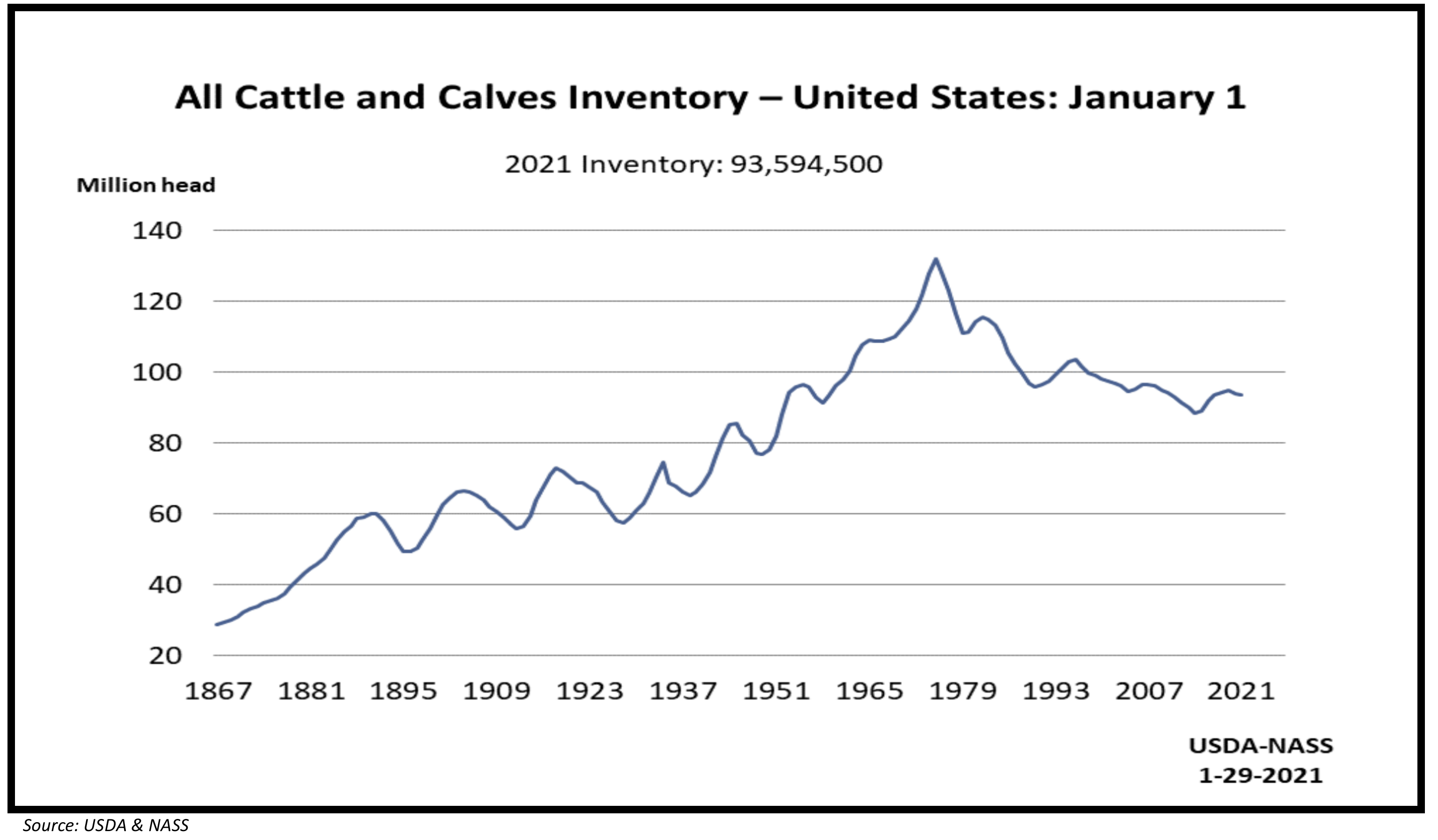}\label{FIG:Production_Three}}
	\subfigure[Farm Production Expenditures.]
	{\includegraphics[width=0.48\linewidth]{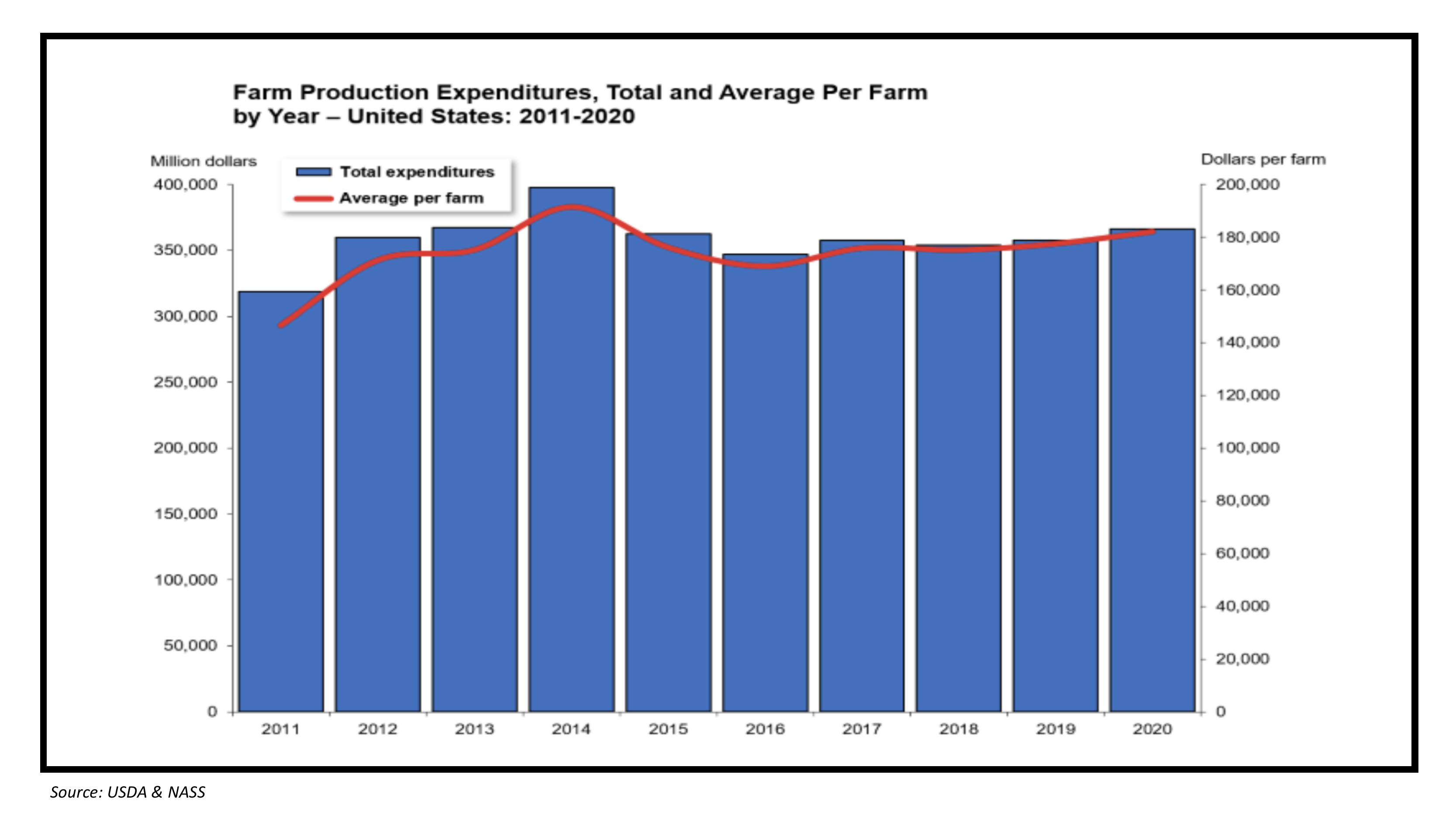}\label{FIG:Production_Four}}
	\caption{Graphs showing Crop yield and Production \cite{Agriculture2021}.}
	\label{fig:Production}
\end{figure*}

Sensors are used to collect data relating to acreage, crop condition, and yield. The amount of crop yield can be calculated by dividing the amount of produce over the harvested area. Crop production can be measured in terms of tonnes per hectare. The \ac{USDA} produces annual reports that include data for yield, acreage, and production for crops, plants, livestock, and animals, along with a Census of Agriculture. Fig.\ref{fig:Production} shows the graphs for some of the crops, livestock, and expenditures of agriculture in the United States for different years. Additionally, the prices for farming, labor, production, and land values are collected monthly and annually \cite{Agriculture2021}.

\subsection{Crop Condition and Soil Moisture}
\begin{figure}[htbp]
	\centering
	\subfigure[Dated November 07, 2021]
	{\includegraphics[width=0.98\textwidth]{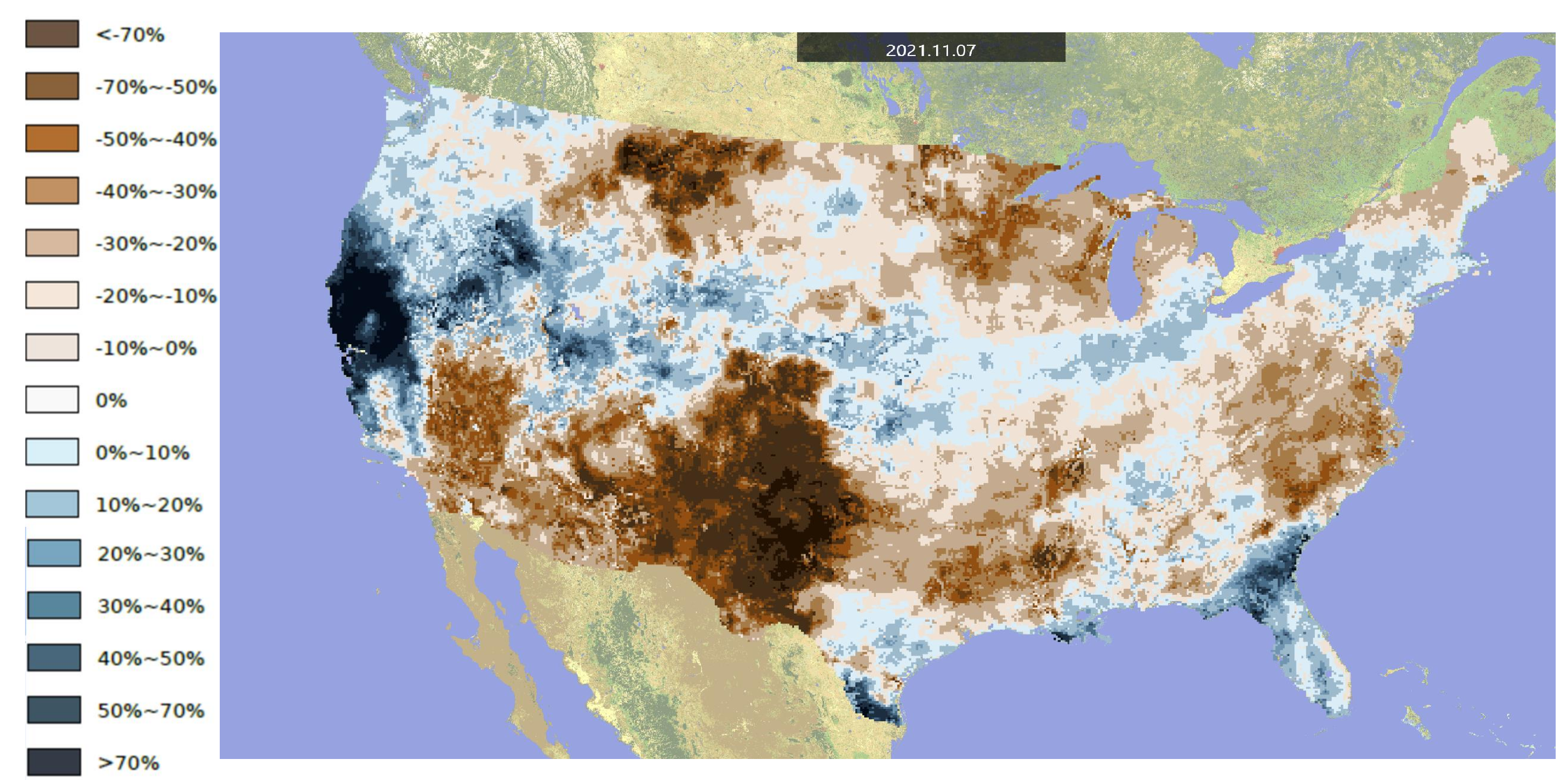}
	\label{fig:CropCASMA}}
	\subfigure[Dated November 30, 2021]
    {\includegraphics[width=0.98\textwidth]{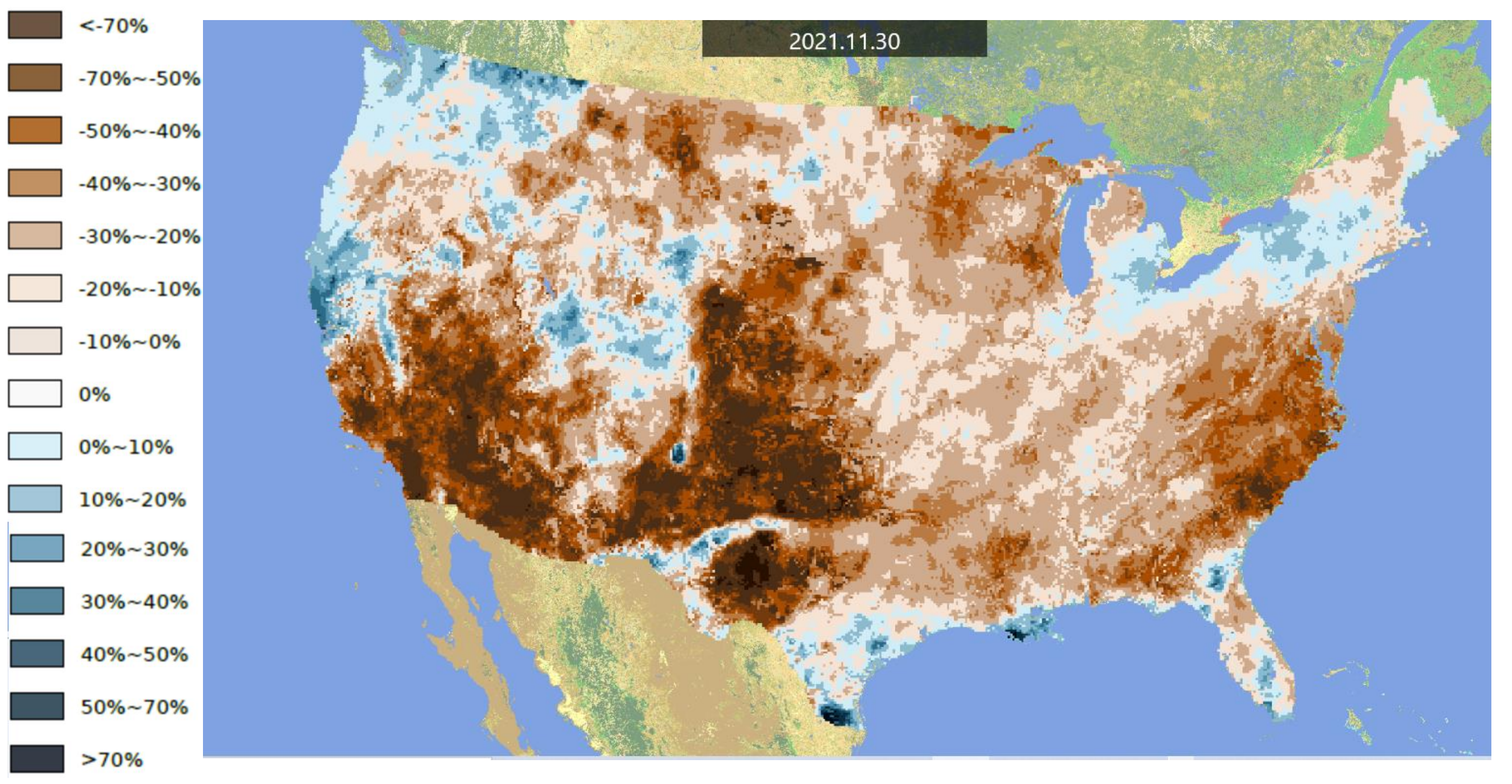}
    \label{fig:CropCASMA_2}}
    \caption{Crop Condition and Soil Moisture in the United States \cite{USDA2021}.}
\end{figure}

Knowledge of soil moisture is a critical factor for the yield and production of crops. In order to execute different agricultural operations easily, data regarding the soil is essential for farmer decision making. \ac{Crop-CASMA} is a web-based geospatial application used to measure soil moisture and vegetation conditions. The data collected is in the form of Geographic Information System mapping format (.gis) \cite{USDA2021}. Fig. \ref{fig:CropCASMA} and \ref{fig:CropCASMA_2} show crop condition and soil moisture analytics in the United States at two different dates.

\subsection{Plant Diseases}
\begin{figure*}[htbp]
	\centering
\subfigure[Healthy Plant Leaves - Apple, Potato, and Peach (From Left to Right)] {\includegraphics[width=0.8\linewidth]{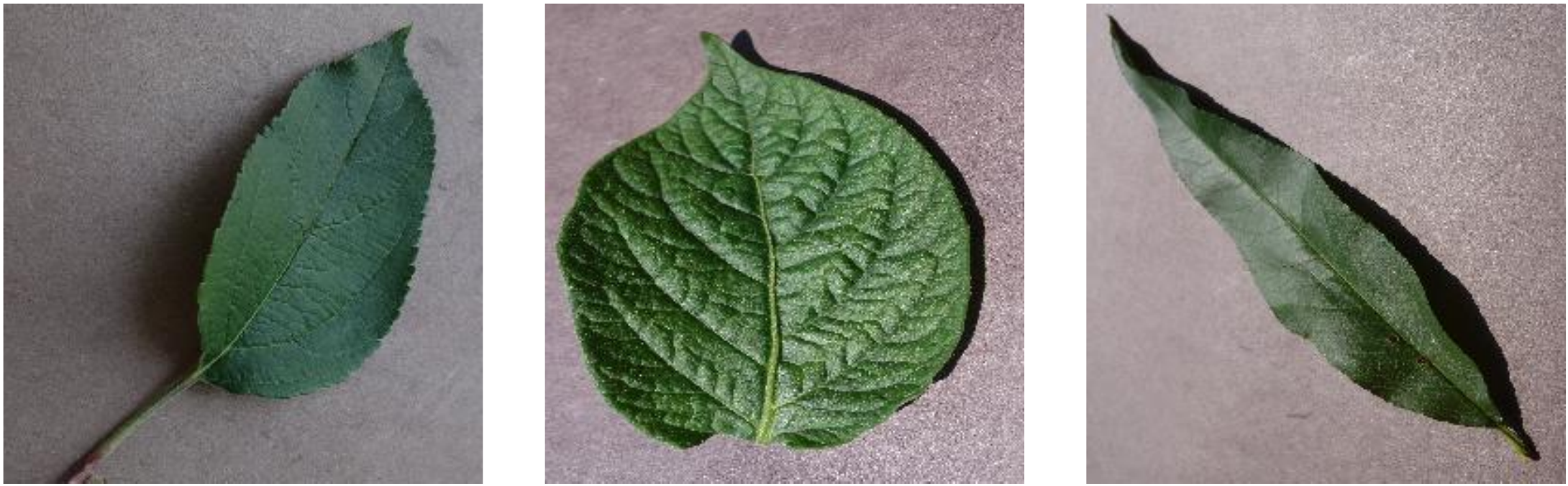}
		\label{FIG:healthy_plant}}
\subfigure[Infected Plant Leaves - Scab Infected Apple,  Late Blight Infected Potato, and  Bacterial Spot Infected Peach (From Left to Right)] {\includegraphics[width=0.8\linewidth]{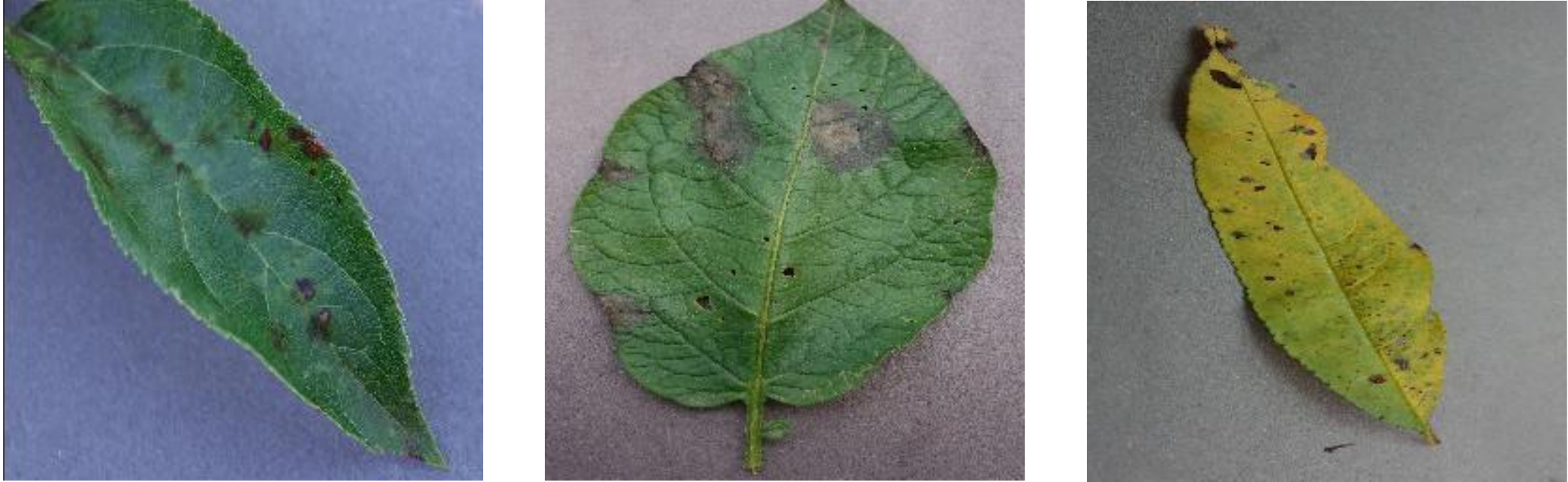}
		\label{FIG:infected_plant}}
\caption{Sample Images from Plant Disease Disease Dataset \cite{Dataset2018}.}
	\label{fig:plant_disease}
\end{figure*}

\begin{figure*}[htbp]
	\centering
\subfigure[Pomegranates of Different Grades of Quality 1 (From Left to Right - Grade 1, Grade 2, Grade 3)] {\includegraphics[width=0.80\textwidth]{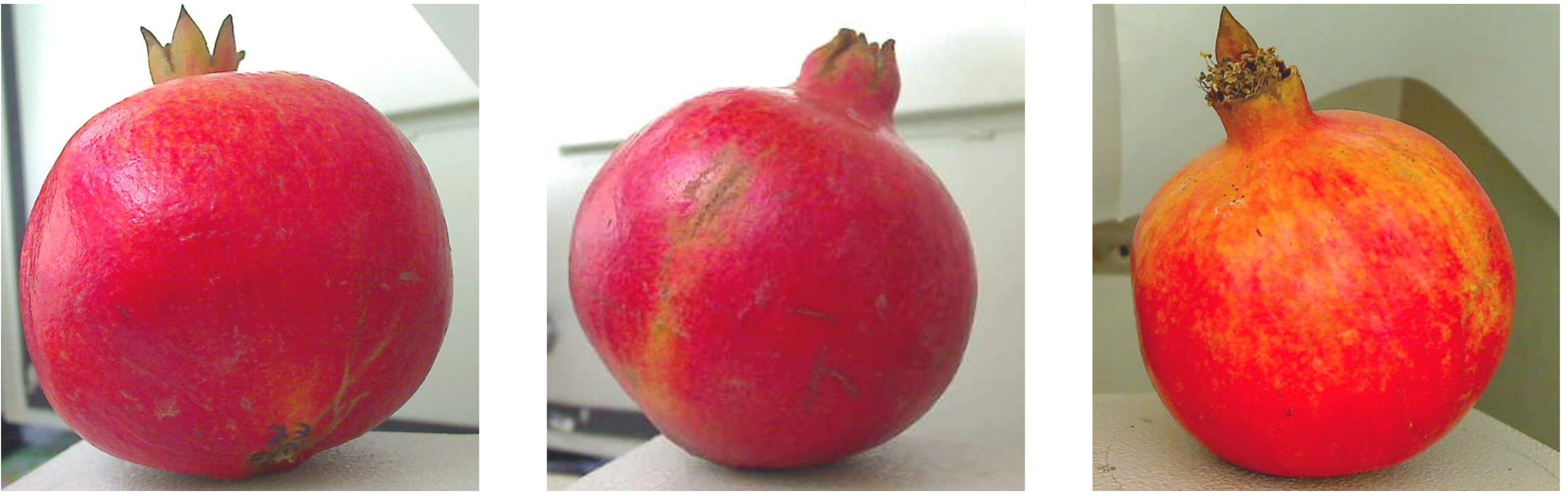}}
\subfigure[Pomegranates of Different Grades of Quality 4 (From Left to Right - Grade 1, Grade 2, Grade 3)] {\includegraphics[width=0.8\linewidth]{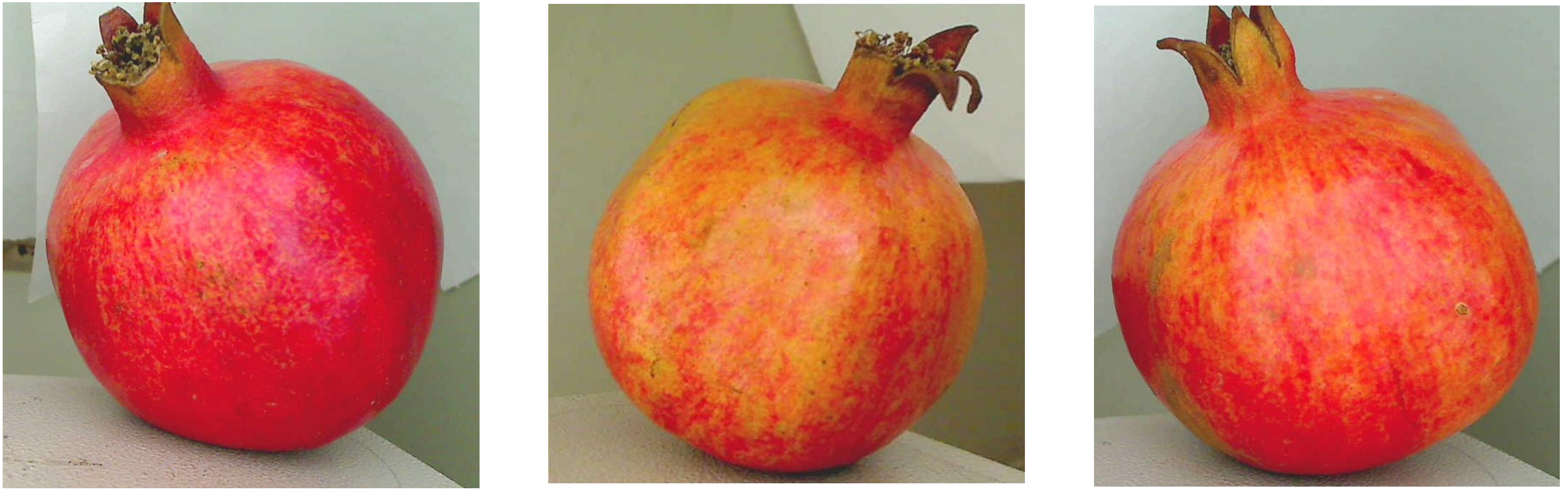}}
    \label{fig:pomegranate}
\caption{Pomegranate Fruit Dataset \cite{pomegranate_fruit_dataset}.}
\end{figure*}

When a plant gets infected with disease, its vital functions are modified and damaged, leading to harmful consumption for individuals. Each plant species has its unique syndrome. Kaggle is a good source of various datasets regarding different plant diseases. Fig. \ref{fig:plant_disease} shows some of the sample images from the Plant Disease dataset \cite{Dataset2018}. Collecting and storing these data can help study, train, and test to improve and impede the diseases in crops. Predicting plant diseases can enhance crop yield and productivity. Fig. \ref{fig:pomegranate} shows some sample images from the Pomegranate Fruit Dataset \cite{pomegranate_fruit_dataset}.

\begin{figure*}[htbp]
	\centering
	\subfigure[Healthy Chinese Cabage Plant] {\includegraphics[width=0.35\linewidth]{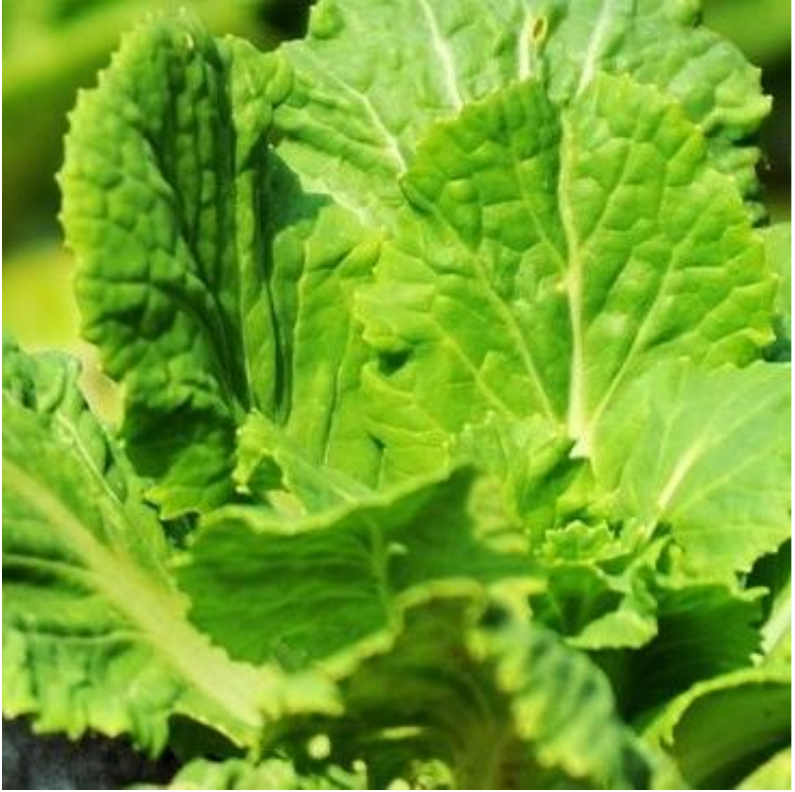}
		\label{FIG:healthy}}
	\subfigure[Infected with Back Moth] {\includegraphics[width=0.35\linewidth]{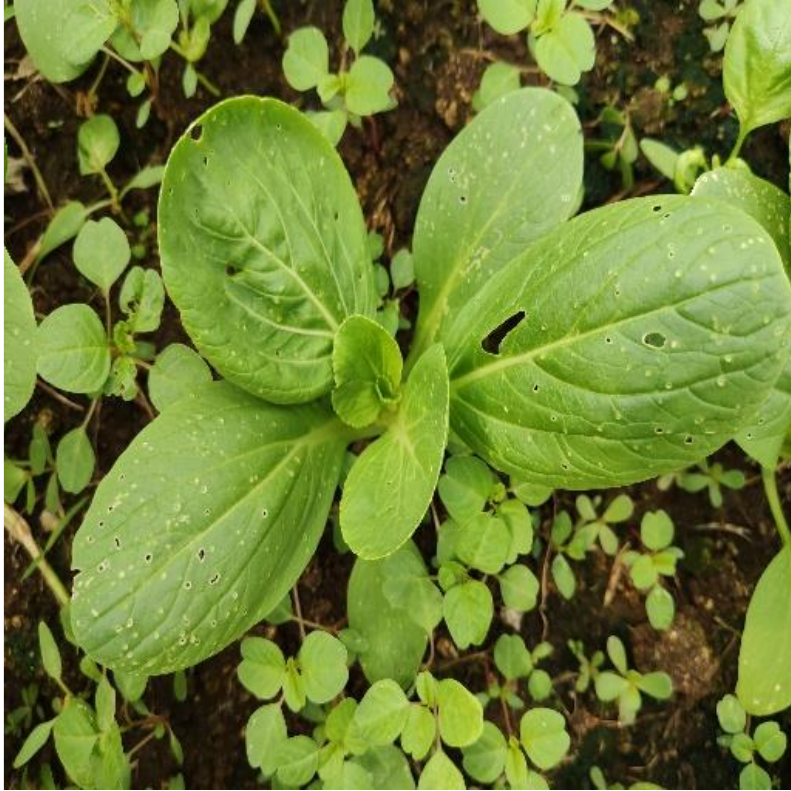}
		\label{FIG:back_moth}}
	\subfigure[Infected with Leaf Miner]
	{\includegraphics[width=0.35\linewidth]{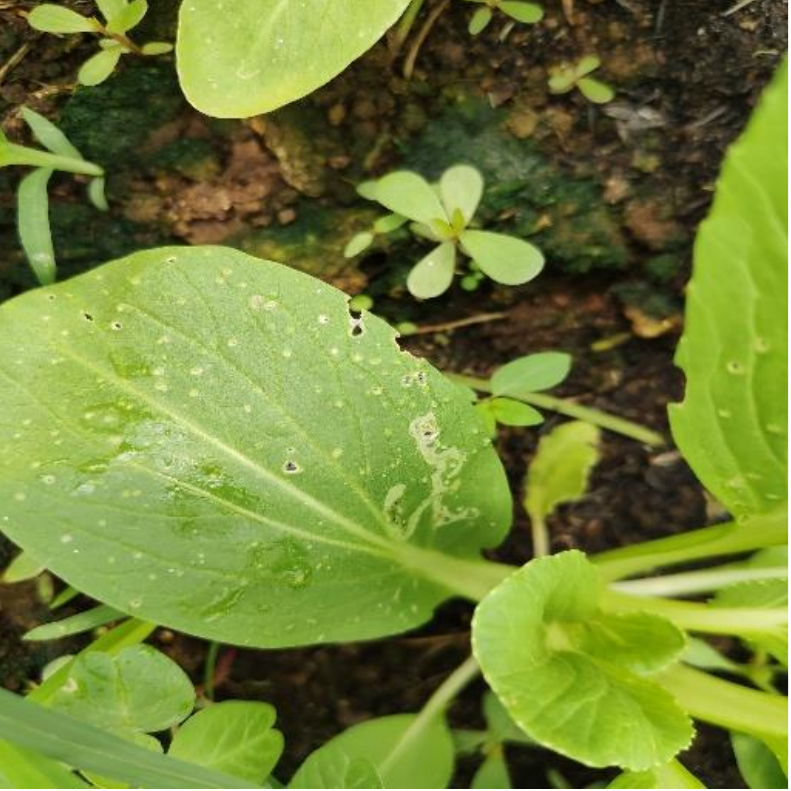}
		\label{FIG:leafminer}}
	\subfigure[Infected with Mildew]
	{\includegraphics[width=0.35\linewidth]{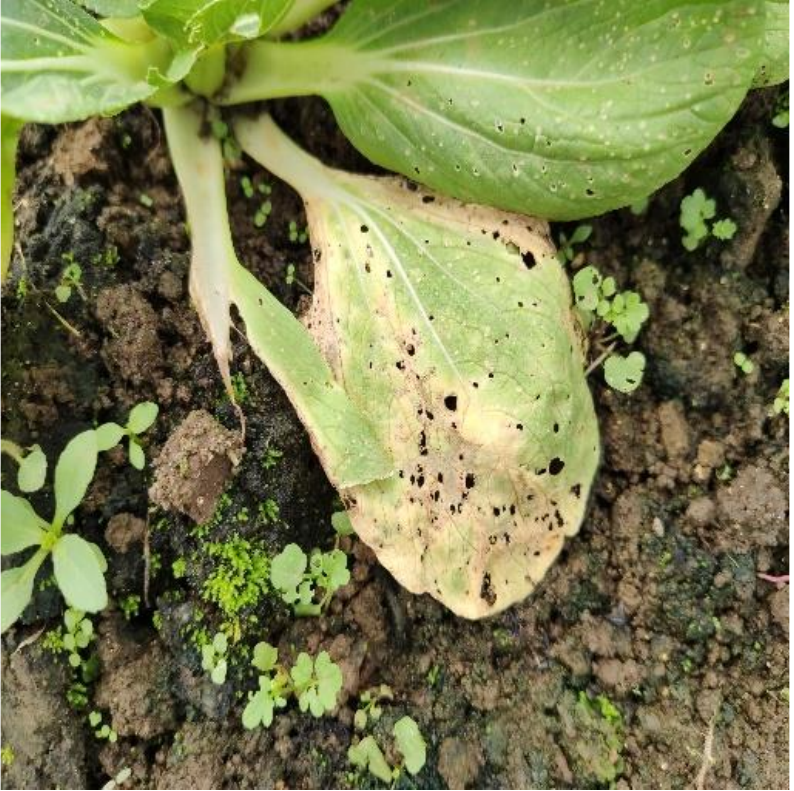}
		\label{FIG:mildew}}
	\caption{Chinese Cabbage Disease Dataset \cite{chinese_cabbage}. }
	\label{fig:chinese_cabbage}
\end{figure*}

Fig. \ref{fig:chinese_cabbage} shows some sample images from the Chinese Cabbage Disease dataset \cite{chinese_cabbage}. 

\subsection{Soil Health and Characterization}
The soil characterization survey is used to give information regarding the properties and features of soil in a specific area. The survey can contain detailed descriptions and soil boundaries that are beneficial to the farmers, estate agents, and engineers.

The \ac{NCSS} provides database reports for soil classification \cite{Survey} along with the pedon number for soil taxonomy. A pedon is a three-dimensional structure of soil that is sufficient to explain the soil's inner composition and can be used for collecting samples for lab analysis. The soil properties at each field, such as available rock fragments, bulk density, moisture, water content, carbon, salt, pH, carbonates, phosphorous, clay, sand, and silt mineralogy, can be obtained from the primary data characterization of the soil. The reports can be seen on-screen or downloaded in text files by giving primary country, state, and county details \cite{Survey}.

\subsection{Pesticide Use in Agriculture}
The primary use of pesticides in agriculture is for controlling weeds, insect infestations, and fungus. However, excessive use of pesticides can destroy other microorganisms necessary for soil health and degrade the quality of groundwater. The \ac{USGS} collects data for the amount of pesticide used in the U.S. on an annual basis in the form of tables, graphs, and maps \cite{SurveyUSGS}. The map provides a more refined picture of estimated pesticide use on agricultural land in terms of pounds per square mile, and the graphs show the estimated usage in millions of pounds for each crop every year. 

\subsection{Water Use in Agriculture}
Water is essential for agriculture. Both surface and groundwater are crucial and are utilized in farming \cite{Survey2015a}. Surface water is formed from natural rivers and lakes; groundwater is found under the earth's surface between rock, soil, and sand cracks. The \ac{USGS} collects total water usage every five years and updates the statistics in billions of gallons per day. The data shows that water use is higher in agricultural areas, including irrigation, livestock, and aquaculture \cite{Survey2015}.

\subsection{Groundwater Nitrate Contamination}
\begin{figure}[htbp]
	\centering
	\includegraphics[width=0.90\textwidth]{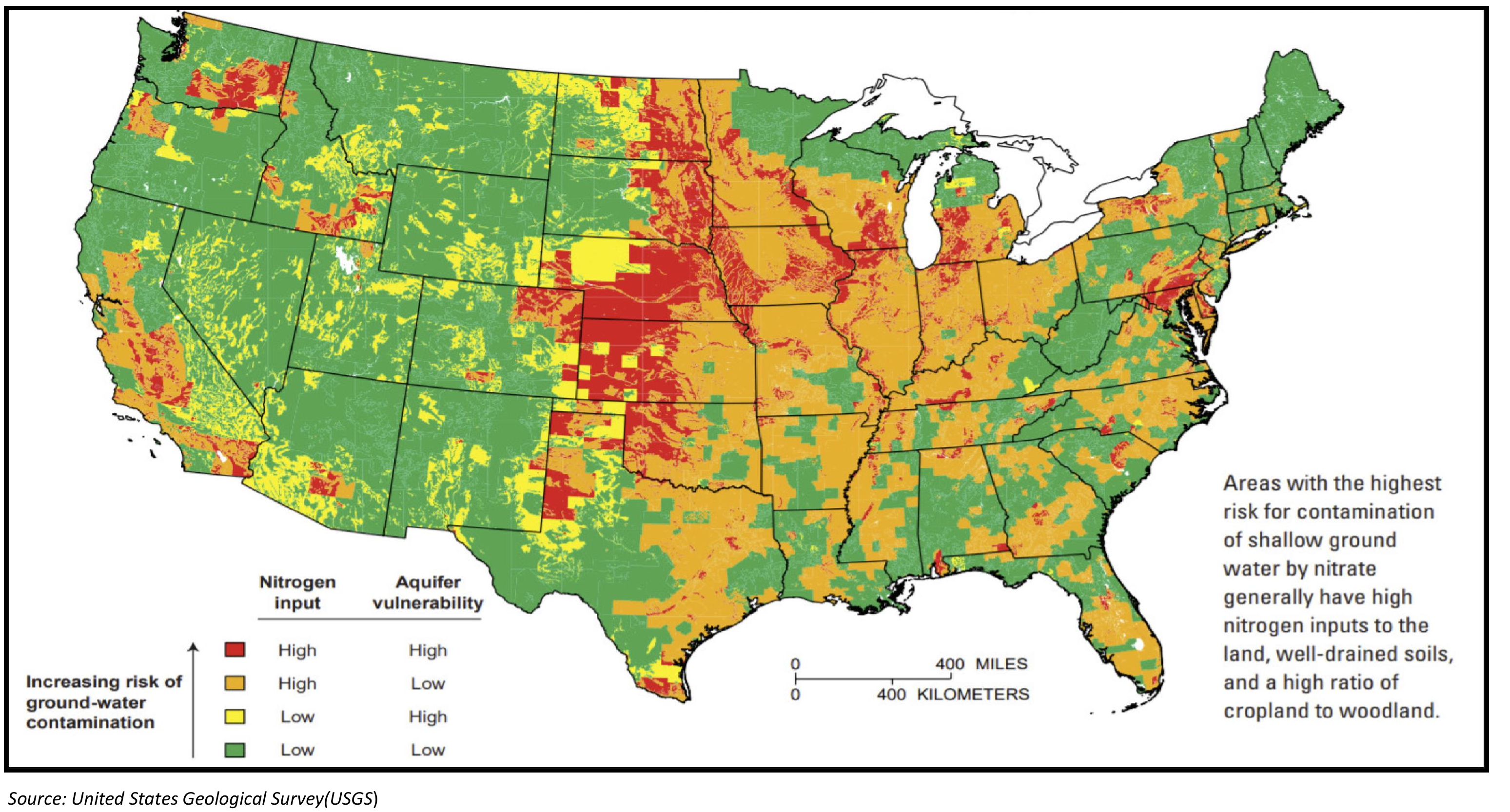}
	\caption{Groundwater Contamination \cite{Surveya}.}
	\label{fig:Water}
\end{figure}

Nitrate is the primary source for the growth of plants and crops. It is an oxidized form of nitrogen, which occurs naturally in the earth, but it can dissipate due to extensive farming. To refill the soil with essential nutrients, nitrogen fertilizers are applied while farming. Nonetheless, these nitrates can be toxic primarily when they enter food, groundwater, and surface water. Fig. \ref{fig:Water} shows the map for contamination of groundwater all over the United States. Collecting data nationwide, the \ac{USGS} has developed a model for estimating groundwater nitrate contamination \cite{Surveya}.

\subsection{Disaster Analysis}

Agriculture is facing threats from uncertain risks and changing landscapes and temperatures. It is necessary to forecast disasters before they occur to know the intensity of these hazards so farmers can be prepared for the worst and plan accordingly. The \ac{USDA} and \ac{NASS} have implemented a research study for disaster analysis assessments in near real-time. To collect the datasets, geospatial techniques and sensors are used in the procedure to estimate the disasters \cite{Agriculture2021a}. One of the example studies for monitoring flooding with the help of Sentinel-1, Synthetic Aperture Radar is given in \cite{8519458}.

%%%%%%%%%%%%%%%%%%%%%%%%%%%%%%%%%%%%%%%%%%%%%%%%%%%
\section{Smart Agriculture Open Research Problems}
\label{Sec:Research_Problems}

%https://ec.europa.eu/info/news/industry-50-towards-more-sustainable-resilient-and-human-centric-industry-2021-jan-07_en

In this section we discuss the open research problems of \textit{Agriculture $4.0$} and \textit{Agriculture $5.0$}. We can divide them into two main sub groups depending on the research focus. 

%https://www.mdpi.com/2076-3417/11/13/5911/htm#B1-applsci-11-05911
\subsection{Technology Perspective}
\begin{figure}[htbp]
	\centering
	\includegraphics[width=0.98\textwidth]{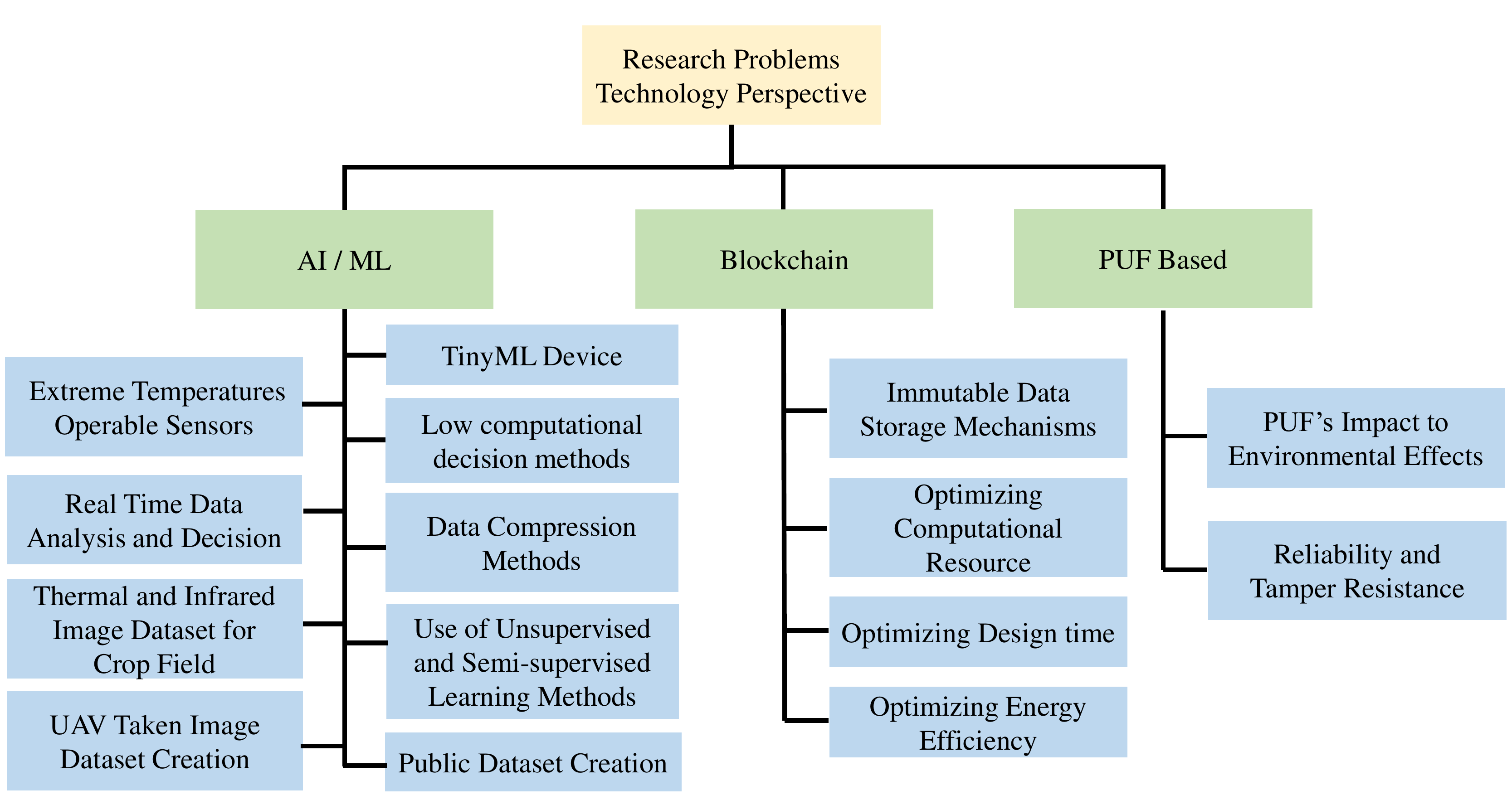}
	\caption{Network and Communication Challenges on Smart Agriculture.}
	\label{fig:tech_probs}
\end{figure}

Smart agriculture faces various challenges as previously mentioned. These challenges need to be addressed by adapting new and existing technologies. Until now most of the smart agriculture \ac{AI} models were cloud based, cloud-edge based, or cloud-fog-edge based. Hardware advancement has boosted the computing paradigm shift. The addition of intelligence to \ac{IoT} devices is the new trend \cite{Signoretti2021Evolving}. Network availability, latency and bandwidth are not anymore barriers in successful, seamless agriculture system operations. This opens up a new avenue for research. Edge AI in smart agriculture is a broad area which will be a hot topic in the near future. Fig. \ref{fig:tech_probs} shows various open research problems in a technology context. Research in the following fields holds much promise:

\begin{itemize}
	\item Low powered and solar powered, low latency TinyML devices.
	\item Low computational decision methods suitable for low powered \ac{IoT} devices.
	\item Sensor technologies operable in extreme temperatures.
	\item Data analytic methods for data compression.
	\item  Quantization and pruning techniques for \ac{AI}/\ac{ML} models.
	\item Unsupervised and semi-supervised learning methods. 
	\item Real time data analysis and decision.
	\item Public dataset creation with sensor data.
	\item \ac{UAV} taken image dataset. 
	\item Thermal and Infrared image dataset for crop field. 
\end{itemize}    

Research areas are not only limited to these. Blockchain based data privacy and integrity and service based smart agriculture applications are other areas to work with:
\begin{itemize}
	\item Blockchain enhanced IoT applications focusing on immutable data storage mechanisms.
	\item Optimizing computational resource, design time, and energy efficiency.
\end{itemize} 

Hardware security is another broad area of research for sustainable Smart Agriculture. The functionality and applications of each \ac{IoT} device in agriculture is unique. Research on \ac{PUF}, which is a hardware fingerprint \cite{Joshi,Labrado} is an important area of research: %developed using the unique internal wiring delays, threshold voltages and other manufacturing variations of each transistor in an Integrated circuit,    and develop a unique response cryptographic key for each and every device . 
\begin{itemize}
	\item PUF's susceptibility to environmental effects like  rainfall, pesticides, fertilizers, and chemicals.
	\item Reliability and tamper resistance of these hardware security modules.
\end{itemize}

\subsection{Network Perspective}

The network component is a very important aspect of smart agriculture which makes use of different \ac{ICT} to interconnect remote devices and make data transfer possible. Budding stage unsecured network layer protocols for limited resource IoT devices have led to various security threats. A classification of research problems which need to be addressed is given in Fig. \ref{fig:NetworkChallenges}.:

\begin{figure}[htbp]
	\centering
	\includegraphics[width=0.98\textwidth]{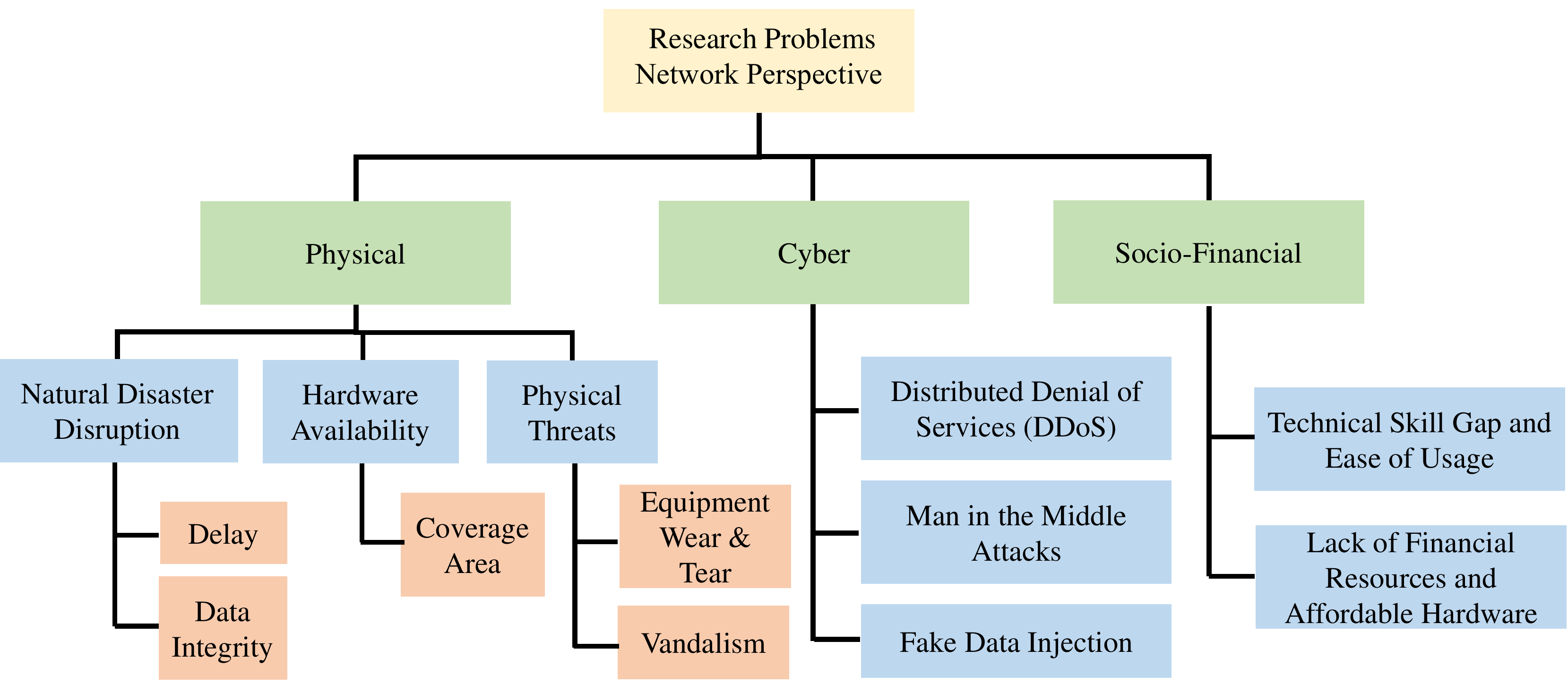}
	\caption{Network and Communication Challenges of Smart Agriculture.}
	\label{fig:NetworkChallenges}
\end{figure}

\begin{itemize}
	\item Providing alternative networking paths which can operate during natural disasters.
	\item Techniques to increase the real-time data operations even when the network is experiencing congestion due to high volumes of transactions.
	\item Robust and resource efficient techniques are still needed to manage data privacy and security challenges. 
	\item Efficient network topologies are needed to maximize usage of the available hardware and increase coverage area to avoid blind spots.
	\item Cost-efficient methods for easy maintenance of network equipment with minimal wear and tear can be challenging. 
	\item Preventive techniques to address physical damages like vandalism by adversaries are much needed. 
	\item Proper routing techniques in the network to avoid network threats like \ac{DDoS} can be an area of interest to work.
	\item Efficient encryption techniques and authentication mechanisms such as hardware assisted authentication are very much needed to be included in network layer to avoid different security threats.
	\item Ease of use and troubleshooting mechanisms can be areas of interest for researchers as this technology is developed for farmers. 
	\item Network equipment is expensive, thus making networking hardware affordable can make the technology more adopted into vast applications in Smart Agriculture.  
\end{itemize}

\section{Conclusions and Future Directions}
\label{Sec:Conclusions}

In today's world, we value more than ever  ``Let food be thy medicine'' % and medicine be thy food” 
%as said by the father of medicine, Greek physician Hippocrates. 
as quality food boosts our immunity. Research on agriculture, food security, and food supply chain has become more relevant. This article provides a detailed survey on the ongoing research trends in smart agriculture. It discusses recent technology trends to challenges and open research problems in this field. The authors believe this work will give an overall idea on technologies, challenges and research problems in smart agriculture. 

Technological advancement along with rapid growth of \ac{ICT} have transformed traditional agriculture to a smart, intelligent, automated agriculture. Smart agriculture reduces the carbon footprint by introducing sustainable, green farming, reducing the use of pesticides and fertilizers, and optimizing the use of natural resources. % it will address some other problems like climate change, disease like cancer. 
Soon the agricultural industry will welcome Agriculture 5.0 \cite{saiz2020smart}. This will raise yields while keeping the system environmentally sustainable. Developing countries will also follow the same trend as developed countries. Humanity will embrace the production and distribution of food in an economically and ecologically efficient  way as never before \cite{fraser2019agriculture}.

%%%%%%%%%%%%%%%%%%%%%%%%%%%%%%%%%%%%%%%
%\input{acronyms.tex}
\section*{List of Acronyms}
\label{sec:acronyms}
\begin{acronym}[NASS]
	\acro{A-CPS}{Agricultural Cyber-Physical Systems}
	\acro{ANN}{Arificial Neural Networks}
	\acro{AI}{Artificial Intelligence}
	\acro{BD}{Big Data}
	\acro{BRT}{Boosted Regression Trees}
	\acro{CNN}{Convolutional Neural Networks}
	\acro{CFS}{Correlation-based Feature Selection}
	\acro{Crop-CASMA}{Crop Condition and Soil Moisture Analytics}
	\acro{CPS}{Cyber-Physical Systems}
	\acro{DDoS}{Distributed Denial-of-Service}
	\acro{DLT}{Distributed Ledger Technology}
	\acro{DNN}{Deep Neural Networks}
	\acro{DoS}{Denial-of-Service}
	\acro{EDC}{Edge Data Centers}
	\acro{EPCIS}{Electronic Product Code Information Services}
	\acro{ERP}{Enterprise Resource Planning}
	\acro{FL}{Fuzzy Logic}
	\acro{GPRS}{Ground Penetrating Radar Services}
	\acro{GPS}{Global Positioning System}
	\acro{GRU}{Gated Recurrent Unit}
	\acro{H-CPS}{Healthcare Cyber-Physical Systems}
	\acro{ICT}{Information and Communication Technologies}
	\acro{IIoT}{Industrial Internet of Things}
	\acro{IoAT}{Internet of Agro-Things}
	\acro{IoMT}{Internet of Medical Things}
	\acro{IoT}{Internet of Things}
	\acro{IPFS}{Interplanetary File System}
	\acro{LPWAN}{Low-Power Wide Area Network}
	\acro{LSTM}{Long Short-Term Memory}
	\acro{LTE}{Long-Term Evolution}
	\acro{M2M}{Machine-to-Machine}
	\acro{MAC}{Media Access Control}
	\acro{ML}{Machine Learning}
	\acro{MLP}{Multi-Layer Perceptrons}
	\acro{NB-IoT}{Narrowband IoT}
	\acro{NASS}{National Agricultural Statistics Service}
	\acro{NCSS}{National Cooperative Soil Survey}
	\acro{NFC}{Near Field Communication}
	\acro{NVDI}{Normalized Difference Vegetation Index}
	\acro{P2P}{Point-to-Point}
	\acro{PBFT}{Practical Byzantine Fault Tolerance}
	\acro{PoS}{Proof-of-Stake}
	\acro{PoW}{Proof-of-Work}
	\acro{PUF}{Physical Unclonable Functions}
	\acro{RFID}{Radio Frequency Identification}
	\acro{RMSE}{Root Mean Square Error}
	\acro{RNN}{Recurrent Neural Network}
	\acro{RPN}{Region Proposal Network}
	\acro{SDN}{Software-Defined Networking}
	\acro{SIL}{Solar Insecticidal Lamps}
	\acro{SPoF}{Single Point-of-Failure}
	\acro{SSD}{Single Seed Descent}
	\acro{SVM}{Support Vector Machines}
	\acro{UAV}{Unmanned Aerial Vehicles}
	\acro{USDA}{U.S. Department of Agriculture}
	\acro{USGS}{U.S. Geological Survey}
	\acro{WSN}{Wireless Sensor Network}
\end{acronym}

%\balance
%%%%%%%%%%%%%%%%%%%%%%%%%%%%%%%%%%%%%%%%%%%%%%%%%%
\bibliographystyle{unsrtnat}
%\bibliography{Bibliography_Smart-Agriculture-Survey}
%\bibliography{arXiv_2022_Smart_Agriculture_Survey_12-06}

%%%%%%%%%%%%%%%%%%%%%%%%%%%%%%%%%%%%%%%%%%%%%%%%%%

\newpage
%\vspace{3cm}
\section*{Authors}

%\begin{biography}
%	{{\includegraphics[width=25mm,height=32mm,clip,keepaspectratio]{AMitra}}}
%\end{biography}

%\vspace{0.5cm}	

\begin{wrapfigure}[11]{l}{0.2\linewidth} 
	\vspace{-\baselineskip}
	\centering
	\includegraphics[height=1.6in]{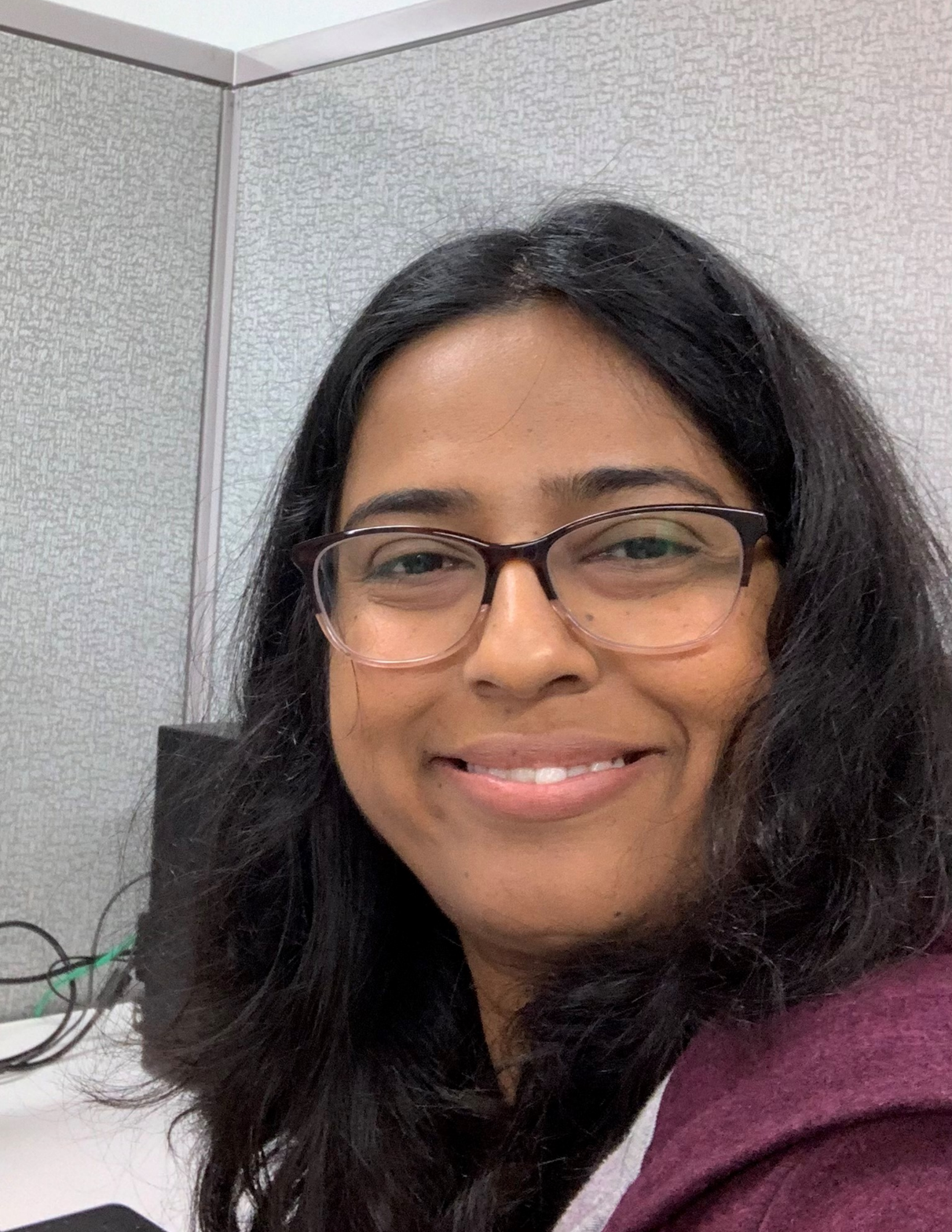}
\end{wrapfigure}
\textbf{Alakananda Mitra} received a Bachelor of Science (Honors) in Physics from Presidency College, University of Calcutta in 2001 and a Bachelor of Technology and Master of Technology in Radiophysics and Electronics from the Institute of Radiophysics and Electronics, University of Calcutta in 2004 and 2006, respectively. She is currently a doctoral student in the research group at Smart Electronics Systems Laboratory (SESL) in the Department of Computer Science and Engineering at the University of North Texas, Denton, USA. Along with her course work, she also works as a Teaching Assistant in the department. She has worked as a Project Linked Personnel at Advanced Computing and Microelectronics Unit in Indian Statistical Institute from 2006 to 2007. Her research interests include artificial intelligence, machine learning, deep learning, edge AI,  and application of AI/ML approaches in multi-media forensics, smart agriculture, and smart healthcare. 

\vspace{2cm} 
\begin{wrapfigure}[11]{l}{0.2\linewidth} 
	%	\vspace{-\baselineskip}
	\centering
	\includegraphics[height=1.6in]{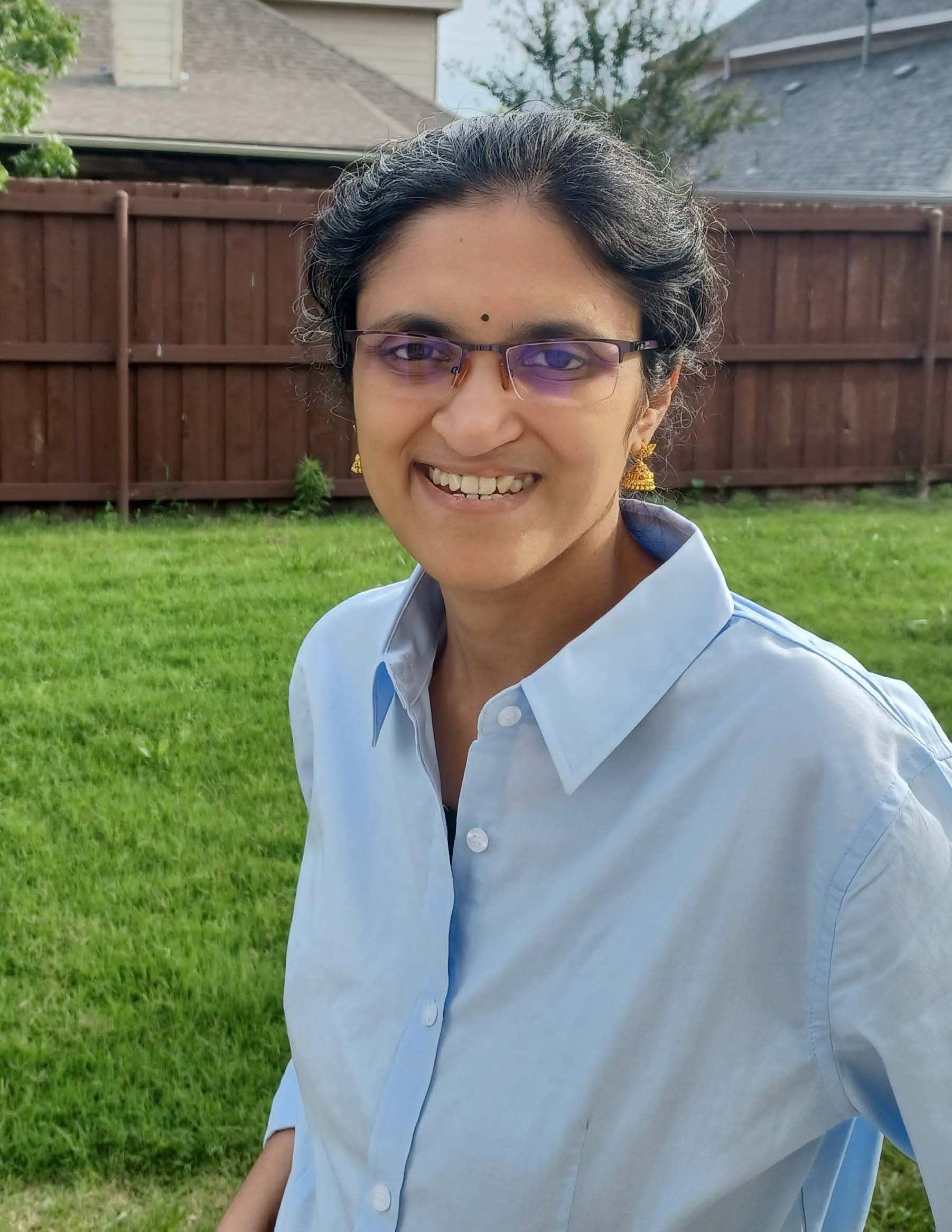}
\end{wrapfigure}
\textbf{Sukrutha L. T. Vangipuram} received a Master of Technology in Computer Science, Jawaharlal Nehru Technological University, Hyderabad in 2012 and Bachelor of Engineering in Information Technology, Osmania University, Hyderabad in 2007. Currently, she is enrolled as a doctoral student in the research group at Smart Electronics Systems Laboratory (SESL) at Computer Science and Engineering at the University of North Texas, Denton, USA. She has worked as an Assistant Professor in the Computer Science Engineering Department, Methodist College of Engineering and Technology, Hyderabad from 2012 -2015 and as a Teaching Assistant in the Information Technology Department in Swamy Vivekananda Institute of Technology Hyderabad, India, from 2007 -2009. Her research interests are Web Programming Services, Services Oriented Architecture, Cloud Computing, Application of Blockchain in Health Care, and Smart Agriculture.

\vspace{2cm} 
\begin{wrapfigure}[7]{l}{0.2\linewidth} 
	\vspace{-\baselineskip}
	\includegraphics[height=1.2in]{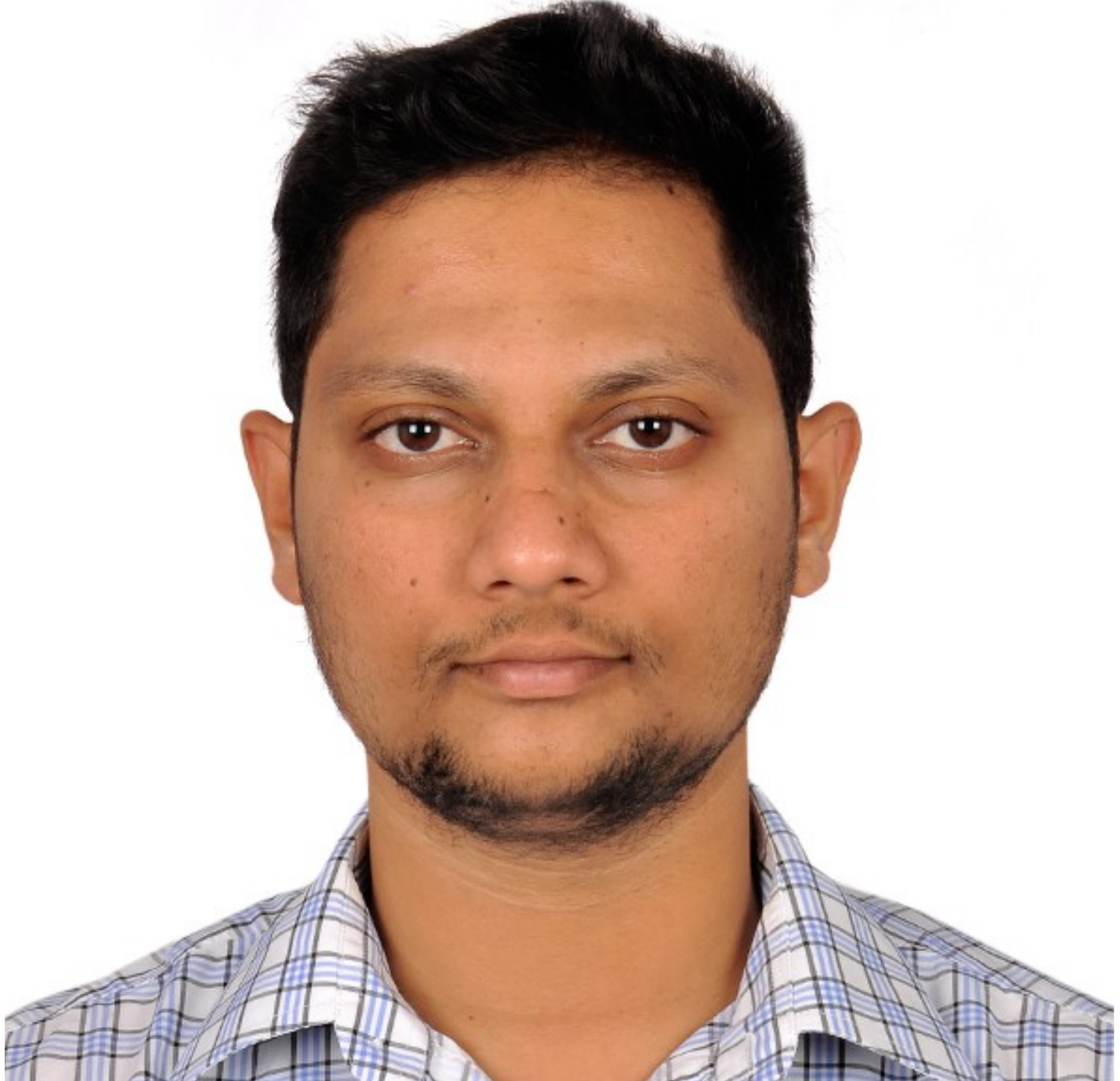}
\end{wrapfigure} 
\textbf{Anand Kumar Bapatla} received a Bachelor’s of Technology (B. Tech) in Electronics and Communication from Gayatri Vidya Parishad College of Engineering, Visakhapatnam, India, in 2014 and an MSCE degree in 2019 from the University of North Texas, Denton, USA. He is currently a Ph.D. candidate in the research group at Smart Electronics Systems Laboratory (SESL) at Computer Science and Engineering at the University of North Texas, Denton, TX. His research interests include smart healthcare and Blockchain applications in Internet of Things (IoT).

\vspace{2cm}
\begin{wrapfigure}[6]{l}{0.2\linewidth}
	\centering
	\vspace{-\baselineskip} 
	\includegraphics[height=1.4in]{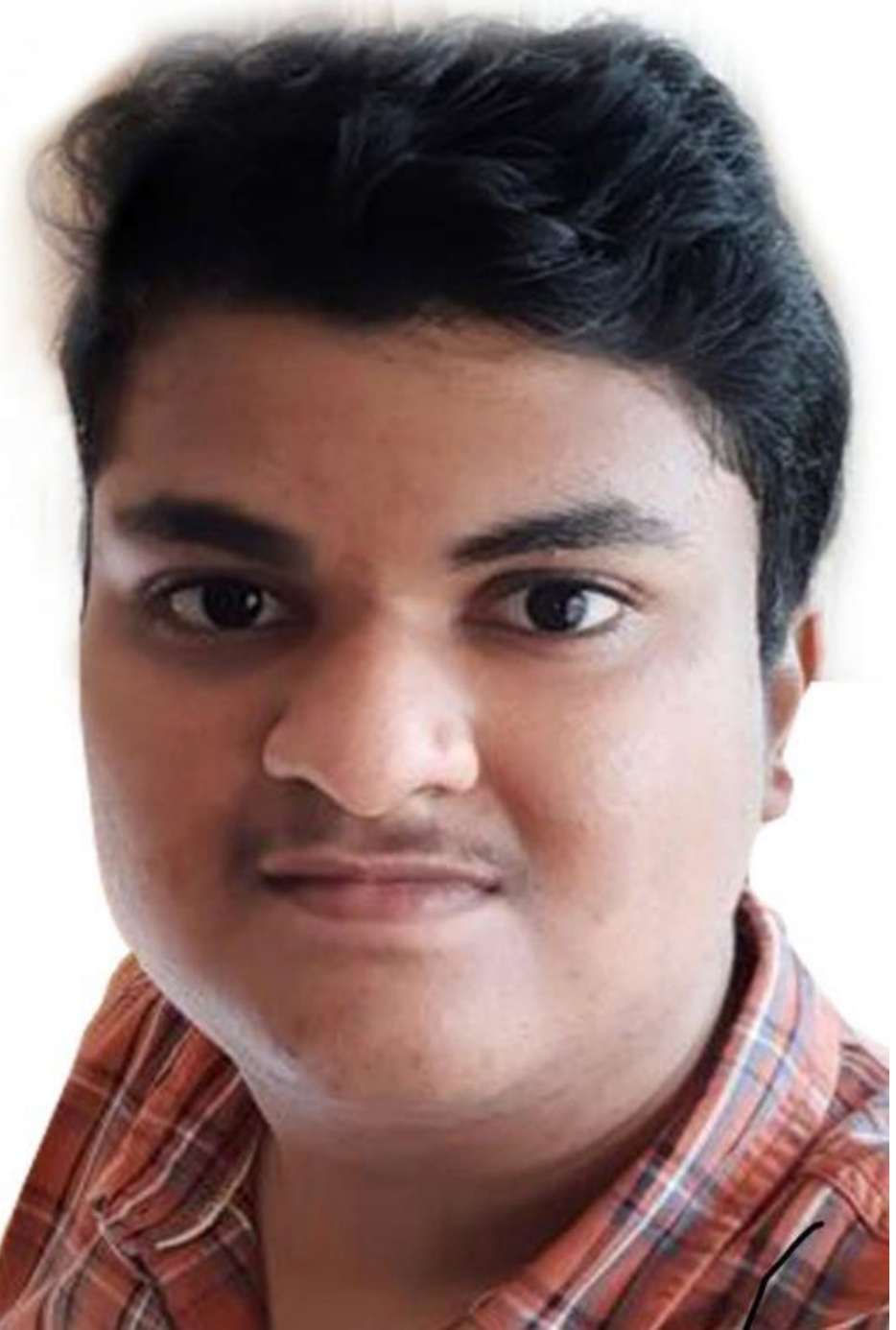}
\end{wrapfigure}
\textbf{Venkata K. V. V. Bathalapalli}
received B.Tech. degree in Electronics and Communication Engineering from Sri Venkateswara 
University, Tirupati, India, in 2020. He is currently pursuing Ph.D. program in Computer Science
and Engineering at Smart Electronics Systems Laboratory, University of North Texas, Denton, TX, USA. 
His research interests are in the areas of Hardware Assisted Security and Blockchain based IoT Device Security 
for Smart Healthcare and Smart Agriculture. 

\vspace{3cm}
\begin{wrapfigure}{l}{0.18\linewidth} 
	\centering
	\includegraphics[height=1.6in]{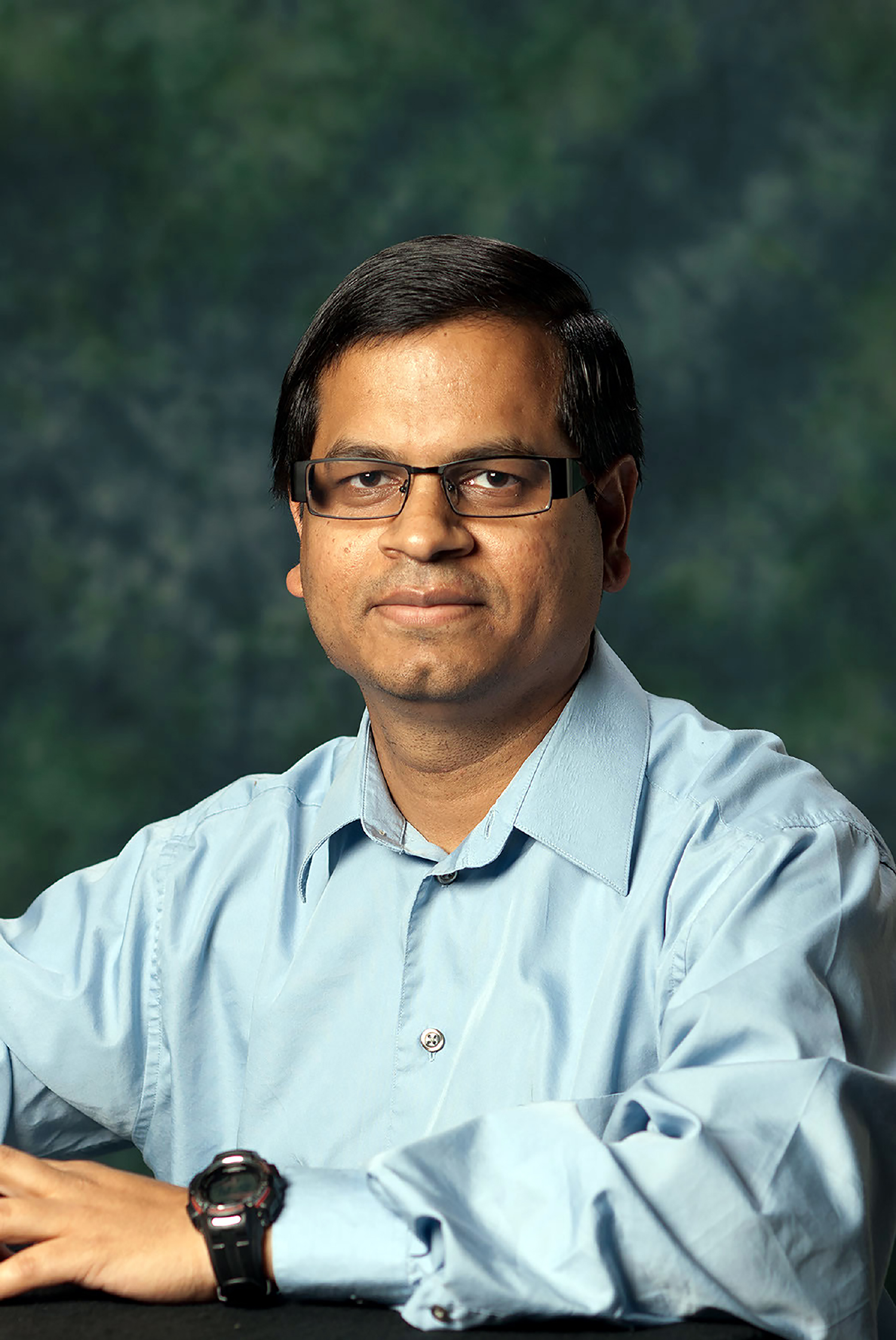}
\end{wrapfigure}
\textbf {Saraju P. Mohanty} received the bachelor’s degree (Honors) in electrical engineering from the Orissa University of Agriculture and Technology, Bhubaneswar, in 1995, the master’s degree in Systems Science and Automation from the Indian Institute of Science, Bengaluru, in 1999, and the Ph.D. degree in Computer Science and Engineering from the University of South Florida, Tampa, in 2003. He is a Professor with the University of North Texas. His research is in “Smart Electronic Systems” which has been funded by National Science Foundations (NSF), Semiconductor Research Corporation (SRC), U.S. Air Force, IUSSTF, and Mission Innovation. He has authored 400 research articles, 4 books, and 7 granted and pending patents. His Google Scholar h-index is 45 and i10-index is 180 with 8500 citations. He is regarded as a visionary researcher on Smart Cities technology in which his research deals with security and energy aware, and AI/ML-integrated smart components. He introduced the Secure Digital Camera (SDC) in 2004 with built-in security features designed using Hardware-Assisted Security (HAS) or Security by Design (SbD) principle. He is widely credited as the designer for the first digital watermarking chip in 2004 and first the low-power digital watermarking chip in 2006. He is a recipient of 13 best paper awards, Fulbright Specialist Award in 2020, IEEE Consumer Electronics Society Outstanding Service Award in 2020, the IEEE-CS-TCVLSI Distinguished Leadership Award in 2018, and the PROSE Award for Best Textbook in Physical Sciences and Mathematics category in 2016. He has delivered 15 keynotes and served on 13 panels at various International Conferences. He has been serving on the editorial board of several peer-reviewed international transactions/journals, including IEEE Transactions on Big Data (TBD), IEEE Transactions on Computer-Aided Design of Integrated Circuits and Systems (TCAD), IEEE Transactions on Consumer Electronics (TCE), and ACM Journal on Emerging Technologies in Computing Systems (JETC). He has been the Editor-in-Chief (EiC) of the IEEE Consumer Electronics Magazine (MCE) during 2016-2021. He served as the Chair of Technical Committee on Very Large Scale Integration (TCVLSI), IEEE Computer Society (IEEE-CS) during 2014-2018 and on the Board of Governors of the IEEE Consumer Technology Society during 2019-2021. He serves on the steering, organizing, and program committees of several international conferences. He is the founding steering committee chair/vice-chair for the IEEE International Symposium on Smart Electronic Systems (IEEE-iSES), the IEEE-CS Symposium on VLSI (ISVLSI), and the OITS International Conference on Information Technology (OCIT). He has mentored 2 post-doctoral researchers, and supervised 13 Ph.D. dissertations, 26 M.S. theses, and 11 undergraduate projects.

\vspace{1.5cm}
\begin{wrapfigure}[10]{l}{0.2\linewidth} 
	\vspace{-\baselineskip}
	\includegraphics[height=1.6in,keepaspectratio]{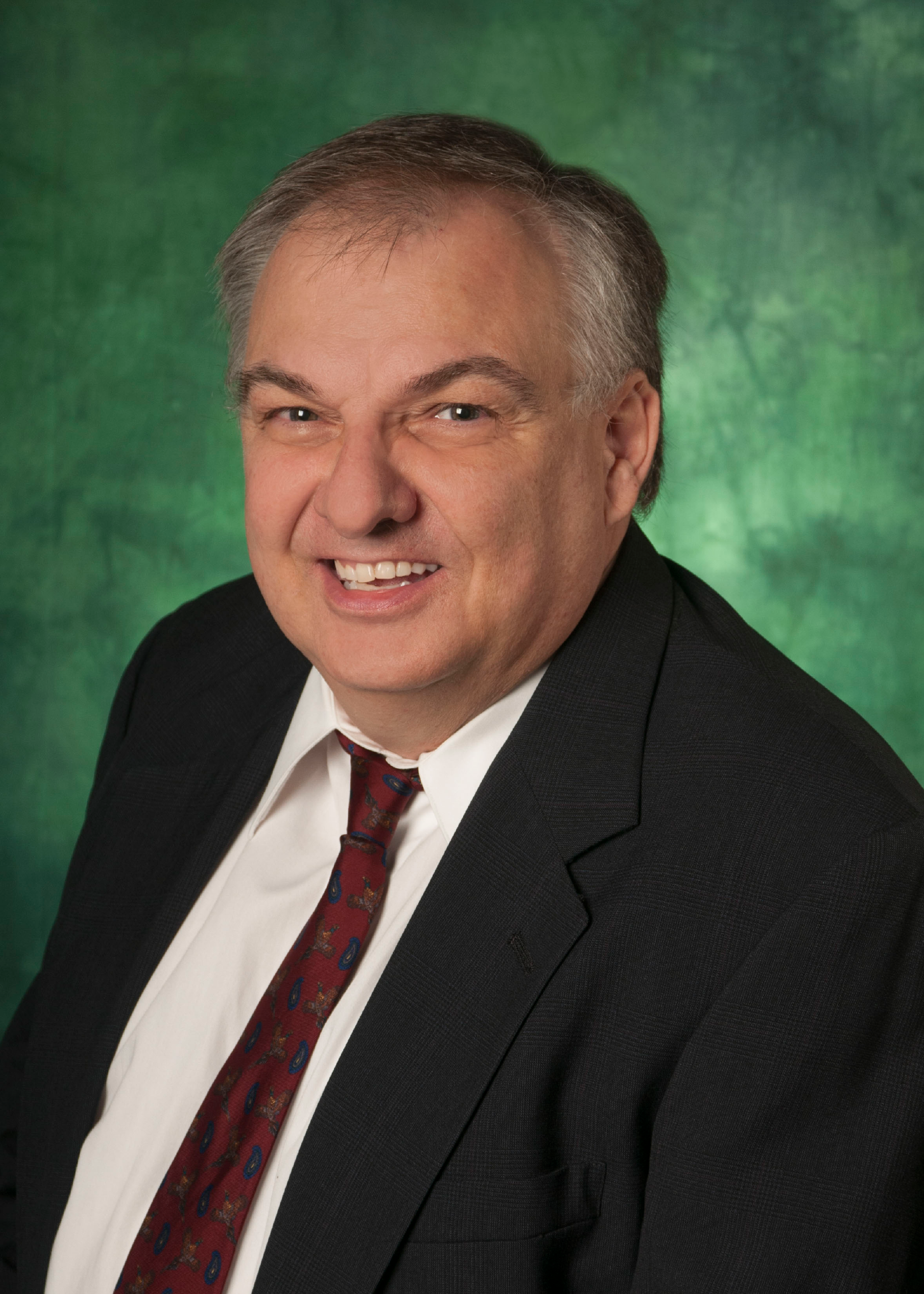}
\end{wrapfigure}
\textbf {Elias Kougianos} received a BSEE from the University of Patras, Greece in 1985 and an MSEE in 1987, an MS in Physics in 1988 and a Ph.D. in EE in 1997, all from Louisiana State University. From 1988 through 1998 he was with Texas Instruments, Inc., in Houston and Dallas, TX. In 1998 he joined Avant! Corp. (now Synopsys) in Phoenix, AZ as a Senior Applications engineer and in 2000 he joined Cadence Design Systems, Inc., in Dallas, TX as a Senior Architect in Analog/Mixed-Signal Custom IC design. He has been at UNT since 2004. He is a Professor in the Department of Electrical Engineering, at the University of North Texas (UNT), Denton, TX. His research interests are in the area of Analog/Mixed-Signal/RF IC design and simulation and in the development of VLSI architectures for multimedia applications. He is an author of over 200 peer-reviewed journal and conference publications.

\vspace{1.5cm}
\begin{wrapfigure}{l}{0.2\linewidth} 
	\vspace{-\baselineskip}
	\includegraphics[height=1.6in,keepaspectratio]{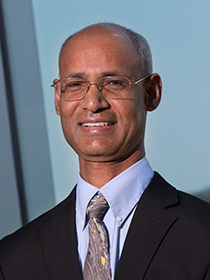}
\end{wrapfigure}
\textbf{Chittaranjan Ray} is a Professor of Civil and Environmental Engineering at the University of Nebraska-Lincoln, and Director of the Nebraska Water Center at University of Nebraska. He has extensive experience in many facets of managing both water quantity and water quality issues, particularly in the areas of chemical and pathogen impacts on ground water quality; flow and transport processes in the vadose zone, technologies for low-cost water supply, and the agriculture-water/energy nexus. He previously served as the interim director of the Water Resources Research Center at University of Hawaii-Manoa. Ray also was Director of the university’s Environmental Center and as Chief Environmental Engineer for the Applied Research Laboratory, a U.S. Navy sponsored facility at University of Hawaii. He has held positions in industry and at the Illinois State Water Survey.

%\end{description}
\end{document}